\documentclass[acmlarge,nonacm]{acmart}
\usepackage{tikz}
\usepackage{amsmath}
\usepackage{lipsum} 
\usepackage{adjustbox}
\usepackage{todonotes}
\usepackage{caption, subcaption}
\usepackage[table,xcdraw]{xcolor}
\usepackage{soul}
\usepackage{tabularx}
\usepackage{booktabs}
\usepackage{longtable}
\usepackage{pdflscape}
\usepackage{float}
\usepackage[ruled, linesnumbered]{algorithm2e}
\SetKwInOut{AlgoInit}{Initialization}
\usepackage{url}
\usepackage{hyperref}
\usepackage{listings}
\usepackage{enumitem}
\usepackage{booktabs}

\newcommand{\nsf}[1]{\href{https://www.nsf.gov/awardsearch/showAward?AWD_ID=#1}{#1}}

\definecolor{paleplum}{rgb}{0.8, 0.6, 0.8}

\lstdefinestyle{mystyle}{
    keywordstyle=\color{purple},
    numberstyle=\color{gray},
    basicstyle=\ttfamily,            
    captionpos=b,                        
    numbers=left,                    
    numbersep=5pt,         
}

\graphicspath{{sections/figures/}}

\begin{document}

\title{Fault Injection Attacks on Q-Learning:\\ Analysis, Vulnerabilities, and Countermeasures}

\author{Tom Slooff}
\affiliation{%
  \institution{Università della Svizzera italiana}
  \city{Lugano}
  \country{Switzerland}
}
\email{tom.slooff@usi.ch}

\author{Anthony Etim}
\affiliation{%
  \institution{Yale University}
  \city{New Haven}
  \state{Connecticut}
  \country{USA}
}
\email{anthony.etim@yale.edu}

\author{Jiaqi Yu}
\affiliation{%
  \institution{Northwestern University}
  \city{Evanston}
  \state{Illinois}
  \country{USA}
}
\email{jiaqi@u.northwestern.edu}

\author{Francesco Regazzoni}
\affiliation{%
  \institution{University of Amsterdam}
  \city{Amsterdam}
  \country{The Netherlands}
}
\affiliation{%
  \institution{Università della Svizzera italiana}
  \city{Lugano}
  \country{Switzerland}
}
\email{f.regazzoni@uva.nl}\email{francesco.regazzoni@usi.ch}

\author{Jakub Szefer}
\affiliation{%
  \institution{Northwestern University}
  \city{Evanston}
  \state{Illinois}
  \country{USA}
}
\email{jakub.szefer@northwestern.edu}

\renewcommand{\shortauthors}{Slooff et al.}

\begin{abstract}
Hardware fault injection attacks threaten machine learning systems, yet their impact on reinforcement learning (RL) has received little attention despite the use of RL in safety-critical applications. This paper presents a security analysis of Q-Learning implementations under voltage and clock glitching attacks on a ChipWhisperer platform, evaluating five variants: Standard, Robust, Resilient, Linear, and Deep Q-Learning. In standard Q-Table implementations, faults corrupt reward computation, action selection, and Q-value updates during learning, and change the selected action during inference. Robust Q-Learning does not prevent these attacks: NaN and infinite values injected during learning persist in the Q-table, and its greedy policy uses the same code as standard Q-Learning. This paper also evaluates a Resilient Q-Learning variant that applies triple modular redundancy with majority voting. It corrects corruption of a single Q-table copy, but inputs shared by the copies, such as the reward, and the voting code itself remain outside its protection. The analysis extends to Linear and Deep Q-Learning, where faults corrupt stored weights during learning and change actions during inference. Moving beyond benchmark environments, this work evaluates an online traffic light controller in a microscopic traffic simulator, where a single injected infinite value degrades accumulated vehicle waiting time by $148\times$ while corrupting under $4\%$ of the Q-table, and where triple modular redundancy corrected the faults injected in our tests, although glitching could still halt the device. Altogether, this work provides a basis for understanding and securing RL deployments against fault injection attacks.
\end{abstract}

\maketitle

\pagestyle{plain}

\section{Introduction}
\label{sec:intro}

Hardware fault injection attacks have long been studied for their detrimental effects on the security and reliability of computing systems~\cite{mishra2025modern}. In recent years, machine learning (ML) implementations have become increasingly prominent targets of such attacks~\cite{ahmadilivani2024}. Demonstrations of bit flip attacks~\cite{Rakin_2019_ICCV} and laser injection attacks~\cite{breier2018practical} have shown that fault injection can significantly degrade model accuracy, cause unpredictable behavior, and enable retrieval of sensitive model parameters~\cite{hector2023}. These vulnerabilities pose serious risks when ML systems perform classification or decision-making tasks in safety-critical applications such as robotics and autonomous driving.

However, the impact of fault injection on reinforcement learning (RL) systems has received little attention, despite RL's growing use in autonomous systems and critical infrastructure. Most of this work targets supervised models~\cite{xu2021security,ahmadilivani2024,Rakin_2019_ICCV,hector2023}. Prior studies of hardware faults in RL include reliability analyses of learning-based navigation systems~\cite{wan2021analyzing,wan2022frlfi}, which evaluate fault tolerance rather than deliberate attacks. Unlike supervised learning models that remain static after training, RL systems rely on continual learning and real-time decision-making, potentially making them more sensitive to hardware faults.

Reinforcement learning has gained widespread adoption in robotics, autonomous driving, and complex decision-making applications. Q-learning and its variants remain foundational RL methods for automated decision-making~\cite{kober2013rlinrobotics,ELALLID20227366}. Q-Table-based Q-learning provides an exact state-action value representation, which makes it useful for understanding learning dynamics. Robust Q-learning targets performance under environmental uncertainty and is not designed to defend against deliberate hardware attacks. Resilient Q-learning, as evaluated in this work, adds triple modular redundancy to tolerate hardware faults. Linear Q-learning enables scalability to larger state spaces while maintaining analytical tractability. Deep Q-learning extends applicability to high-dimensional environments through neural network approximation. These Q-learning variants represent different design trade-offs and collectively form the focus of this security analysis.
This gap in understanding how deliberate hardware faults affect RL performance and security motivates the present work.

This paper analyzes the effects of hardware fault injection attacks on RL implementations, focusing on online Q-learning variants. The evaluation uses the Frozen Lake environment from the Gymnasium library~\cite{towers2024gymnasium}, a standard benchmark for RL testing, and an online traffic-signal controller in the SUMO microscopic traffic simulator~\cite{SUMO2018}. This work demonstrates the vulnerabilities of standard RL implementations to hardware faults and provides a basis for developing countermeasures.

\subsection{Contributions}

This work makes the following contributions toward understanding RL vulnerabilities to both voltage and clock glitching fault injection attacks:

\begin{enumerate}
    \setlength{\itemsep}{0pt}
    \setlength{\parskip}{0pt}
    \item Analysis of fault injection attacks on the reward computation, action selection, and learning updates of Q-Table-based Q-learning, as well as its computations during inference, showing that each is vulnerable to at least one of voltage or clock glitching.
    \item Evaluation of Robust Q-learning showing that algorithmic robustness to environmental uncertainty does not prevent targeted hardware attacks.
    \item Evaluation of Resilient Q-learning using triple modular redundancy, showing that majority voting corrects corruption of a single Q-table copy, while inputs shared by the copies and the voting code itself remain outside its protection.
    \item Analysis of Linear Q-learning revealing similar vulnerabilities to tabular methods, with additional challenges in identifying consistent fault models due to increased computational complexity.
    \item Evaluation of Deep Q-learning showing that faults during inference transiently change the selected action without updating stored weights, and that faults during training can corrupt the stored weights.
    \item Evaluation beyond benchmark environments: an online traffic light controller in a microscopic traffic simulator, where a single injected infinite value degrades accumulated vehicle waiting time by $148\times$, where our searches found no voltage-glitch configuration that reproducibly forces an action in the two-action greedy policy, and where triple modular redundancy corrected the faults injected in our tests, although glitching could still halt the device.
\end{enumerate}

\section{Background}
\label{sec:bg}

This section provides background on fault injection attacks using voltage and clock glitching, focusing on attacks targeting machine learning algorithms. Table~\ref{tab_prior_work} in the Appendix summarizes existing work on fault injection attacks against different ML algorithms; none of the listed prior works targets RL.

\subsection{Voltage Glitching Attacks}
\label{sec_background_voltage_glitching}

Voltage glitching attacks introduce brief, controlled disturbances in a device's power supply to induce faults in hardware or software execution. Unlike electromagnetic interference~\cite{EM_fault_example}, clock glitching~\cite{clock_glitch}, or laser-induced faults~\cite{breier2018practical}, voltage glitching exploits transient power variations to cause instruction skips, data corruption, or security mechanism bypasses~\cite{chen2021voltpillager}. These attacks have been demonstrated against cryptographic systems, exposing secret keys and corrupting security protocols~\cite{barenghi2012fault}, and fault injection more broadly has been demonstrated against ML models~\cite{ahmadilivani2024,Rakin_2019_ICCV,breier2018practical,hector2023,xu2021security}.

\subsection{Clock Glitching Attacks}
\label{sec_background_clock_glitching}

Clock glitching attacks manipulate a device's clock signal by introducing timing violations that cause instruction skips, incorrect operations, or system crashes~\cite{clock_glitch}. Attackers can exploit these timing faults to manipulate control flow~\cite{tehranipoor2023clock}, extract sensitive information, or compromise encryption keys~\cite{lee2025clock}. While clock glitching requires access to the clock distribution network, making it more difficult to deploy than voltage glitching, it provides more predictable and consistent fault effects by directly manipulating operation timing rather than data values~\cite{yuce2018fault}.

\subsection{Existing Fault Injection Attacks on ML}

Fault injection attacks target ML systems by introducing deliberate faults during training or inference to produce incorrect predictions or classifications. These attacks exploit vulnerabilities in ML model implementations with potential consequences in safety-critical applications such as autonomous vehicles and medical diagnostics.

Remote attacks include Rowhammer-induced bit flips, which have been used to manipulate model weights in DeepHammer and DeepVenom~\cite{yao2020deephammer, yao2024deepvenom}, demonstrating that small numbers of bit flips can drastically reduce model accuracy or implant backdoors. Terminal Brain Damage~\cite{hong2019terminal} explored the effects of random bit corruptions on neural network weights, biases, and activations. Dynamic voltage and frequency scaling (DVFS) has been exploited to create transient faults in GPU-based ML accelerators~\cite{sun2023lightning}. Training-assisted bit-flip attacks~\cite{dong2023one} have demonstrated that bit flips during training can create models easily convertible to malicious variants at deployment.

Physical attacks include clock glitches targeting convolutional layer operations~\cite{liu2020imperceptible}, fault injection in quantized neural network accelerators~\cite{khoshavi2020fiji}, laser fault injection disrupting activation functions~\cite{breier2018practical}, and fault injection degrading model accuracy and causing misclassifications~\cite{hou2021physical,etim2026fault}. These studies focus primarily on supervised learning models. For RL, hardware faults have been studied from a reliability perspective, for example in learning-based navigation systems~\cite{wan2021analyzing,wan2022frlfi}, but to our knowledge deliberate voltage and clock glitching of RL implementations has received little attention.
\section{Introduction to Reinforcement Learning}

\begin{figure}[t!]
    \centering
    \includegraphics[width=.5\linewidth]{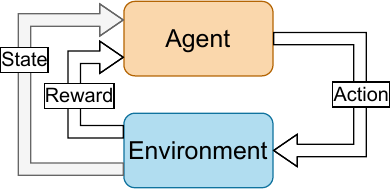}
    \caption{\small Typical Reinforcement Learning (RL) loop.}
    \label{fig:rl}
\end{figure}

\begin{table}[t]
\centering
\caption{\small Comparison of Q-learning variants used in this work.}
\label{tab_qlearning_comparison}
\small
\centering
\begin{tabular}{l|c|c}
\toprule
\textbf{Approach} & \textbf{Storage} & \textbf{Built-in Protection} \\
\midrule
Q-Table & Table & None \\
Robust Q-Learning & Table & Environmental noise \\
Resilient Q-Learning & 3$\times$ Table & Single-copy Q-table faults \\
Linear Q-Learning & Weights & None \\
Deep Q-Learning & NN Weights & None \\
\bottomrule
\end{tabular}
\end{table}

Reinforcement learning (RL) is a machine learning paradigm where an agent learns to maximize cumulative reward through interactions with an environment, as shown in Figure~\ref{fig:rl}. Unlike supervised learning, the agent is not provided with optimal actions but must discover effective strategies through trial and error. This makes RL suitable for applications where optimal behavior is unknown or where collecting labeled training data is infeasible. In particular, RL has been applied in robotics~\cite{kober2013rlinrobotics}, autonomous driving~\cite{ELALLID20227366}, traffic signal control~\cite{NOAEEN2022116830}, neural architecture search~\cite{JAAFRA201957}, and large language models~\cite{bai2022traininghelpfulharmlessassistant}.

RL typically uses a \textit{greedy policy} that selects the action that maximizes the expected cumulative reward in the current state, defined as $a^* = \arg\max_a Q(s,a)$, where $Q(s,a)$ represents the expected total reward from taking action $a$ in state $s$ and following the policy thereafter. During training, agents typically use an $\epsilon$-greedy policy that balances exploration (random actions with probability $\epsilon$) and exploitation (greedy actions with probability $1-\epsilon$). During inference, a pure greedy policy ($\epsilon = 0$) is used to always select the action with maximum expected reward, directly deriving the optimal action from learned Q-values without exploration. This work evaluates the Q-Learning variants summarized in Table~\ref{tab_qlearning_comparison} and described below.

\subsection{Q-Table-based Q-Learning}

Tabular, or Q-Table-based, Q-learning stores exact state-action values in a table and can be efficiently implemented on resource-constrained devices~\cite{robust-qlearning, tlc_nonstationarity}. After taking action $a$ in state $s$, receiving reward $r$, and reaching state $s'$, the agent updates the table as
\begin{equation}\label{eq:q_update}
Q(s,a) \leftarrow (1-\alpha)\,Q(s,a) + \alpha\left(r + \gamma \max_{a'} Q(s',a')\right),
\end{equation}
where $\alpha$ is the learning rate and $\gamma$ the discount factor. Hardware implementations of tabular Q-learning have been demonstrated on embedded systems~\cite{8937555}, and hybrid offline-online approaches have been proposed~\cite{pmlr-v139-mitchell21a}. This work implements standard Q-Table-based Q-learning as a baseline for~evaluation.

\subsection{Robust Q-Learning}
\label{sec:robust_rl}

Prior work has addressed environmental noise and uncertainty in RL through robust RL~\cite{6788393,pmlr-v97-tessler19a}, fault-tolerant control~\cite{JIANG201260,electronics9091513, patton2021fault, AMIN201958}, and adversarial RL~\cite{pmlr-v70-pinto17a,8794293}. These approaches target environmental noise, actuator faults, and sensor errors, but do not consider deliberate hardware attacks.

Robust Q-learning~\cite{robust-qlearning} optimizes for the worst case over an uncertainty set of transition models, constructed with an $R$-contamination model. The original algorithm minimizes cost, using $V_t(s)=\min_{a} Q_t(s,a)$ with a maximization over states as the worst case. This work implements the reward-based form, to evaluate whether robustness to environmental noise provides security against targeted hardware fault attacks. At each learning step the state values are recomputed as $V(s)=\max_{a} Q(s,a)$ for all states, and the update is
\begin{equation}\label{eq:robust_update}
Q(s,a) \leftarrow (1-\alpha)\,Q(s,a) + \alpha\left(r + \gamma R \min_{\tilde{s}} V(\tilde{s}) + \gamma (1-R)\, V(s')\right),
\end{equation}
where $R$ is the robustness parameter. The worst case enters through the minimization over all states, so the update depends on the state values of the whole table rather than only on the next state.

\subsection{Resilient Q-Learning}
\label{sec:resilient_rl}
This work also evaluates a Resilient Q-learning variant that applies triple modular redundancy~\cite{lyons1962tmr} to the Q-table. The Q-table is replicated three times, and majority voting is applied during action selection and when updating modified entries, which prevents a single corrupted copy from changing the policy. This is a classical redundancy-based fault tolerance technique~\cite{lyons1962tmr,JIANG201260,electronics9091513, patton2021fault, AMIN201958} and is not specific to RL; we use it as a baseline countermeasure.

\subsection{Linear Q-Learning}
\label{sec_linear_q_learning_rl}

Linear Q-learning~\cite{melo2007q} approximates state-action values using a weighted sum of features rather than maintaining explicit table entries. Each state-action pair is represented as a feature vector with learned weights, enabling generalization across similar states and reducing memory requirements. The algorithm updates weights based on temporal difference errors between predicted and observed values. This work implements Linear Q-learning with 16, 8, and 4 features to evaluate how function approximation affects vulnerability to fault injection attacks explored in this work.

\subsection{Deep Q-Learning}
\label{sec:deep_q_learning}

Deep Q-learning replaces tabular value storage with a neural network that approximates the action-value function~\cite{mnih2013playing,mnih2015human}. The network takes state representations as input and outputs estimated values for all actions. This work implements a compact Deep Q-learning agent where the discrete observation space is encoded as a one-hot vector and processed through a small fully connected network with one hidden layer and four output neurons. This design fits within the memory and computational constraints of the fault injection platform while capturing the generalization properties of deep value functions. The same $\epsilon$-greedy exploration used in tabular and linear implementations is applied, but updates now adjust network weights rather than table entries or linear coefficients.

\section{Threat Model}
\label{sec:threat}
\begin{figure}[t]
    \centering
    \includegraphics[width=0.75\linewidth]{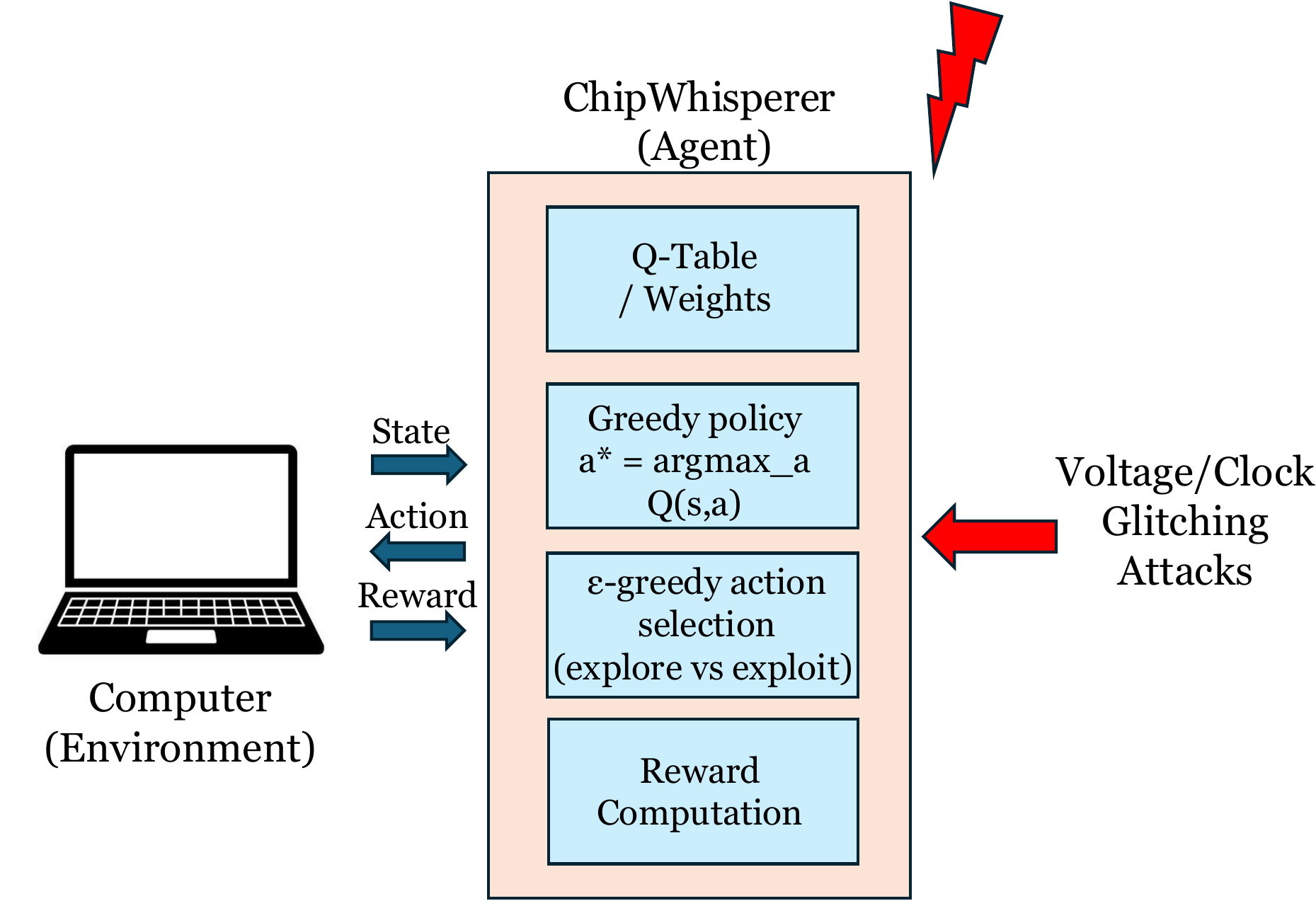}
    \caption{\small Fault injection attack evaluation setup, threat model and types of fault attacks considered in this work, overlaid on the RL loop to demonstrate where and how the attacker can affect RL. Note that for some Q-learning variant implementations the reward is sent from the environment, while for others the agent computes the reward~itself. The agent runs on the target microcontroller that the ChipWhisperer controls.}
    \label{fig:fault-setup}
\end{figure}

This work evaluates physical fault injection attacks using voltage and clock glitching techniques. The threat model assumes an attacker with physical access to the device, a plausible scenario for robotics, autonomous vehicles, and other deployed systems where physical security cannot be guaranteed~\cite{szefer2018principles}. Figure~\ref{fig:fault-setup} illustrates the threat model, showing how fault attacks map onto components of the RL loop.

\textbf{Attack Platform.} Experiments use the ChipWhisperer~\cite{o2014chipwhisperer} platform, specifically the ChipWhisperer Husky, to inject voltage faults and clock glitches with precise timing and parameter control. The RL agent executes on a microcontroller under ChipWhisperer control, enabling systematic evaluation of fault effects across different timing and fault parameters.

\textbf{Attacker Knowledge.} We assume the attacker knows that online RL is deployed on the target device and may know the specific RL variant (Q-Table-based, Linear, Deep Q-Learning, etc.). The attacker can identify and target specific functions within the implementation, such as reward computation, action selection, or value updates. This assumes white-box knowledge of the implementation but does not require knowledge of learned parameters.

\textbf{Triggering and Timing.} In our experiments the firmware raises a trigger signal around the targeted function, and each glitch is timed as a number of clock cycles after this trigger (External Offset). This gives a level of synchronization that an unmodified device would not provide. A real attacker would need to derive timing from another observable event, such as communication with the agent, and to profile an identical device in advance. 

\textbf{Number of Faults.} Our voltage glitching experiments apply one glitch event, which can contain several pulses (Repeat), per targeted execution. An attacker who corrupts several Q-table copies of the Resilient implementation in the same update is not evaluated.

\textbf{Attack Goals.} The attacker aims to disrupt the agent's control policy through one of two strategies: (a) injecting faults during training to corrupt the learning process and produce suboptimal policies, or (b) injecting faults during inference to manipulate action selection and directly control agent behavior.

\textbf{Attack Targets During Learning.}
The attacker can target the following components during online learning:

\begin{itemize}
\setlength{\itemsep}{0pt}
\setlength{\parskip}{0pt}
    \item \textit{Reward Computation}: Faults in reward calculation corrupt the training signal, which can cause the agent to learn incorrect state-action values and converge to suboptimal policies.
    \item \textit{Action Selection}: Faults in the exploration-exploitation strategy (e.g., $\epsilon$-greedy) bias which state-action pairs are explored, potentially preventing discovery of optimal behaviors.
    \item \textit{Learning Step}: Faults during Q-value updates directly corrupt stored values (Q-Table entries, linear weights, or neural network parameters), which can persistently degrade the learned policy.
\end{itemize}

\textbf{Attack Targets During Inference.}
During deployment, the attacker targets:

\begin{itemize}
\setlength{\itemsep}{0pt}
\setlength{\parskip}{0pt}
    \item \textit{Greedy Policy Evaluation}: Faults during action selection cause the agent to deviate from the optimal policy. A greedy policy selects the action maximizing expected reward, $a^* = \arg\max_a Q(s,a)$. Faults can force incorrect actions without corrupting stored values, enabling transient control over agent behavior.
\end{itemize}
\section{Evaluation Setup}
\label{sec:exp}

All RL implementations are written in C and were developed for this work following standard descriptions of each algorithm; the Deep Q-Learning agent uses the AIfES framework~\cite{wulfert2024aifes} for its neural network primitives. Experiments use a ChipWhisperer Husky\footnote{NAE-CWHUSKY - \url{https://www.newae.com/products/nae-cwhusky}} with the CW312T-SAM4S\footnote{CW312T-SAM4S - \url{https://chipwhisperer.readthedocs.io/en/latest/chipwhisperer-target-cw308t/CW312T_ATSAM4S/README.html}} target board. The C implementations include all core RL algorithm functionality plus convenience functions for setting and retrieving Q-Table values (for tabular methods, including standard, Robust, and Resilient Q-Learning) or weights (for Linear and Deep Q-Learning). These convenience functions enable rapid agent initialization and targeted fault injection on specific functions. The ChipWhisperer Python library configures the hardware, programs the target, and manages communication between the host computer and target device. RL environments execute on a separate computer and communicate with the target using the SimpleSerial~protocol.

Figure~\ref{fig:fault-setup} shows the experimental setup. The agent executes on the target device while the environment runs on the host computer. For tabular implementations, reward calculation occurs on the agent device, requiring only state and action to be communicated between devices. For Linear Q-Learning,
state features and rewards are sent to the agent. The Deep Q-Learning firmware implements the environment on the device for training, and the host sends the state for inference. Voltage and clock glitching faults are injected during agent execution.

\subsection{Fault Parameters and Search Space}

Each fault injection experiment searches over the following parameters to identify effective attacks:

\begin{itemize}
\setlength{\itemsep}{0pt}
\setlength{\parskip}{0pt}
    \item \textbf{Width}: Duration of a single glitch pulse, in ChipWhisperer phase steps.
    \item \textbf{Offset}: Offset within a clock cycle from the rising edge, in phase steps.
    \item \textbf{External Offset}: Number of clock cycles from trigger to glitch.
    \item \textbf{Repeat}: Number of glitch pulses (voltage faults only).
\end{itemize}

\begin{table}[t!]
\caption{\small Fault parameter search space in evaluation setup. Width and Offset are in phase steps and External Offset in clock cycles. Each experiment searched within these ranges, often over narrower bounds and with its own step size.}
    \centering
    \setlength{\tabcolsep}{6pt}
    \small
    \renewcommand{\arraystretch}{1.2}
    \begin{tabular}{r|c|c|c}
        \toprule
        \textbf{Parameter} & \textbf{Minimum} & \textbf{Maximum} & \textbf{Step Size} \\ 
        \midrule
        Width            & 0   & 4400 & 1--200 \\
        Offset           & 0   & 4400 & 1--200 \\
        External Offset  & 0   & -  & 1--200 \\
        Repeat           & 1   & 12   & 1   \\
        \bottomrule
    \end{tabular}
    \label{tab:fault_search_space}
\end{table}

Table~\ref{tab:fault_search_space} defines the overall fault parameter search space; each experiment searched within these ranges, often over narrower bounds. Optuna~\cite{optuna}, an optimization framework, searches this space to maximize a success metric. The ChipWhisperer divides each clock cycle into $4592$ phase steps, enabling sub-cycle glitch granularity. Width and Offset values in close proximity produce nearly identical fault effects, so each successful glitch is characterized by a specific External Offset, a specific Repeat value, and a range of Width and Offset values rather than a single precise pair. The maximum External Offset depends on the target function's execution time in clock cycles. Several searches use adaptive samplers and each experiment has its own fault criterion, so the fault counts reported in this paper describe what each search found; they are not comparable estimates of how vulnerable each implementation is. Clock glitching parameter spaces are shown in Appendix~\ref{sec:appendix_cglitch_params}.

\subsection{Frozen Lake Environment}

Experiments use the Frozen Lake environment\footnote{Frozen Lake - \url{https://gymnasium.farama.org/environments/toy_text/frozen_lake/}} from the Gymnasium library~\cite{towers2024gymnasium}. Figure~\ref{fig:fl_policy_normal} shows a $4\times4$ grid where the agent must navigate from the starting position (blue cell) to the goal (green cell) while avoiding holes (red cells). The discrete state space assigns a unique number to each cell. The action space consists of four actions: $\{\leftarrow, \downarrow, \rightarrow, \uparrow\}$. The agent receives a reward of 1 upon reaching the goal and 0 for all other transitions. Black arrows illustrate an optimal policy. Unless otherwise specified, experiments use the $4\times4$ Frozen Lake map with learning parameters $\alpha = 0.7$ and $\gamma = 0.9$.

\section{Faulting Standard Q-Learning}
\label{sec:fault_staandard_qlearning}
\begin{figure*}[t]
    \centering
    \begin{subfigure}[t]{0.24\textwidth}
        \centering
        \includegraphics[width=0.93\textwidth]{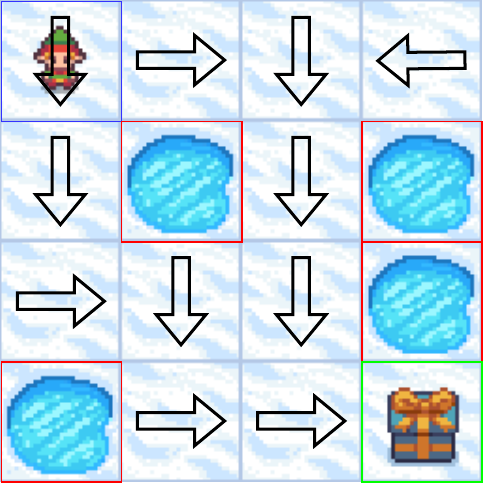}
        \caption{Normal Policy}
        \label{fig:fl_policy_normal}
    \end{subfigure}
    \hfill
    \begin{subfigure}[t]{0.24\textwidth}
        \centering
        \includegraphics[width=0.93\textwidth]{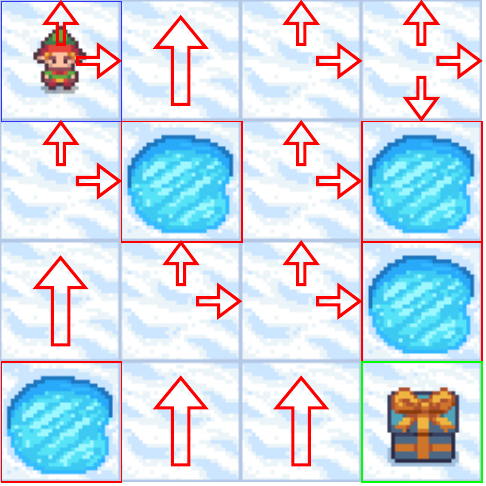}%
        \caption{Clock Glitching the \\Inference}
        \label{fig:fl_policy_cglitch_policy}
    \end{subfigure}
    \hfill
    \begin{subfigure}[t]{0.24\textwidth}
        \centering
        \includegraphics[width=0.93\textwidth]{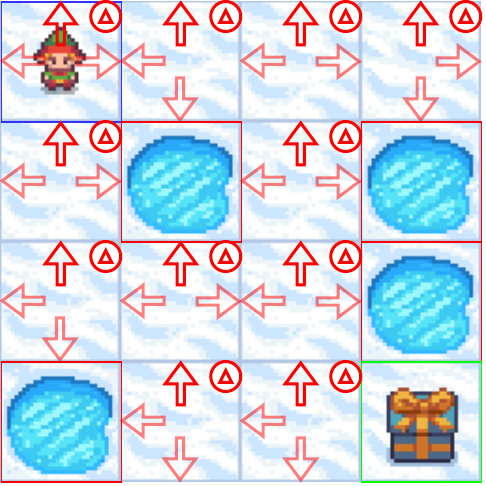}
        \caption{Voltage Glitching the Inference}
        \label{fig:fl_policy_vglitch_policy}
    \end{subfigure}
    \hfill
    \begin{subfigure}[t]{0.24\textwidth}
        \centering
        \includegraphics[width=0.93\textwidth]{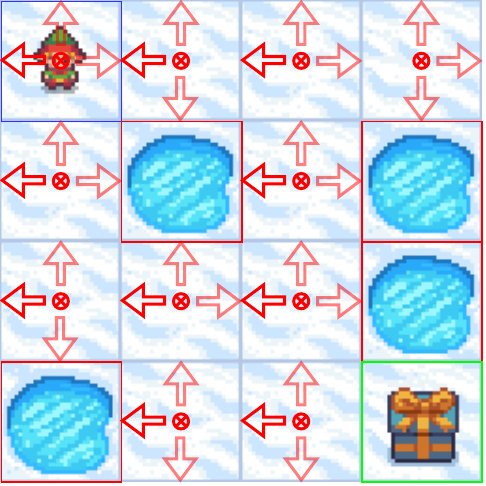}
        \caption{Voltage Glitching the Learning}
        \label{fig:fl_policy_vglitch_learn}
    \end{subfigure}
    \caption{Agent policy after glitching inference and learning components (standard Q-Learning implementation). The delta symbol indicates an out-of-bounds action, the cross indicates a NaN or Inf value.}
    \label{fig:fl_tab_policy_glitched}
\end{figure*}

This section evaluates fault injection attacks on all components of standard Q-Table-based Q-Learning implementations. Figure~\ref{fig:fl_tab_policy_glitched} illustrates the impact of successful attacks on learned policies and how the agent will behave after being corrupted by the faults (during learning or inference). The most common fault results are indicated by the opaque arrows.

\textbf{Attacks on Reward Computation During Learning.}
In Frozen Lake, rewards are represented by an array with value $1$ for the goal state and $0$ elsewhere. Recall that for tabular implementations, reward calculation occurs on the agent device. Learning relies on the direct array lookup during each transition. Faults are injected at this lookup step.
Some voltage fault configurations consistently produce altered reward values, most commonly yielding approximately $2.123$. Separate faults produce IEEE-754 $NaN$ values and large negative values ($<-1\text{e}37$), which propagate into the Q-Table through the update equation. The successful fault parameters are visualized in Figure~\ref{fig:appendix_reward_v_space} (Appendix~\ref{sec:appendix_vglitch_params}). These faults do not modify the stored reward array, so each fault only affects the update that uses the faulted value; however, the value written by that update stays in the Q-Table until later updates change it, and a $NaN$ does not recover through further updates.
Clock glitching produces no observable effect, consistent with the function's simplicity: reward computation consists solely of a direct memory load with no arithmetic or control-flow operations sensitive to timing violations. While clock glitching could in principle cause incorrect memory addressing, the single-cycle nature of this load operation provides insufficient timing margin for clock faults to take effect.

\textbf{Attacks on Action Selection During Learning.}
The $\epsilon$-greedy policy generates a random number in $[0,1]$ and compares it to epsilon. Values less than or equal to epsilon trigger random action selection; otherwise, the greedy policy determines the action. Experiments evaluate two settings: $\epsilon=1$ (always random) and $\epsilon=0$ (always greedy).
Voltage faults produce two main effects. Some faults maintain correct branch execution but disrupt the device random number generation, causing unexpected random actions. Others bypass the epsilon comparison and enter the wrong branch: the most consistent configuration we tested takes the greedy branch in 17 of 20 attempts at $\epsilon=1$ and the random branch in 2 of 20 attempts at $\epsilon=0$. Forcing the greedy branch would reduce exploration and could trap the agent in local minima during early training, although this effect on the learned policy is not evaluated here.
Clock glitching yields more consistent control. The implementation evaluates the random-action branch first and the greedy-action branch second, each containing a return instruction. Glitching either the branch or return instruction in the first branch forces the policy to always explore (random) or always exploit (greedy) in our tests.

\textbf{Attacks on Learning Step During Learning.}
Q-Table updates are evaluated by comparing faulted Q-Tables to reference Q-Tables. Voltage faults cause both overvaluation and undervaluation of state-action pairs, leading to suboptimal action selection. Several faults inject NaN or Inf values into the Q-Table. These values propagate through computations without causing crashes. Positive Inf values cause policies to prefer actions reaching the faulted state, while negative Inf has the opposite effect. NaN values cause all comparisons to return false, leading to action avoidance except for action 0 (LEFT), which serves as the baseline in greedy action selection. Positive Inf values, and NaN values in action 0, propagate to other Q-Table cells through subsequent updates, which degrades the policy in the states that reach them.

Clock glitching experiments initialize the Q-Table with the reference Q-Table (optimal learned values). Faults target the function calculating updated cell values. Clock glitching also produces a diverse set of faults (Figure~\ref{fig:c1_learning_step_distribution} in Appendix~\ref{sec:appendix_cglitch_params} shows successful clock glitch fault distributions for External Offsets up to 200 of the 749 cycles of the update). In this range External Offset is the key parameter determining fault effectiveness. With correct External Offset, a range of Width and Offset values produce successful faults, and for a single External Offset value the tested Width and Offset combinations yield the same glitched return value.
Overall, clock glitching manipulates Q-Table values in both positive and negative directions and can produce NaN or Inf values, with the same propagation effects as voltage glitching.

\textbf{Attacks on Greedy Policy During Inference.}
The greedy policy calculates the action with highest value for a given state. Figure~\ref{fig:fl_policy_normal} shows the reference policy for the $4\times4$ Frozen Lake environment. Multiple optimal policies exist, all guiding the agent along the shortest path to the goal.

Voltage glitching produces diverse effects on greedy policy evaluation (searched parameter space in Figure~\ref{fig:appendix_inference_v_space}, Appendix~\ref{sec:appendix_vglitch_params}). Several faults force specific action outputs. The most common target is the UP action: of the 37 fault configurations we examined in detail on states 0, 4, and 8, nine produce UP in 92--100\% of the outputs that did not crash, and three of these produce UP in all 20 attempts on each of these states. Other faults target different actions with lower consistency (40-60\%). When the targeted action is not output, faults may result in normal policy execution or device crashes.

Some voltage faults consistently return specific actions only in certain states, ranging from nearly all states to single states. Some faults return out-of-bounds actions ($a \notin [0,3]$). A significant portion of voltage faults exhibit inconsistent behavior, producing different actions in different states with mixed behaviors across the state space.

Clock glitching can change the output action through appropriate fault parameter selection. The available faulted output action depends on both fault parameters and the correct optimal action. Similar to learning step attacks, given correct External Offset, a wide range of Width and Offset values succeed in generating an attack. Also, similar to voltage attacks, in addition to faults that produce incorrect action, we observe faults that generate out-of-bounds action values of $a \notin [0,3]$.

\subsection{Key Findings}
Each Q-Table-based Q-Learning component we tested is vulnerable to at least one of the two glitching methods, during training and inference. Voltage glitching corrupts reward values ($\approx2.123$, NaN, $<-1\text{e}37$); the stored rewards are unchanged, so each fault affects one update, but the value written by that update stays in the Q-Table until later updates change it. Clock glitching had no observed effect on this memory load. Action selection attacks bypass $\epsilon$-greedy logic, with voltage faults taking the greedy branch in 17 of 20 attempts at $\epsilon=1$ and clock glitching forcing always-explore or always-exploit behavior. Learning step attacks write corrupted values directly into the table: both attack types inject NaN or Inf values, and positive Inf and NaN in action 0 propagate through subsequent updates (External Offset is the critical clock glitching parameter). Inference attacks change the selected action without corrupting stored values, with voltage glitching forcing specific actions on many or few states ($\uparrow$: 92--100\% of non-crashing outputs on the states examined in detail), and clock glitching selecting the faulted action through the External Offset. Faults in the learning step persist in the table, and in our experiments clock glitches produced more repeatable outcomes than voltage glitches.
\section{Faulting Robust Q-Learning}

\begin{figure*}[t!]
    \centering
    \begin{subfigure}[t]{0.24\textwidth}
        \centering
        \includegraphics[width=0.93\textwidth]{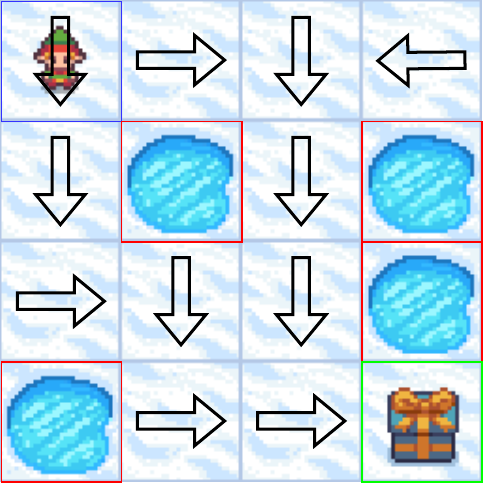}
        \caption{Normal Policy}
        \label{fig:fl_robust_policy_normal}
    \end{subfigure}
    \hfill
    \begin{subfigure}[t]{0.24\textwidth}
        \centering
        \includegraphics[width=0.93\textwidth]{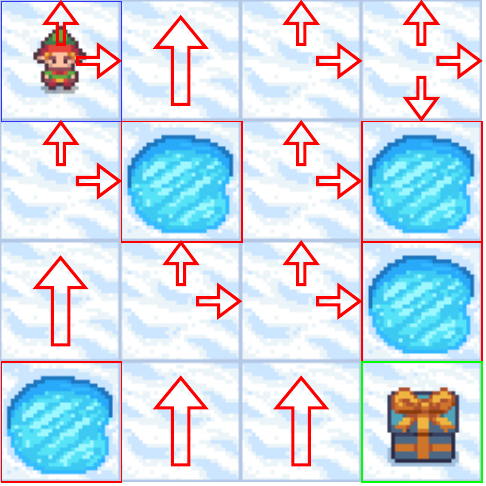}
        \caption{Clock Glitching the \\Inference}
        \label{fig:fl_robust_policy_cglitch}
    \end{subfigure}
    \hfill
    \begin{subfigure}[t]{0.24\textwidth}
        \centering
        \includegraphics[width=0.93\textwidth]{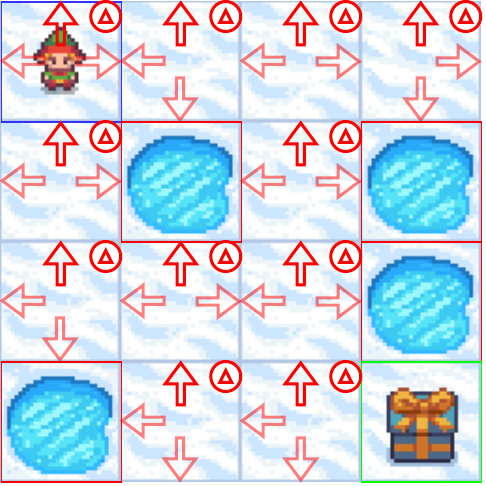}
        \caption{Voltage Glitching the Inference}
        \label{fig:fl_robust_policy_vglitch}
    \end{subfigure}
    \hfill
    \begin{subfigure}[t]{0.24\textwidth}
        \centering
        \includegraphics[width=0.93\textwidth]{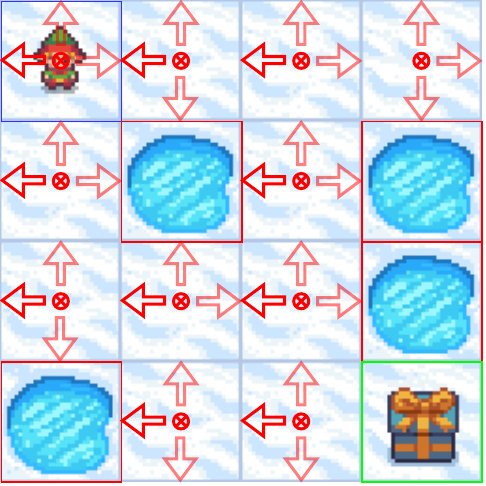}
        \caption{Voltage Glitching the Learning}
        \label{fig:fl_robust_policy_vglitch_learn}
    \end{subfigure}
    \caption{Agent policy examples after glitching inference and learning components (Robust Q-Learning implementation). The delta symbol indicates an out-of-bounds action, the cross indicates a NaN or Inf value.}
    \label{fig:fl_robust_policy_glitched}
\end{figure*}

This section evaluates Robust Q-Learning~\cite{robust-qlearning}, which extends standard Q-Learning with the state values $V(s)$ and the minimization over states in Equation~\ref{eq:robust_update} (Section~\ref{sec:robust_rl}) to handle environmental uncertainty. Experiments focus on the learning step, where this algorithmic modification occurs, and the greedy policy. The agent learns using $\alpha=0.8$, $\gamma=0.96$, and $R=0.2$ (robustness parameter); \cite{robust-qlearning} also evaluates $R=0.2$ on Frozen Lake. Figure~\ref{fig:fl_robust_policy_glitched} illustrates the impacts of the attacks during learning and inference.

\textbf{Attacks on Reward Computation During Learning.} Reward computation is implemented identically to standard Q-Learning. Experiments are not repeated; results from standard Q-Learning from prior section apply directly.

\textbf{Attacks on Action Selection During Learning.} Action selection is implemented identically to standard Q-Learning. Experiments are not repeated; results from standard Q-Learning from prior section apply directly.

\begin{figure}[t]
    \begin{subfigure}{\linewidth}
        \centering
        \includegraphics[width=0.7\linewidth]{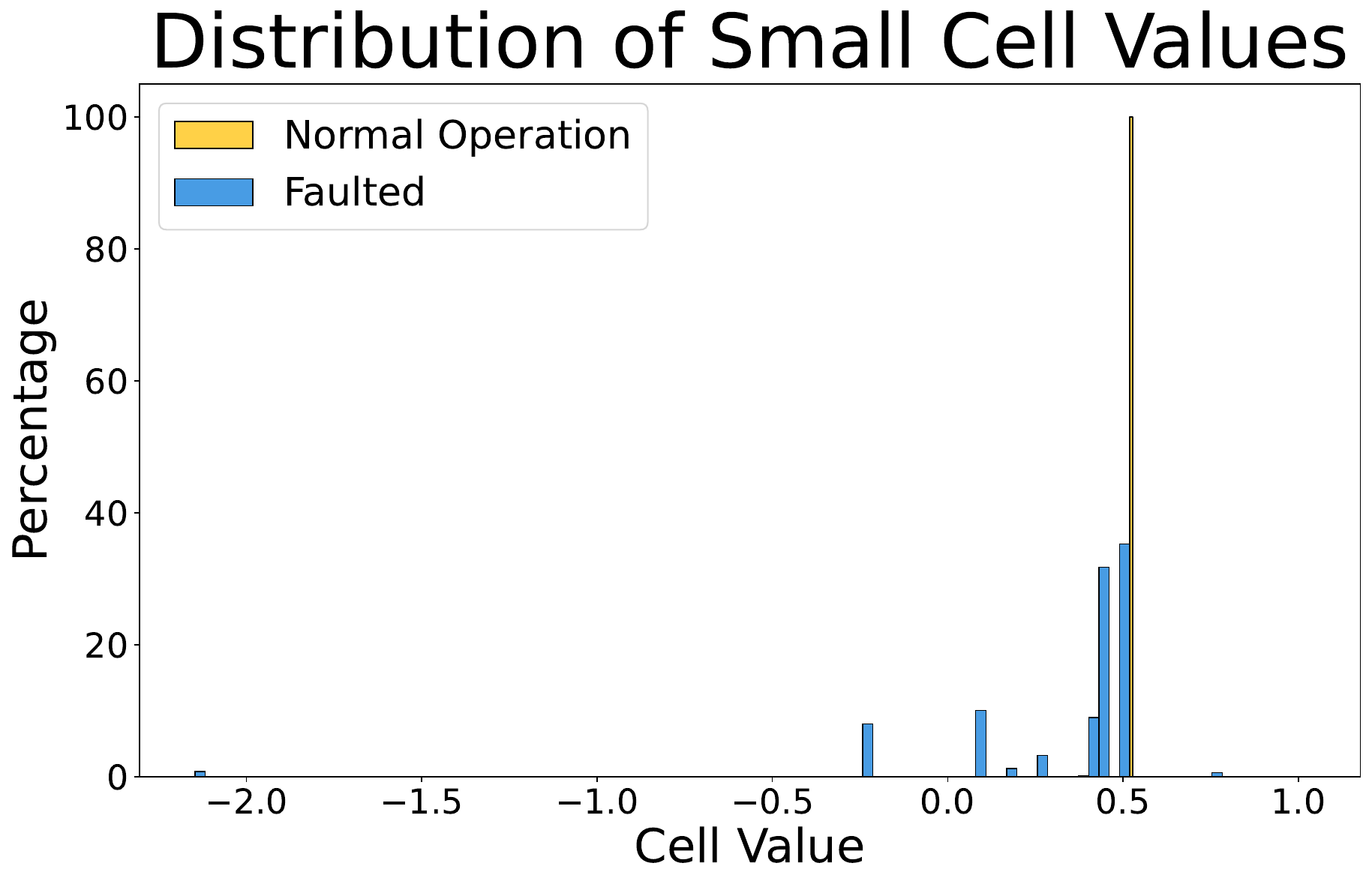}
        \caption{Small value range}
        \label{fig:fl_robust_update_dist_small}
    \end{subfigure}

    \begin{subfigure}{\linewidth}
        \centering
        \includegraphics[width=0.7\linewidth]{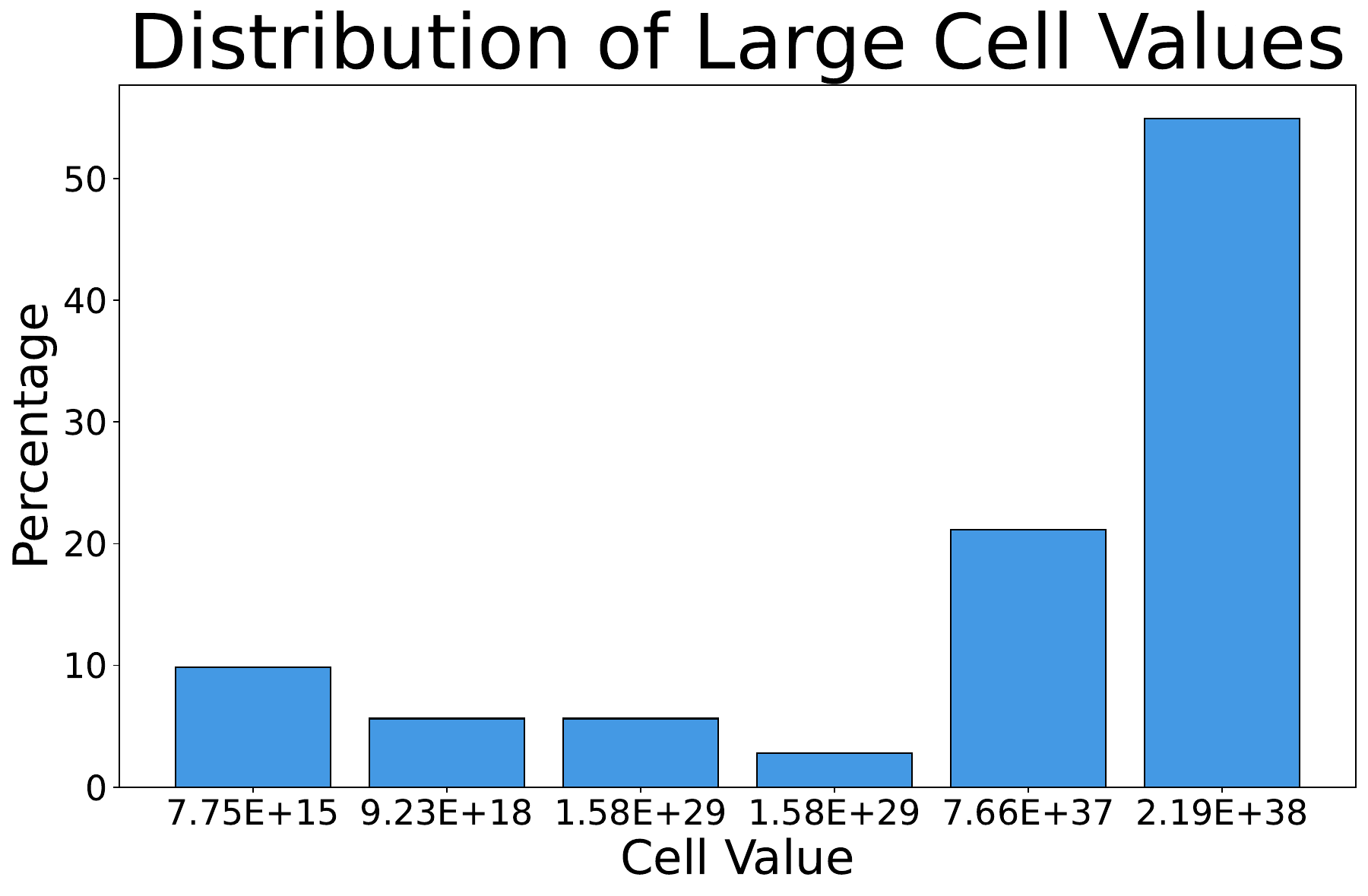}
        \caption{Large value range}
        \label{fig:fl_robust_update_dist_large}
    \end{subfigure}
    \caption{Corrupted Q-value distribution for Robust Q-Learning}
    \label{fig:fl_robust_update_dist}
\end{figure}

\textbf{Attacks on Learning Step During Learning.}
Voltage faults inject diverse corrupted values into the Q-Table (Figure~\ref{fig:fl_robust_update_dist} shows the distribution; searched parameter space in Figure~\ref{fig:appendix_robust_update_v_space}, Appendix~\ref{sec:appendix_vglitch_params}). Many faulted values remain close to the correct value (Figure~\ref{fig:fl_robust_update_dist_small}), while others are corrupted to extremely large magnitudes (Figure~\ref{fig:fl_robust_update_dist_large}), with values spanning from $10^{15}$ to $10^{38}$ and including Inf.
Both $NaN$ and $Inf$ values result from faults. The robust update reads the state values of all states through the minimization in Equation~\ref{eq:robust_update}, whereas standard Q-Learning reads only the next state. Because $V(s)$ takes the maximum over actions, a single $-Inf$ cell does not change $V(s)$ unless every action of that state is $-Inf$. If a state value does become $-Inf$, every subsequent update uses it, regardless of the state being updated. We did not measure whether this makes corruption spread faster than in standard Q-Learning, and faults in the rebuild of the $V$ array, which runs outside the trigger window, were not evaluated.

Clock glitching on Robust Q-Learning produces faults similar to those observed in the unprotected Q-Learning (Figure~\ref{fig:c1_robust_learning_step_distribution} in Appendix~\ref{sec:appendix_cglitch_params}). For the External Offsets shown, which cover up to 200 of the 1{,}727 cycles of the update, the number of successful faults is close to that of the standard implementation. Within this range the $V$ array does not prevent clock glitches from disturbing the update. The remainder of the update and the rebuild of the $V$ array are not covered by these results.

\textbf{Attacks on Greedy Policy During Inference.}
The greedy policy implementation is identical to standard Q-Learning. Repeated experiments confirm similar vulnerability. Voltage faults most commonly force the $\uparrow$ action with high effectiveness (similar to standard implementation). Other actions and out-of-bounds actions are also returned but with lower consistency.

\subsection{Key Findings}
Robust Q-Learning does not prevent hardware fault attacks despite its algorithmic robustness to environmental uncertainty. The state values and the minimization over states are designed for environmental noise and do not stop voltage or clock glitches from corrupting the update, which still produce NaN and Inf values. Once a state value is corrupted to $-Inf$, every later update uses it, but a single $-Inf$ cell does not reach the state values, and we did not measure whether corruption spreads faster than in standard Q-Learning. Inference uses the same code as standard Q-Learning and remains vulnerable in the same way. These results show that algorithmic robustness to environmental noise does not by itself provide security against deliberate hardware attacks.
\section{Faulting Resilient Q-Learning}
\label{sec:fault_resilient_qlearning}

\begin{figure*}[t!]
    \centering
    \begin{subfigure}[t]{0.24\textwidth}
        \centering
        \includegraphics[width=0.93\textwidth]{frozen_lake_policy_normal.drawio.pdf}
        \caption{Normal Policy}
        \label{fig:fl_resilient_policy_normal}
    \end{subfigure}
    \hfill
    \begin{subfigure}[t]{0.24\textwidth}
        \centering
        \includegraphics[width=0.93\textwidth]{frozen_lake_policy_normal.drawio.pdf}%
        \caption{Clock Glitching the Inference}
        \label{fig:fl_resilient_policy_cglitch}
    \end{subfigure}
    \hfill
    \begin{subfigure}[t]{0.24\textwidth}
        \centering
        \includegraphics[width=.93\textwidth]{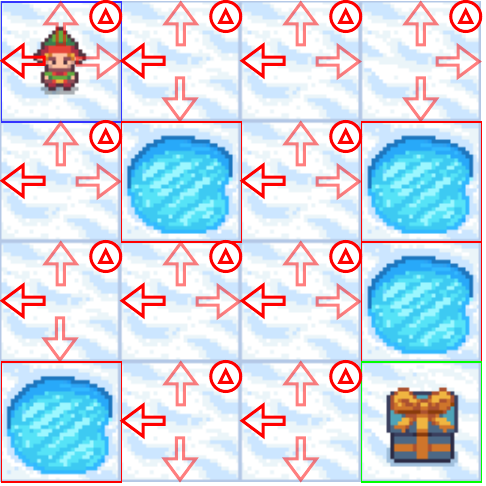}
        \caption{Voltage Glitching the Inference}
        \label{fig:fl_resilient_policy_vglitch_inf}
    \end{subfigure}
    \hfill
    \begin{subfigure}[t]{0.24\textwidth}
        \centering
        \includegraphics[width=.93\textwidth]{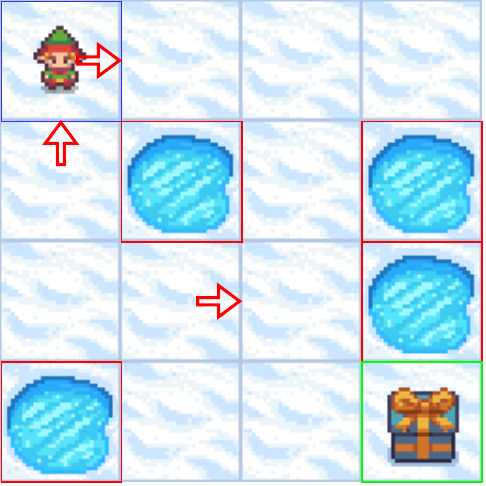}
        \caption{Voltage Glitching the Learning}
        \label{fig:fl_resilient_policy_vglitch_learn}
    \end{subfigure}
    \caption{Examples of the agent policy after glitching the inference and learning (Resilient Q-Learning implementation). The delta symbol indicates an out-of-bounds action, the cross indicates a NaN or Inf value.}
    \label{fig:fl_resilient_policy_glitched}
\end{figure*}

In this section we evaluate the resilient Q-Learning implementation, an implementation which provides some resilience against hardware faults. Two distinct countermeasures are added to the standard Q-Learning implementation and an evaluation is performed on each component. 
The first countermeasure is the triple replication of the Q-Table combined with majority voting during action selection. By maintaining three independent copies of the Q-Table and selecting actions based on majority agreement, a faulted value in a minority of the Q-Tables does not affect the policy of the agent.
The second countermeasure is the post-update fault correction mechanism. Immediately after each Q-Table update, a majority vote is performed on the just-updated cell across all three copies to synchronize them to the majority value. This corrects single-copy corruption of that cell before it can propagate to future updates.
The overhead of the introduced countermeasures is shown in Table~\ref{tab:resilient-overhead} in the Appendix. Both countermeasures protect the stored Q-values. Values read once and used for all three copies, such as the reward and the state and action indices, are outside this protection, and the voting code itself runs once. Figure~\ref{fig:fl_resilient_policy_glitched} shows examples of the agent policy under fault.

\textbf{Attacks on Reward Computation During Learning.}
The reward computation in this implementation is performed identically to the standard Q-Learning implementation, so we do not repeat our experiments. The reward is read once and the same value is used to update all three copies. A real-valued reward fault therefore produces the same wrong value in every copy, which the majority vote cannot detect. Non-finite rewards produce non-finite values in all copies, which the correction resets to 0. 

\textbf{Attacks on Action Selection During Learning.}
The action selection in this implementation is performed similarly to the standard Q-Learning implementation, adapted to use multiple Q-Tables as opposed to one. In the standard implementation, faults on action selection only transiently affect the random number generation or branch selection and do not alter the Q-Table, so we choose not to repeat the experiments for the Resilient Q-Learning implementation.

\textbf{Attacks on Learning Step During Learning.}
We separately evaluated the two countermeasures of the resilient Q-Learning.

\textit{Faulting The Q-Table Update:}  When faulting the Q-Table update, there are faults which changed values in the Q-Table, including infinite values: 58 of the 21{,}150 trials in Figure~\ref{fig:appendix_resilient_update_v_space}, each a full training run with the glitch armed on every update, ended with a table different from the reference, 6 of them with infinite values. A value corrupted in one copy is overwritten by the post-update correction, and majority voting in the policy calculation prevents a minority of corrupted copies from changing the selected action. A real-valued error that reaches all copies decays as the cell is updated again, because each update keeps only a $(1-\alpha)$ fraction of the previous value. Non-finite values in two or three copies of a cell never compare as equal in the voting code, so the correction resets that cell to 0 the next time it is updated; until then, the cell can still affect the policy. We evaluated a single glitch per update and did not evaluate an attacker who corrupts several copies in the same update.

\textit{Faulting The Fault Correction:}  When faulting the fault correction itself, some faults successfully triggered a cell reset in unnecessary cases. This results in a specific cell value being set to 0 in all Q-Tables. The reset cell is then relearned through later updates.

The two countermeasures of resilient Q-Learning were also evaluated with clock glitching, over External Offsets up to about 200 (Figure~\ref{fig:c1_resilient_learning_step_distribution}), which covers only the start of the 1{,}938-cycle update. In this range we observed fewer successful faults than for the standard and the Robust Q-Learning, although the range covers a smaller share of the update here than in the standard implementation (200 of 749 cycles). In both faulting the Q-Table update function and faulting the fault correction, a total of only four  four of about 200 External Offset values, and which output was faulted values of the External Offset led to a faulty output, all of which were $Inf$ values. Within the tested configurations, no valid incorrect action values were produced via clock glitching, though additional configurations beyond our search space could potentially yield different results.

\textbf{Attacks on Greedy Policy During Inference.}
Voltage glitches were applied to the greedy policy including the majority voting, which resulted in several faults being found. However, the faults are less consistent than for the other implementations and result in more device crashes. Interestingly, the most consistent fault for this implementation results in the action $\leftarrow$ (LEFT) as opposed to $\uparrow$ (UP) as for the other implementations.

Clock glitching of the greedy policy of the Resilient implementation was generally unsuccessful, so Figure~\ref{fig:fl_resilient_policy_cglitch} shows the normal policy.

\subsection{Key Findings}
The countermeasures correct corruption of a single Q-table copy during learning and prevent a minority of corrupted copies from changing the selected action. Our voltage glitching search found 58 update faults in 21{,}150 trials for this implementation, but it uses different bounds and fault criteria from the searches for the other implementations, so we do not report a reduction factor. The protection has limits: values shared by the copies, such as the reward, are outside it, multi-copy corruption was not evaluated, and voltage glitches still change the selected action during inference.

\section{Faulting Linear Q-Learning}

In addition to tabular based Q-Learning implementations we evaluate an implementation based on linear function approximation. Linear function approximation provides several benefits compared to tabular methods, particularly in environments where states are not univariate discrete and tabular methods therefore incur a loss of information through discretizations. Furthermore this approximation method is more generalized: weights are shared across states. This allows this method to potentially learn faster, and to be applied to other environments. For these experiments we continue to use the Frozen Lake environment with differing maps, to investigate if such implementation can be similarly faulted and what the behavior of the agent is under fault. 

Linear function approximation replaces the Q-Table with a linear function, calculated in our implementation according to the following formula: 

\begin{equation}\label{formula:linear_q_value}
    Q(s,a) = \sum^{n-1}_{i=0} w^a_i f_i(s)
\end{equation}

\noindent Where $n$ is the number of features in the implementation, $f_i(s)$ is feature $i$ of state $s$, and $w^a_i$ the weight corresponding to feature $i$ and action $a$. 

We utilize action-agnostic features to minimize the amount of data transfer through the serial port but create separate weight vectors for each action. In our experiments we investigate across varying number of features on different maps. The hyperparameter $\alpha$ of the Q-Learning (see Equation~\ref{eq:linear_update}) is adapted by experiment such that the agent can converge to a stable policy (e.g., $\alpha=0.7$ for 16 features, $\alpha=0.001$ for 8 features, and $\alpha=0.01$ for 4 features).

\subsection{Linear Q-Learning with 16 Features}
\label{sec:linear_16}

\begin{figure}
    \centering
     \begin{subfigure}[t]{0.32\linewidth}
    \includegraphics[width=\linewidth]{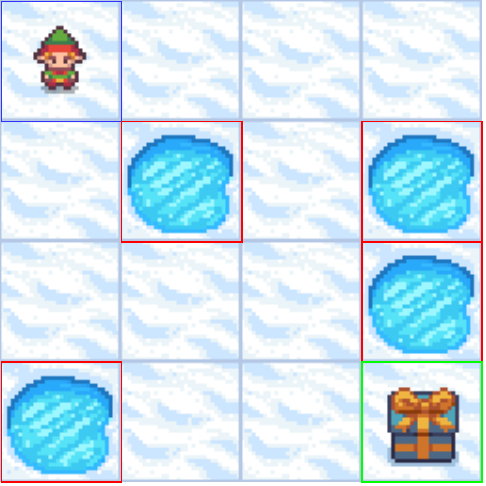}
    \caption{16 Features Map}
    \label{fig:fl_16features}
    \end{subfigure}
    \hfill
    \begin{subfigure}[t]{0.32\linewidth}
    \includegraphics[width=\linewidth]{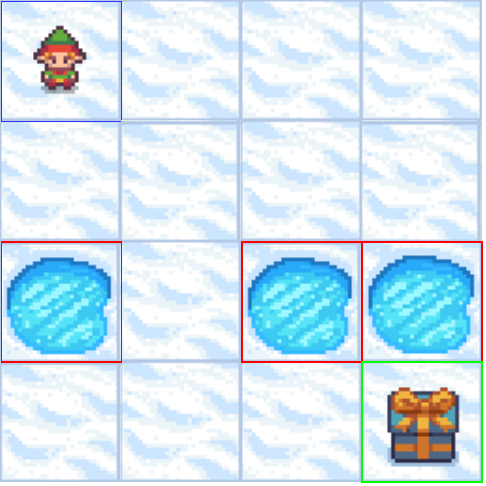}
    \caption{8 Features Map}
    \label{fig:fl_8features}
    \end{subfigure}
    \begin{subfigure}[t]{0.32\linewidth}
    \includegraphics[width=\linewidth]{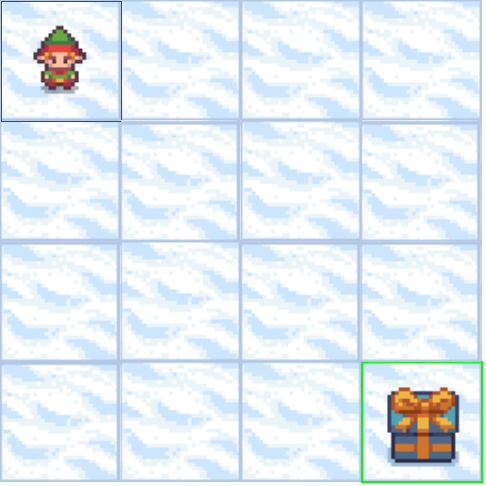}
    \caption{4 Features Map}
    \label{fig:fl_4features}
    \end{subfigure}
    \caption{Frozen Lake maps used for each experiment.}
\label{fig:fl_maps}
\end{figure}

First we create an implementation with an identical number of stored values as in the Q-Table, i.e. $n = 16$. This is achieved by calculating the one-hot encoding of the state value, i.e. to represent state $0$ we use a binary vector of length $16$ with all $0$'s except index $0$ which has a value of $1$. The experiments are performed on the same $4 \times 4$ map as in the tabular experiments, visualized in Figure~\ref{fig:fl_16features}. 

\textbf{Faulting The Learning.} The learning, i.e. updating of the weights, in linear function approximation is performed according to the formula:

\begin{equation} \label{eq:linear_update}
    w^a_i = w^a_i + \alpha (r + \gamma \max_{a'} Q(s', a') - Q(s,a)) f_i(s)
\end{equation}

\noindent Where $s'$ is the state reached from $s$ after taking action $a$, $\alpha$ the learning rate, $\gamma$ the discount factor, and $r$ the reward. 

\begin{figure}[t]
\centering
    \includegraphics[width=0.75\linewidth]{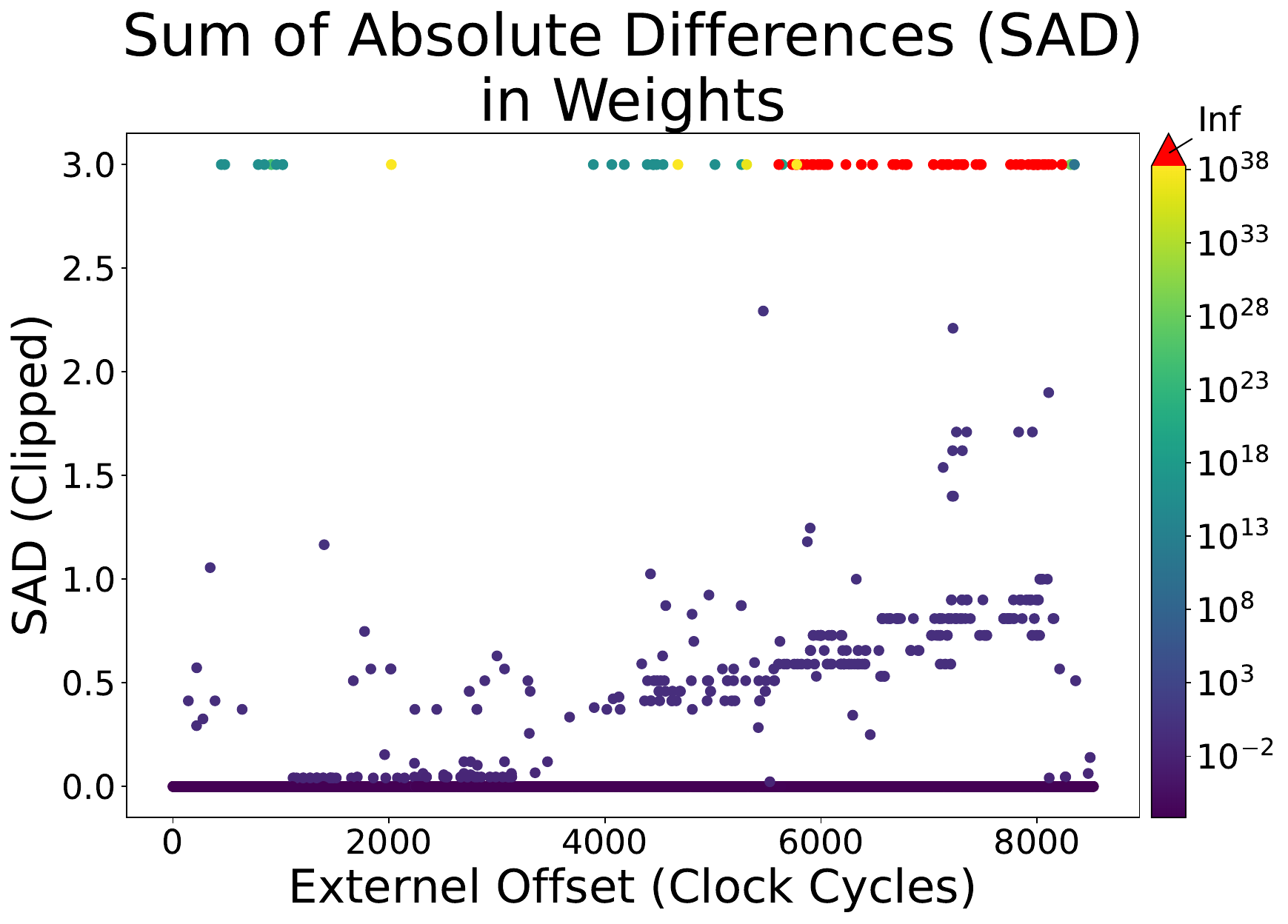}
    \caption{Impact of Faults during Learning Relative to External Offset}
    \label{fig:linear-learning-fault-diffs}
\end{figure}

Figure~\ref{fig:fl_linear_update_v_space} in Appendix~\ref{sec:appendix_vglitch_params} shows the successful fault configurations, with Figure~\ref{fig:linear-learning-fault-diffs} highlighting the difference in weight values as a result of voltage faults at different clock cycles, measured by the sum of absolute differences (SAD) between faulted and reference weights. This figure shows that the largest weight differences occur when faulting the later stages of the function. The faulty weight values are similar to those observed in the tabular Q-Learning implementations, with weights reaching large ($> 1e38$) and Infinite values.

\begin{figure}[t]
    \centering
    \includegraphics[width=0.85\linewidth]{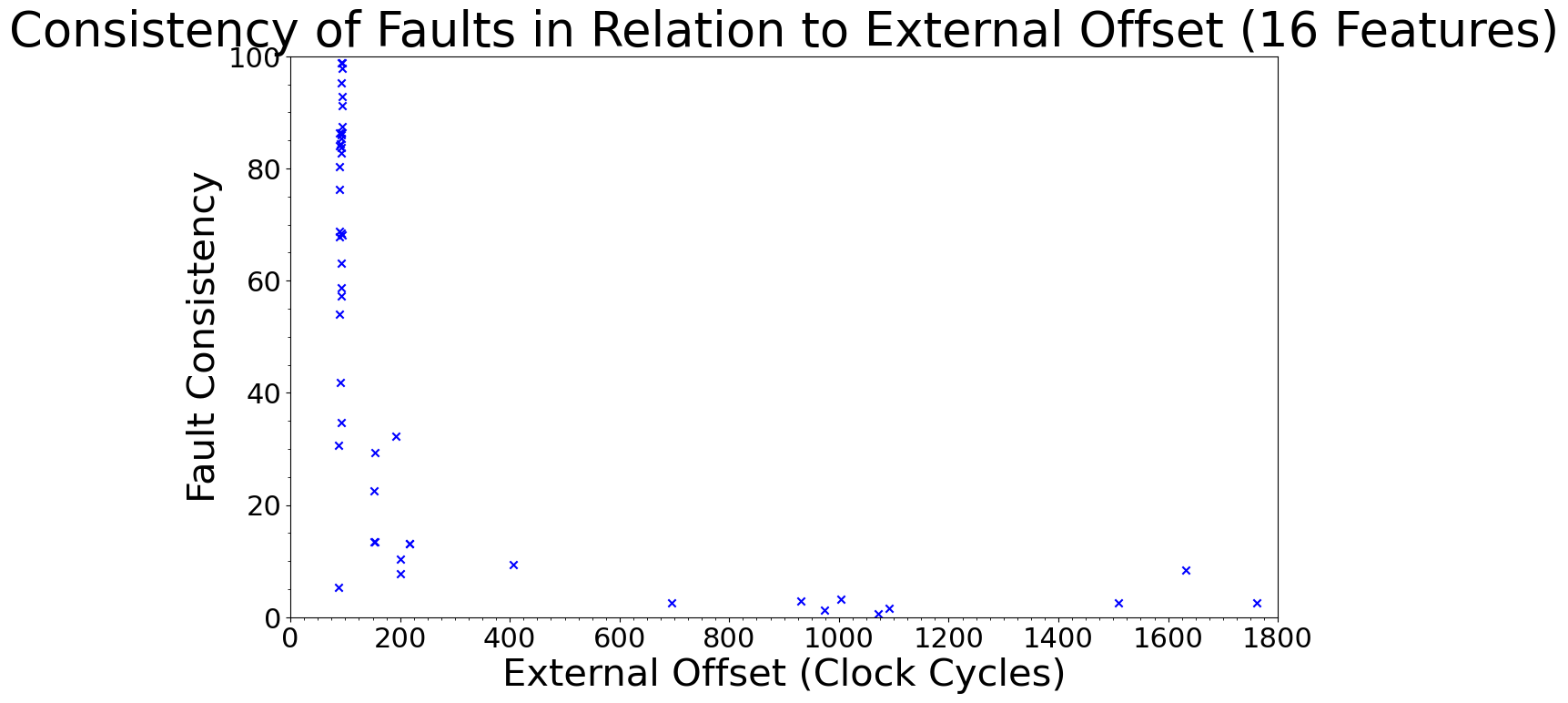}
    \caption{Fault consistency in relation to the external offset.}
    \label{fig:fault_consistency_ext_offset}
\end{figure}

\textbf{Faulting The Inference.} 
For voltage attacks, similar fault results were seen as for the tabular implementations, although with lesser consistency. Since in this implementation $Q(s,a)$ is computed as opposed to looked up from memory, the timing between the Q value comparisons is longer and more variable. Additionally, the timing of this implementation is not constant because it depends on the state. This makes later operations harder to target with a single External Offset. However, consistent faults can be found in the early computation.
This pattern can be clearly seen in Figure~\ref{fig:fault_consistency_ext_offset}, which shows several consistent faults which only exist in the first 100 clock cycles, whereas the later faults are much less consistent. These first 100 cycles fall before the first dot product in Table~\ref{tab:pipeline_linear_infer}; we did not trace what these faults affect.

With clock glitching, for the External Offsets shown in Figure~\ref{fig:c1_linear_policy_distribution} in the Appendix (up to 200 of the 4{,}562 cycles), there are more successful attacks compared to those of the Q-Table implementations. There were also more crashes, possibly due to more calls to nested functions in the attacked function.

\textbf{Fault Impact.} To evaluate the fault impact, we simulate the insertion of faulty values into the weight vector. We insert each of ten values $\{-\infty, -10^{30}, -10, -1, 0, 1, 10, 10^{30}, \mathrm{NaN}, \infty\}$ at 12 to 13 weight locations, one at a time. These values were chosen to span the magnitudes observed during the hardware fault injection experiments on the learning function; they are not exact fault outputs. After each insertion we execute at most 600 episodes (i.e. agent lifetimes) of learning. An agent is deemed to have ``recovered'' when it reaches any optimal policy.

\begin{figure}[t!]
    \centering
    \includegraphics[width=0.7\linewidth]{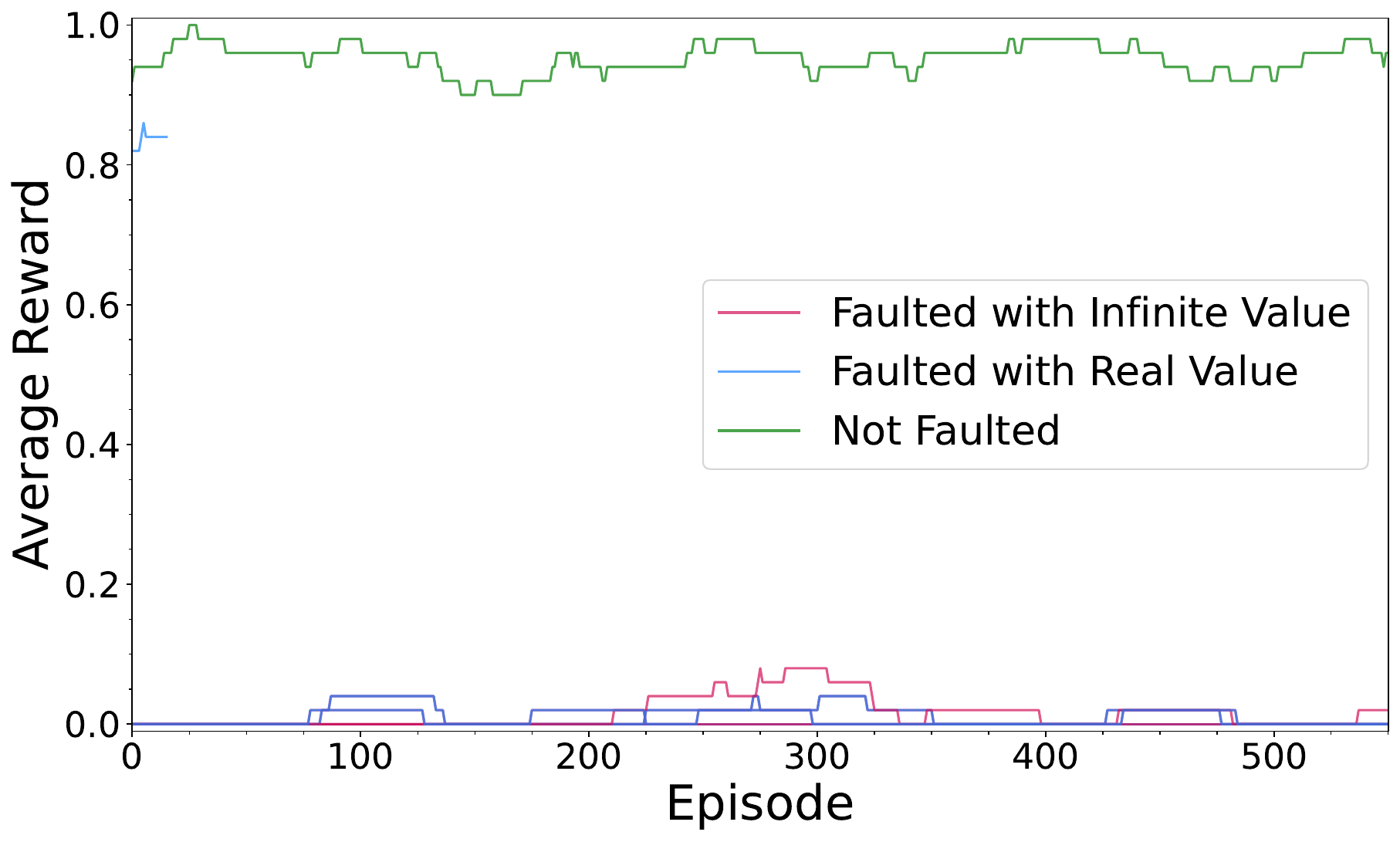}
    \caption{Performance during recovery for 16 Features-based Linear Q-learning, for insertions that changed the policy. A run stops when the agent recovers, so later episodes show only runs that have not recovered.}
    \label{fig:fl_recovery_l_16}
\end{figure}

Figure~\ref{fig:fl_recovery_l_16} shows the average reward obtained during the recovery of the implementation, i.e. from the moment a faulty weight value has been inserted into the agent, for insertions that changed the policy. Of the 91 real-valued insertions, 68 left the policy optimal. The remaining 23 all recovered, usually quickly (median 3 episodes, mean 76.4, maximum 469). Of the 39 non-real insertions, 9 left the policy optimal and 30 changed it; 28 of these did not recover within 600 episodes, and their performance falls to what can be roughly expected from a random policy based on the $\epsilon$ hyperparameter. The other 2 recovered within one episode.

\begin{figure}
    \centering
     \begin{subfigure}[t]{0.32\linewidth}
    \includegraphics[width=\linewidth]{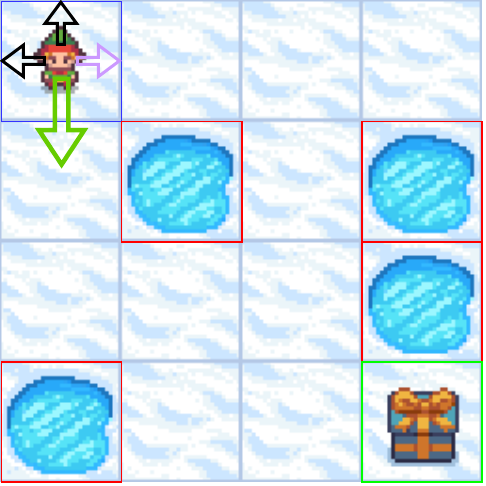}
    \caption{Agent chooses action}
    \label{fig:linear_fault_16_1}
    \end{subfigure}
    \hfill
    \begin{subfigure}[t]{0.32\linewidth}
    \includegraphics[width=\linewidth]{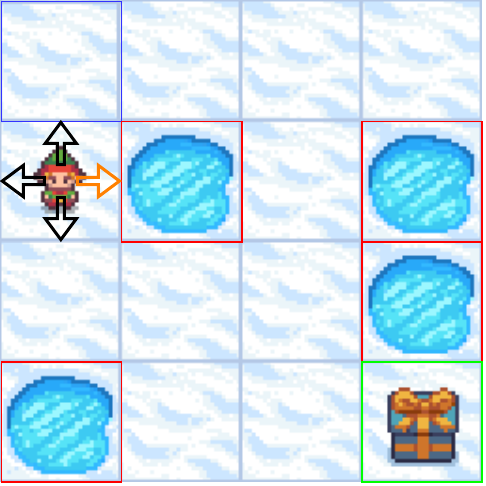}
    \caption{Weight update}
    \label{fig:linear_fault_16_2}
    \end{subfigure}
    \begin{subfigure}[t]{0.32\linewidth}
    \includegraphics[width=\linewidth]{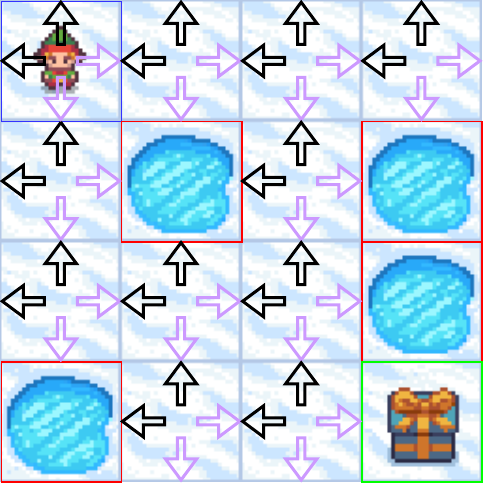}
    \caption{Converged action evaluations}
    \label{fig:linear_fault_16_3}
    \end{subfigure}
    \caption{Evolution of the action evaluation of the agent. The colors of the arrows signify the evaluation of the action as follows: black = real value, \textcolor{paleplum}{pink} = NaN value, \textcolor{orange}{orange} = Inf value.}
\label{fig:linear_fault_16}
\end{figure}

\textit{Non-Real Faults.} A non-real value (i.e. Inf or NaN) that changes the policy typically leaves all weights for the faulted action as $NaN$ values. For Inf values specifically, the faulted weight may not immediately compromise the agent's decision in states where the corresponding feature is zero (due to the multiplication $0 \times \mathrm{Inf} = \mathrm{NaN}$), which can result in the value propagating to other action weights. NaN values, once introduced, cause all comparisons to return false, preventing the affected actions other than the default action from being selected by the greedy policy.
This can be illustrated in the following example, which is visualized in Figure~\ref{fig:linear_fault_16}: faulting the weight of action $2$ ($\rightarrow$) for feature $4$ (i.e. state = $4$ for the one-hot encoding) to a value of Inf. When the agent starts in state $0$, this does not result in action $2$ being evaluated as Inf, since only feature $0$ has a non-zero value, but as being a $NaN$ due to the multiplication $0 \times \mathrm{Inf}$.  This results in the normal optimal decision to move down, reaching state $4$. After taking this action the agent will update the weights for going down in state $0$ based on the evaluation of state $4$. Now the Infinite weight value becomes relevant since it results in an evaluation of Infinite, which subsequently results in weight values for action $1$ (down) getting Infinite values. Following this, the first update of the weights of an action containing an Infinite value will result in a $NaN$ due to the computation $Inf - Inf$. Any comparison between a real float and a $NaN$ value returns $False$, which means these actions are no longer chosen by the greedy policy, and they also will not propagate to other action weights. For the above example this means the weights for actions $1$ and $2$ will converge to $NaN$ values, making the agent unable to reach the goal state through greedy action selection.

Finally, the most impactful fault location is any fault in the weight for the default action. In this implementation this is action  $0$ ($\leftarrow$), against which the other actions are compared to determine the optimal choice. When one of these weights is faulted with a non-real value, this quickly results in this action being evaluated as NaN, which for the default action results in it being greedily chosen. It being greedily chosen also means that its value is used to update the weights of other actions (which will still be chosen due to the epsilon-greedy policy during learning), and thus the NaN value propagates to all weights of the agent. 

\subsection{Linear Q-Learning with 8 Features}

To further understand the effect of faults on a function-approximated implementation, the number of features is reduced by half to a total of $8$ features. Additionally, a different map is used in these experiments which challenge the agent to move through a specific cell. The features are given below, and the map is visualized in Figure~\ref{fig:fl_8features}. Note, all features are normalized to the range $[0, 1]$.

\begin{enumerate}
\setlength{\itemsep}{0pt}
\setlength{\parskip}{0pt}
    \item \textit{Feature 1:} Bias, this feature is always set to be 1
    \item \textit{Feature 2:} X coordinate
    \item \textit{Feature 3:} Y coordinate
    \item \textit{Feature 4:} Distance to the end goal
    \item \textit{Feature 5:} Distance to the nearest hole
    \item \textit{Feature 6:} Number of actions available
    \item \textit{Feature 7:} Binary feature indicating if $X=1$
    \item \textit{Feature 8:} Binary feature indicating if $Y=Y_{goal}$
\end{enumerate}

\textbf{Faulting The Learning.}
The learning in this implementation uses the same code as the 16 feature version with fewer features, so we do not repeat the experiments to find faults.

\textbf{Faulting The Inference.}
The inference in this implementation uses the same code as the 16 feature version with fewer features, so we do not repeat the experiments to find faults.

\textbf{Fault Impact.}
To understand the impact of faults on this implementation we insert the same set of values, using the same procedure to study the recovery of the agent.

\begin{figure}[t!]
    \centering
    \includegraphics[width=0.7\linewidth]{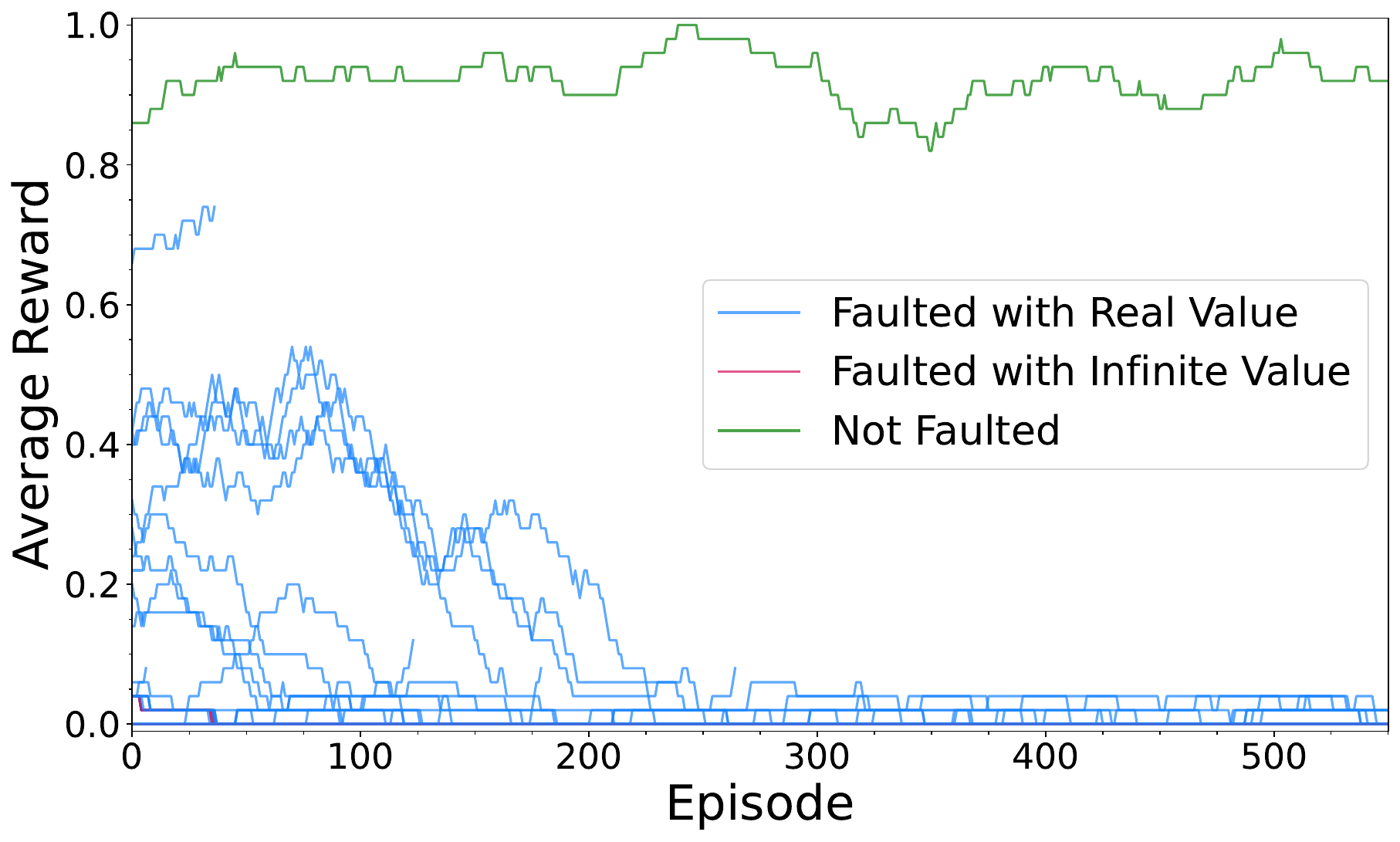}
    \caption{Performance during recovery for 8 Features-based Linear Q-learning, for insertions that changed the policy. A run stops when the agent recovers, so later episodes show only runs that have not recovered.}
    \label{fig:fl_recovery_l_8}
\end{figure}

The performance of the agent after injection of faults is visualized in Figure~\ref{fig:fl_recovery_l_8}. Of the 91 real-valued insertions, 35 left the policy optimal and 56 changed it. Only 26 of those 56 recovered within 600 episodes (median 24.5 episodes, mean 50.3), compared with all 23 for the 16 feature implementation (Table~\ref{tab:recovery_comparison}). One possible contributor is the construction of the features: with one-hot encoding a faulted weight only affects one (state, action) pair, whereas here features are shared across states. However, the two experiments also use different maps and learning rates ($\alpha=0.7$ and $\alpha=0.001$), so we cannot attribute the difference to feature sharing alone. Non-real values rarely recover in either implementation (Table~\ref{tab:recovery_comparison}).

\begin{table}[h]
\centering
\caption{Recovery after inserting one faulty weight value. Agents are given at most $600$ episodes after insertion and are deemed recovered once they reach any optimal policy. Affected insertions are those that changed the policy. Mean and median are over recovered runs only. All recoveries from non-real values ($\mathrm{NaN}$, $\pm\infty$) took one episode. The three experiments use different learning rates and maps, so the differences cannot be attributed to the feature encoding alone.}
\label{tab:recovery_comparison}
\begin{adjustbox}{max width=\linewidth}
\small
\begin{tabular}{r|l|l|l|r|r|r|r}
\toprule
\textbf{Features} & \textbf{Encoding} & $\boldsymbol{\alpha}$ & \textbf{Map} & \textbf{Real: affected} & \textbf{Real: recovered} & \textbf{Episodes (mean / median)} & \textbf{Non-real: recovered / affected} \\
\midrule
16 & one-hot & 0.7 & standard & 23 & 23 (100\%) & 76.4 / 3.0 & 2 / 30 \\
8  & shared  & 0.001 & custom holes & 56 & 26 (46\%) & 50.3 / 24.5 & 4 / 29 \\
4  & shared  & 0.01 & no holes & 48 & 35 (73\%) & 37.1 / 8.0 & 4 / 25 \\
\bottomrule
\end{tabular}
\end{adjustbox}
\end{table}

\subsection{Linear Q-Learning with 4 Features}

\begin{figure}[t!]
    \centering
    \includegraphics[width=0.7\linewidth]{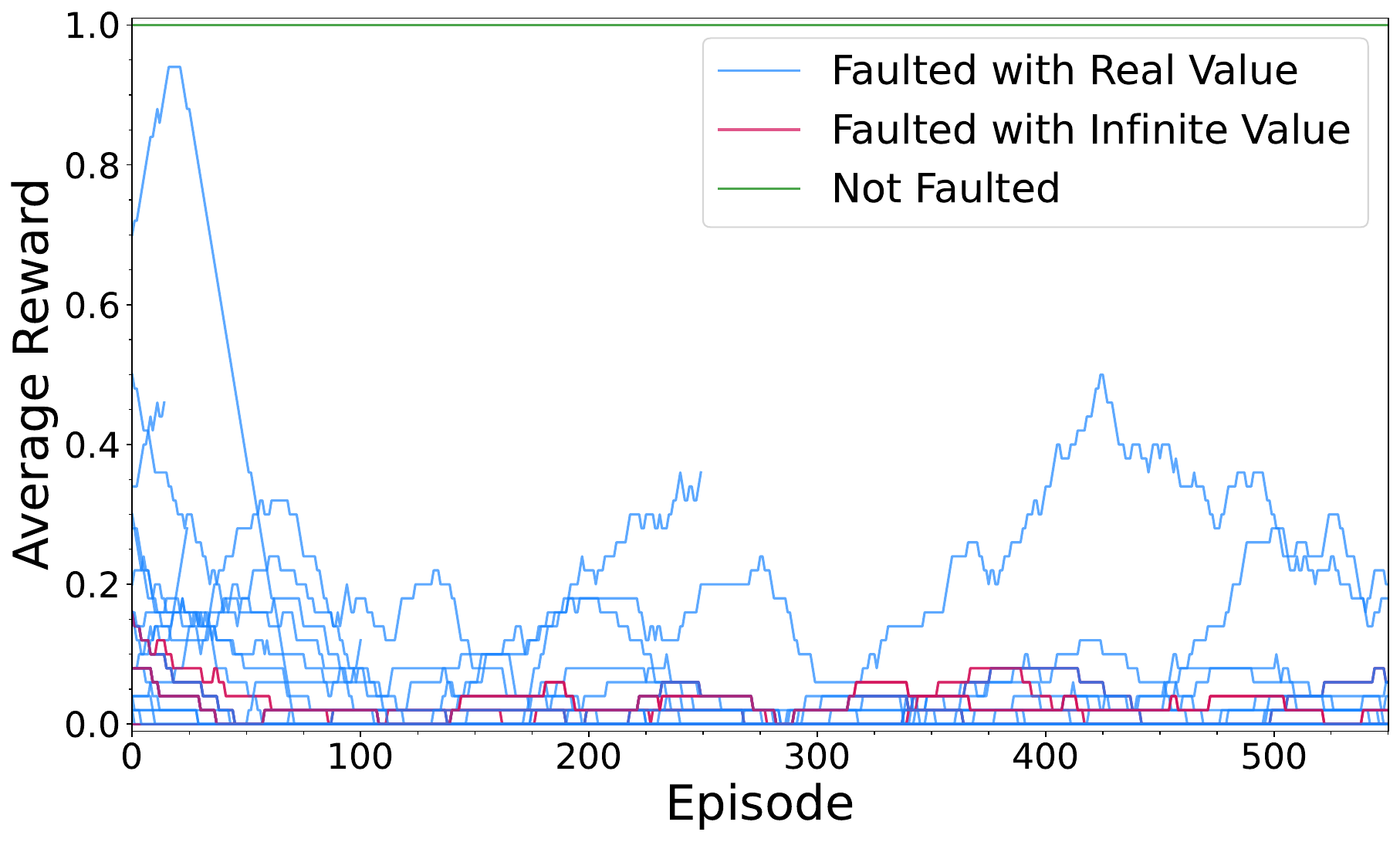}
    \caption{Performance during recovery for 4 Features-based Linear Q-learning, for insertions that changed the policy. A run stops when the agent recovers, so later episodes show only runs that have not recovered.}
    \label{fig:fl_recovery_l_4}
\end{figure}

Lastly the amount of features is halved again to a total of $4$ features. For this implementation another map is used for the experiments, as $4$ features do not provide enough information to the agent to be able to navigate to the end goal while avoiding holes. Therefore in these experiments a simple $4\times4$ map without any holes is used. The map is visualized in Figure~\ref{fig:fl_4features}. The $4$ features are a subset of the features used in the $8$-feature implementation, describing a state at a cell in the map as shown below. All features are normalized to the range $[0, 1]$.

\begin{enumerate}
\setlength{\itemsep}{0pt}
\setlength{\parskip}{0pt}
    \item \textit{Feature 1:} Bias, this feature is always set to be 1
    \item \textit{Feature 2:} X coordinate
    \item \textit{Feature 3:} Y coordinate
    \item \textit{Feature 4:} Distance to the end goal
\end{enumerate}

\textbf{Faulting The Learning.}
The learning in this implementation uses the same code as the 16 feature version with fewer features, so we do not repeat the experiments to find faults.

\textbf{Faulting The Inference.}
The inference in this implementation uses the same code as the 16 feature version with fewer features, so we do not repeat the experiments to find faults.

\textbf{Fault Impact.}
To study the impact of the faults on this implementation we use the same procedure used to study the impact of the 8 feature implementation.

Figure~\ref{fig:fl_recovery_l_4} shows how the agent performs after the insertion of a faulty value into one of the weights. For this implementation a map is considered without holes, meaning that episodes can only end by reaching the goal or after $99$ steps. Of the 84 real-valued insertions, 36 left the policy optimal and 48 changed it; 35 of those 48 recovered (median 8 episodes, mean 37.1) and 13 did not recover within 600 episodes. Of the 36 non-real insertions, 25 changed the policy and 21 of those did not recover, with performance conforming to what can be expected of a random policy based on the given $\epsilon$.

\subsection{Key Findings.} The linear function approximation allows this implementation to generalize across states, which benefits normal operation. Faults still corrupt stored weights with both real and non-real values. Real-valued insertions that changed the policy always recovered with 16 one-hot features (23 of 23), but less often with shared features (26 of 56 with 8 features, 35 of 48 with 4 features). These experiments use different maps and learning rates, so the differences cannot be attributed to feature sharing alone. Non-real values that changed the policy rarely recovered in any configuration (2 of 30, 4 of 29, and 4 of 25).

\section{Faulting Deep Q-Learning}
\label{sec:dqn_fault_results}

In addition to the tabular and linear Q-Learning implementations, we evaluate a Deep Q-Network (DQN) agent~\cite{mnih2015human}, which uses a neural network to approximate the action-value function $Q(s,a)$, as introduced in Section~\ref{sec:deep_q_learning}. DQN improves the stability of value-based deep reinforcement learning by combining (i) \emph{experience replay}, where transitions $(s,a,r,s',\texttt{done})$ are stored in a replay buffer and training is performed on uniformly sampled mini-batches, and (ii) a separate \emph{target network} that is used to compute learning targets.

\begin{figure}[t!]
    \centering
     \begin{subfigure}[t]{0.32\linewidth}
    \includegraphics[width=\linewidth]{frozen_lake_policy_normal.drawio.pdf}
    \caption{Normal Policy}
    \label{fig:deep_q_no_fault}
    \end{subfigure}
    \hfill
    \begin{subfigure}[t]{0.32\linewidth}
    \includegraphics[width=\linewidth]{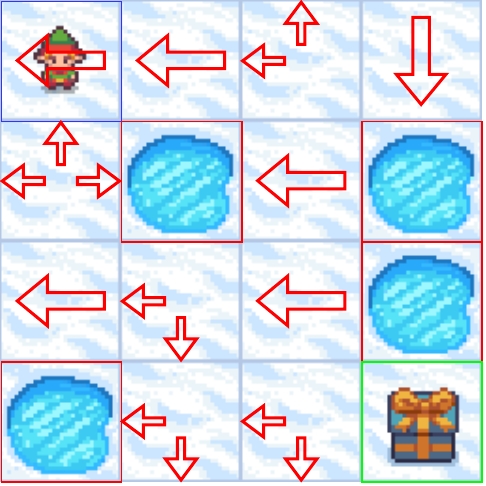}
    \caption{Voltage Glitching the Inference}
    \label{fig:deep_q_fault}
    \end{subfigure}
    \hfill
    \begin{subfigure}[t]{0.32\linewidth}
    \includegraphics[width=\linewidth]{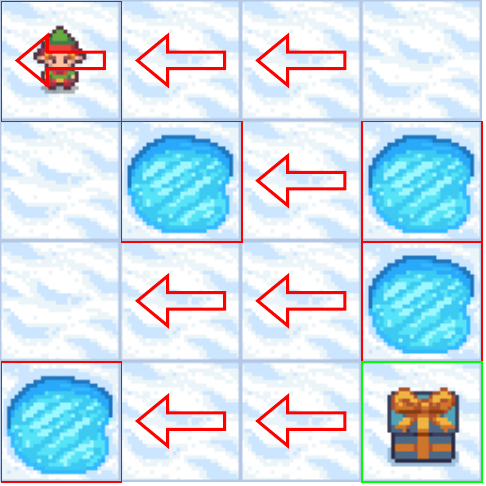}
    \caption{Clock Glitching the Inference}
    \label{fig:frozen_lake_cg_deep_q}
    \end{subfigure}
    \caption{Examples of the agent policy after glitching the inference (Deep Q-Learning implementation).}
\label{fig:deep_qlearning_states_no_fault_and_fault_inference}
\end{figure}

We implement the DQN algorithm in C, including the replay buffer, $\epsilon$-greedy action selection with linearly decayed $\epsilon$, computation of learning targets, and periodic synchronization of the target network. Neural network inference and training primitives are provided by the AIfES embedded ML framework~\cite{wulfert2024aifes}. The agent operates on a deterministic $4 \times 4$ FrozenLake environment, which the firmware implements on the device for training. The discrete state index $s \in \{0,\dots,15\}$ is encoded as a one-hot vector $\mathbf{x}\in\mathbb{R}^{16}$ and passed through a small fully connected network consisting of one hidden dense layer and ReLU activation, followed by a linear output layer with four neurons producing Q-values for the four actions.

Training uses \emph{temporal-difference (TD) learning}: after taking action $a$ in state $s$ and observing reward $r$ and next state $s'$, the current estimate $Q_{\text{policy}}(s,a)$ is updated toward a one-step bootstrapped target that uses the network’s estimate of future value. Concretely, for each sampled transition we compute the one-step TD target
\[
y = 
\begin{cases}
r, & \text{if } \texttt{done}=1 \\
r + \gamma \max_{a'} Q_{\text{target}}(s',a'), & \text{otherwise},
\end{cases}
\]
which corresponds to a TD error $\delta = y - Q_{\text{policy}}(s,a)$.
We then minimize mean-squared error between the predicted Q-vector and a target vector in which only the entry for the chosen action is replaced by $y$ (leaving the other action values unchanged). Optimization uses Adam~\cite{kingma2014adam} as implemented in AIfES, and the target network is periodically synchronized by copying the policy network parameters. For the inference experiments, the network is first trained without faults until it converges to a near-optimal policy and the resulting parameters are then frozen.

\begin{figure}[t]
    \centering
    \includegraphics[width=0.75\linewidth]{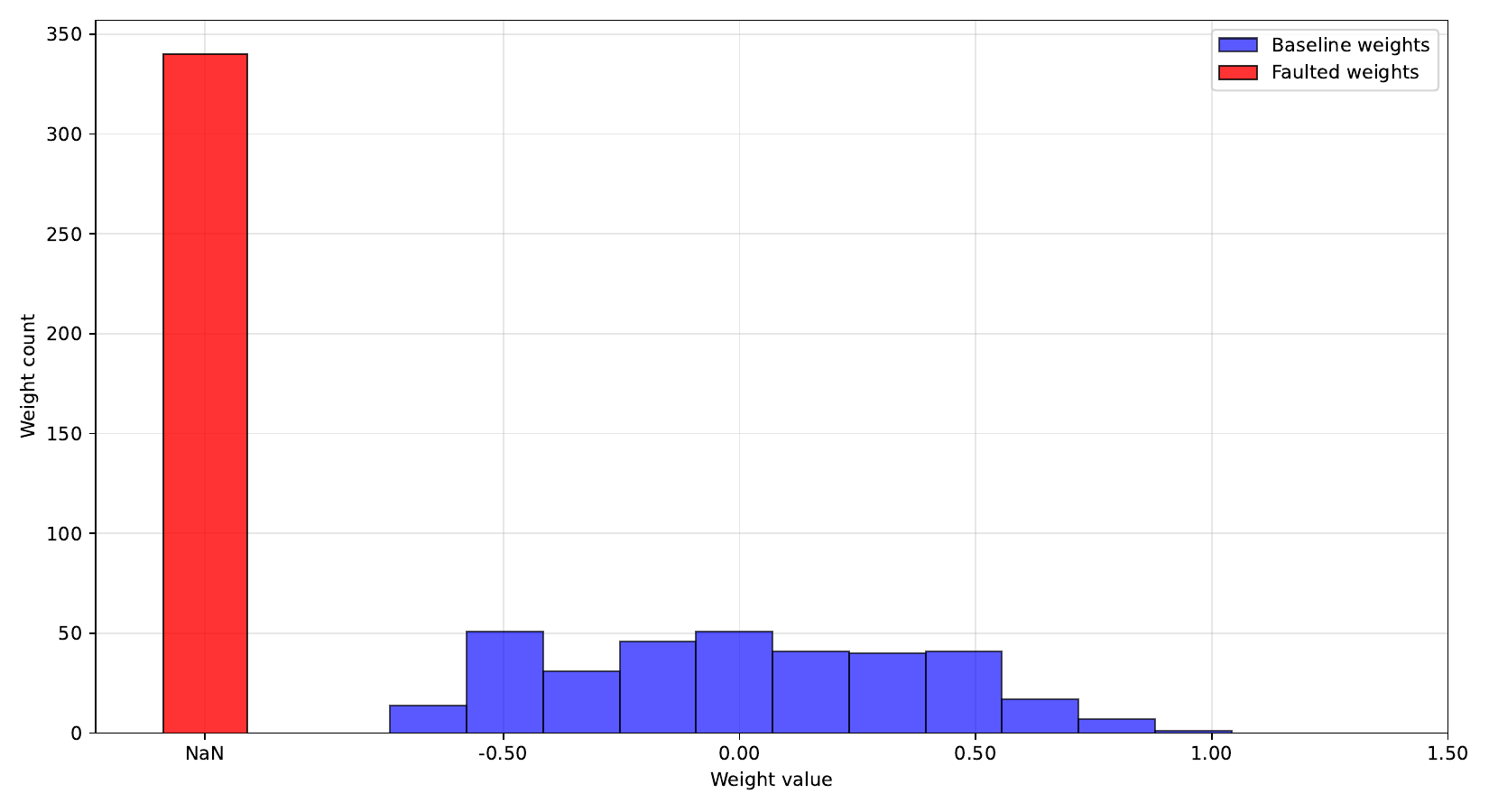}
    \caption{Baseline and faulted weights for target episode 199, best trial. All 340 faulted weights are NaN.}
\label{fig:deepq_training_faulted_weights_episode_199}
\end{figure}

\begin{figure}[t]
    \centering
    \includegraphics[width=0.75\linewidth]{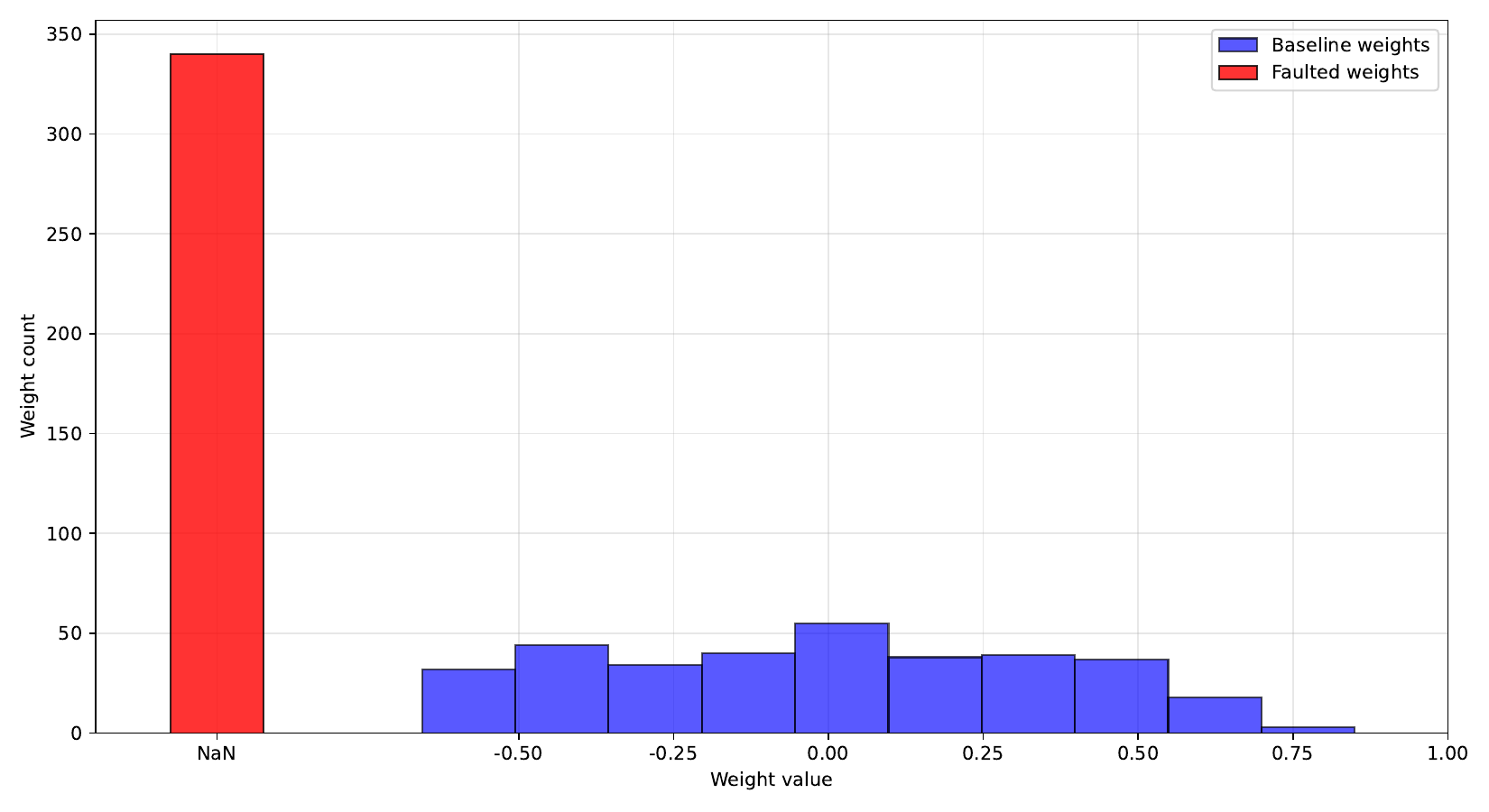}
    \caption{Baseline and faulted weights for target episode 100, best trial. All 340 faulted weights are NaN.}
\label{fig:deepq_training_faulted_weights_episode_100}
\end{figure}

\begin{figure}[t]
    \centering
    \includegraphics[width=0.75\linewidth]{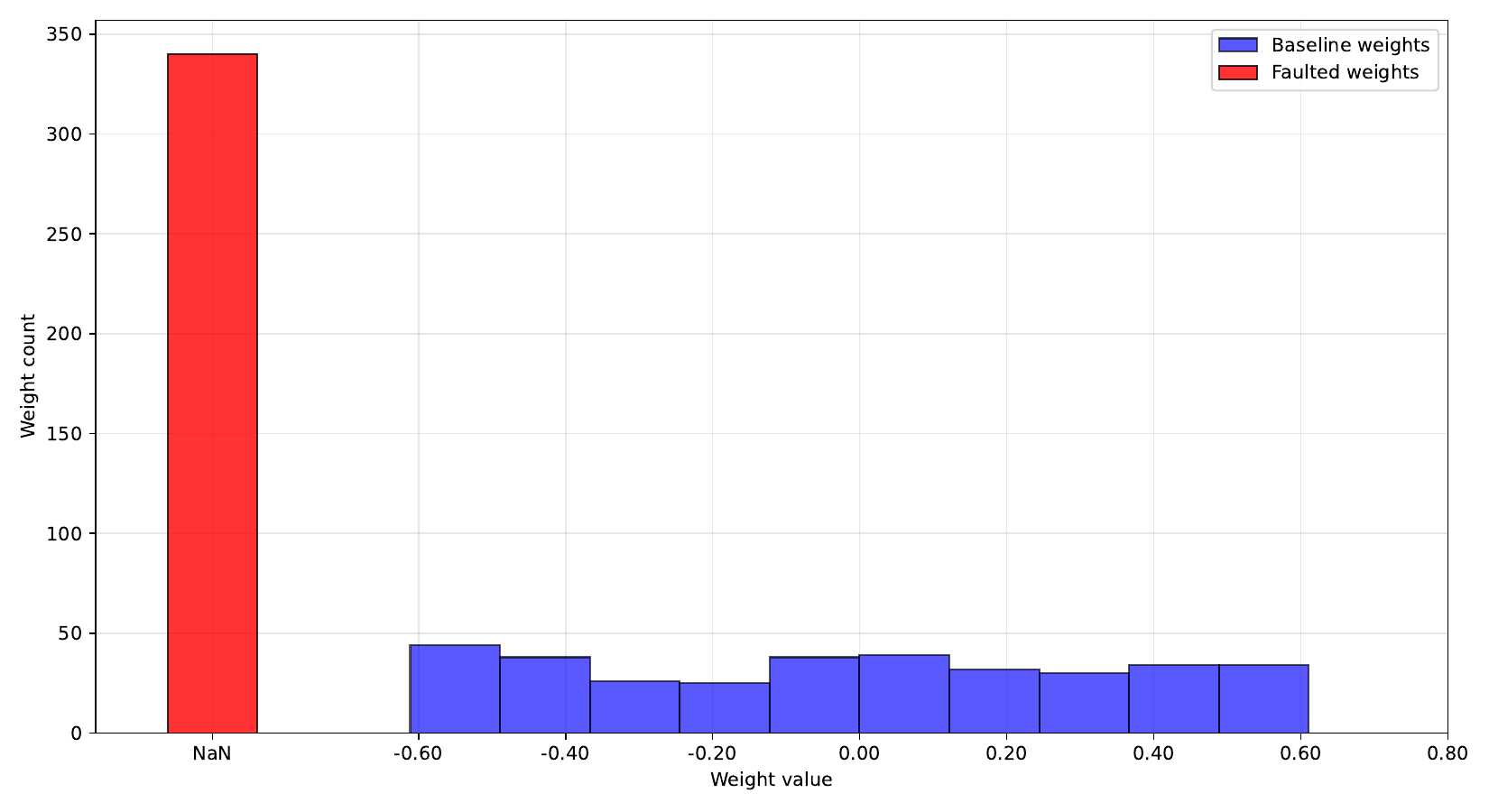}
    \caption{Baseline and faulted weights for target episode 1, best trial. All 340 faulted weights are NaN.}
\label{fig:deepq_training_faulted_weights_episode_1}
\end{figure}

\textbf{Faulting The Learning.}
To study faults during training, we load the weights saved before a target episode (1, 100, or 199) onto the device, run a single training step on a batch recorded from that episode, and inject a voltage glitch during the optimizer's weight update. We then compare the resulting weights with a clean reference snapshot for that episode. Optuna searched Width and Offset from 2400 to 2800 in steps of 100, External Offset from 0 to 200, and Repeat from 1 to 5, with three attempts per configuration. Weights differed from the reference in at least one attempt in 2{,}972 of 10{,}000 trials at episode 1, 3{,}458 of 10{,}000 at episode 100, and 3{,}128 of 10{,}673 at episode 199. In the best trial at each target episode, all 340 weights became NaN (Figures~\ref{fig:deepq_training_faulted_weights_episode_199}, \ref{fig:deepq_training_faulted_weights_episode_100}, and~\ref{fig:deepq_training_faulted_weights_episode_1}).

\textbf{Faulting The Inference.}
Voltage glitches were injected targeting the forward pass used to compute $Q(s,a)$ for the current state. In the absence of faults, the learned policy is effectively deterministic, with the agent always selecting the optimal action in each non-terminal state on the $4 \times 4$ Frozen Lake map. With faults, successful attack configurations were found. Optuna was used again to perform the search, separately for each state. Searches over the full inference window used a 200-cycle External Offset grid and found at least one incorrect action in 454 of 103{,}437 trials (0.44\%). Focused searches at single-cycle resolution found them in 7{,}753 of 262{,}433 trials (2.95\%) for External Offsets up to 200 and in 2{,}362 of 110{,}000 trials (2.15\%) for External Offsets from 1000 to 2000, with individual searches ranging from about 0.1\% to about 9\%. These rates come from adaptive searches over different bounds, so they are not directly comparable with those of the other implementations. A 10{,}000-trial focused search took about 4 to 8 hours for the states we timed (0 and 1). The resulting faulty actions are shown in Figure~\ref{fig:deep_q_fault}. For example, in the start state ($s = 0$), the agent should move $\downarrow$ but moves $\leftarrow$ under faults. The weights are not updated during these experiments, so we treat these as transient errors. Most of the recorded faults (7{,}753 of 10{,}569) are at External Offsets up to 200, where 56\% of trials were concentrated; these offsets fall before or at the very start of the first dense layer (Table~\ref{tab:pipeline_dqn_infer}), so these faults may not act by directly perturbing the computed $Q$-values. The searches tried 1{,}358 of about 32{,}000 External Offsets, so large parts of the inference window remain unexplored at single-cycle resolution.

Useful clock glitches were also harder to identify on the deep Q-Learning agent than on the tabular and linear Q-Learning implementations. The inference procedure of the deep network spans many more clock cycles, which makes it harder to identify an External Offset that reliably produces a useful fault. One possible additional factor, which we did not test, is that small perturbations to internal parameters or intermediate values may not change the final action output.

\begin{figure}[t]
    \centering
    \includegraphics[width=0.75\linewidth]{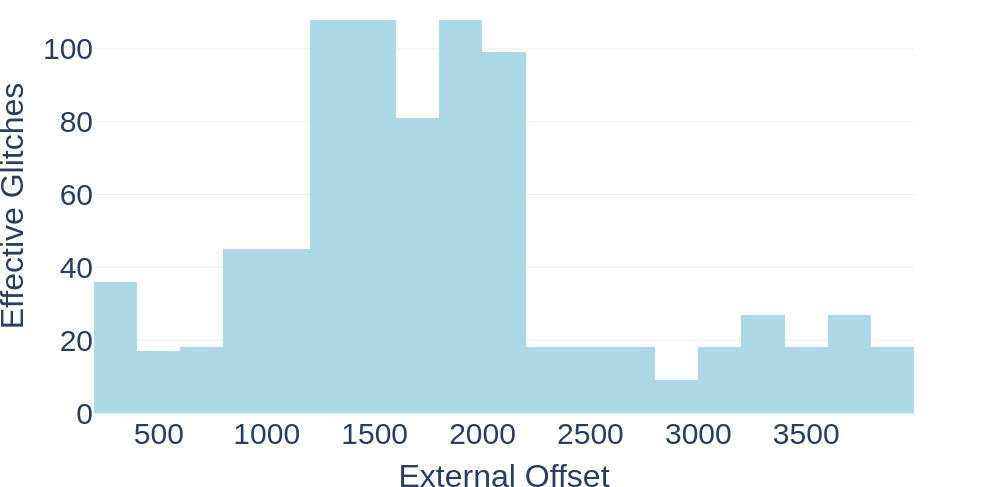}
    \caption{Distribution of effective clock glitching External Offsets on the Deep Q-Learning inference.}
    \label{fig:c1_glitch_qlearning_deep_histo}
\end{figure}

Figure~\ref{fig:c1_glitch_qlearning_deep_histo} shows a histogram of the number of successful glitches as a function of the External Offset, for offsets up to about 4000 of the 32{,}182 cycles of inference. In earlier experiments on tabular and linear Q-Learning, we mainly targeted the first $400$ clock cycles of the evaluation function and were able to obtain multiple hundreds of useful parameter sets that produced successful glitches (e.g., Figure~\ref{fig:c1_learning_step_distribution} in Appendix~\ref{sec:appendix_cglitch_params}). For the deep Q-Learning, successful glitches are most frequent between roughly 1200 and 2200 cycles. Later parts of inference are not covered by this sweep.

\subsection{Key Findings.}
Experiments on Deep Q-Learning show that replacing tabular or linear value functions with a neural network does not remove the agent's vulnerability to voltage glitch faults. Inference-time faults change the selected action in a small fraction of trials without updating the stored weights, and training-time faults can corrupt the stored weights: in the best trial at each target episode, all 340 weights became NaN. Because the searches explored only part of the inference window and used different bounds from those of the other implementations, these results do not show whether Deep Q-Learning is more or less robust than tabular or linear Q-Learning.

\section{Faulting the Traffic Light Controller}
\label{sec:fault_tlc}
The evaluations so far use Frozen Lake, a small discrete benchmark. To assess whether the vulnerabilities carry over to a larger and more safety-relevant control task, we now target online traffic light control (TLC), a widely studied application of reinforcement learning~\cite{NOAEEN2022116830}. We use SUMO-RL~\cite{sumorl}, a Gymnasium-compatible library built on the SUMO microsimulator~\cite{SUMO2018}, and a tabular Q-learning agent following recent TLC formulations~\cite{tlc_ql_ref, tlc_nonstationarity}: the state is the discretized queue of each lane of the intersection, the actions are the signal phases, and the reward is the change in cumulative vehicle waiting time between successive actions. The agent uses $\alpha=0.1$, $\gamma=0.99$, and $\epsilon=0.05$, starts from a pre-trained Q-table, and learns online during 20{,}000-second episodes with a decision every 5 seconds. The agent runs on the ChipWhisperer target exactly as before, while the environment executes on the host and exchanges states and rewards over SimpleSerial. One difference from the Frozen Lake implementations is relevant to the threat model: the reward is computed by the environment and transmitted to the device, so there is no on-chip reward lookup to fault. The reward-computation attack of Section~\ref{sec:fault_staandard_qlearning} therefore does not apply here, and we evaluate the learning step, the greedy policy used at inference, and the resilient (triple-redundant) countermeasure.

\subsection{Faulting the Learning}
\begin{figure}[t]
  \centering
  \includegraphics[width=0.75\linewidth]{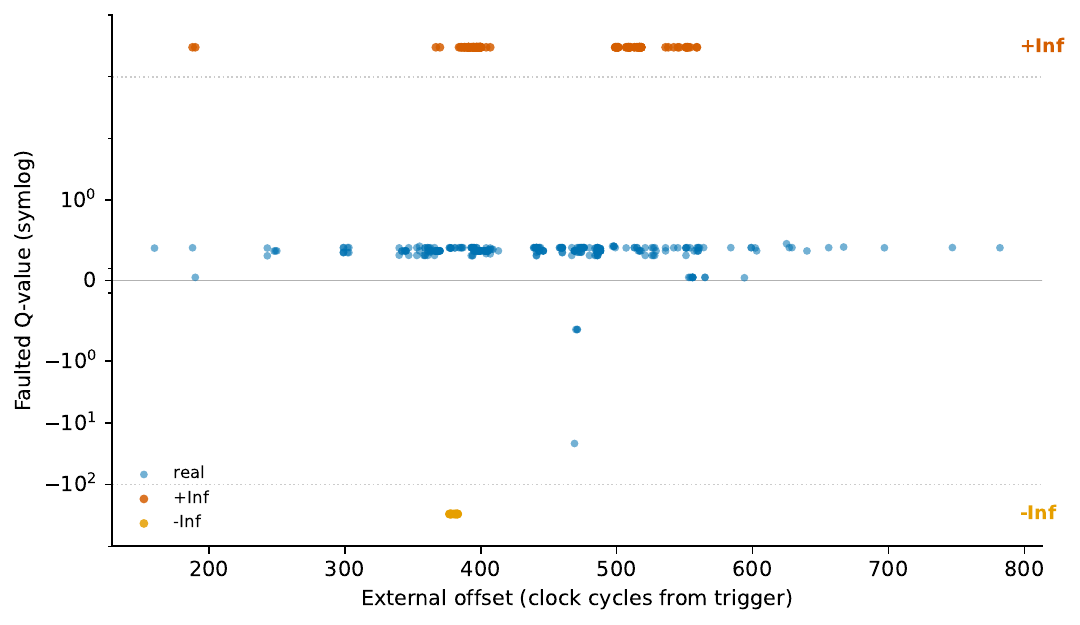}
  \caption{Faulted Q-value from voltage glitching the TLC Q-table update, versus
  external offset. Real faults cluster near the correct value with a tail to
  $-21.7$; $+\infty$/$-\infty$ are pinned as sentinel rows (symlog). No $\mathrm{NaN}$
  values were observed.}
  \label{fig:tlc-fault-value}
\end{figure}

\begin{figure}[t]
  \centering
  \includegraphics[width=0.75\linewidth]{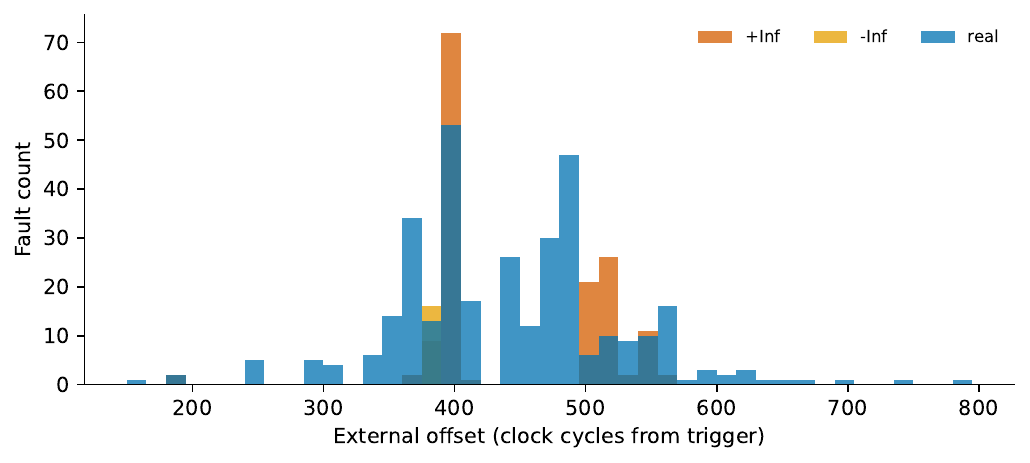}
  \caption{Successful update-fault count versus external offset, by type. $+\infty$
  faults occupy two narrow bands ($\approx$380--410 and $\approx$500--560 cycles);
  $-\infty$ faults occupy a single $\approx$6-cycle window (377--383).}
  \label{fig:tlc-fault-count}
\end{figure}

\begin{figure*}[t]
  \centering
  \includegraphics[width=\linewidth]{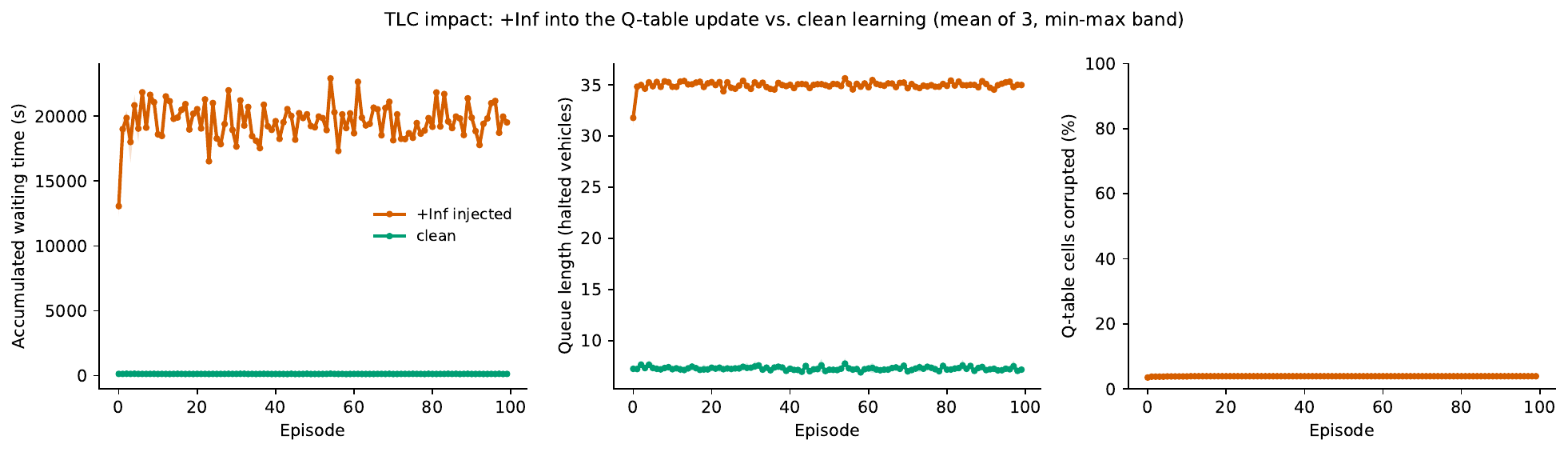}
  \caption{Impact of a single $+\infty$ injected into the TLC Q-table update
  (mean of 3 repeats, min--max band). Left: accumulated waiting time. Middle:
  queue length. Right: fraction of Q-table cells corrupted. The fault is injected
  in episode~0; the corrupted agent's waiting time and queue jump immediately and do not
  recover during the run, while the clean agent, which starts from the same pre-trained table, stays flat. Only $\sim$4\% of the table is
  corrupted.}
  \label{fig:tlc-impact}
\end{figure*}

\begin{figure}[t]
  \centering
  \includegraphics[width=0.75\linewidth]{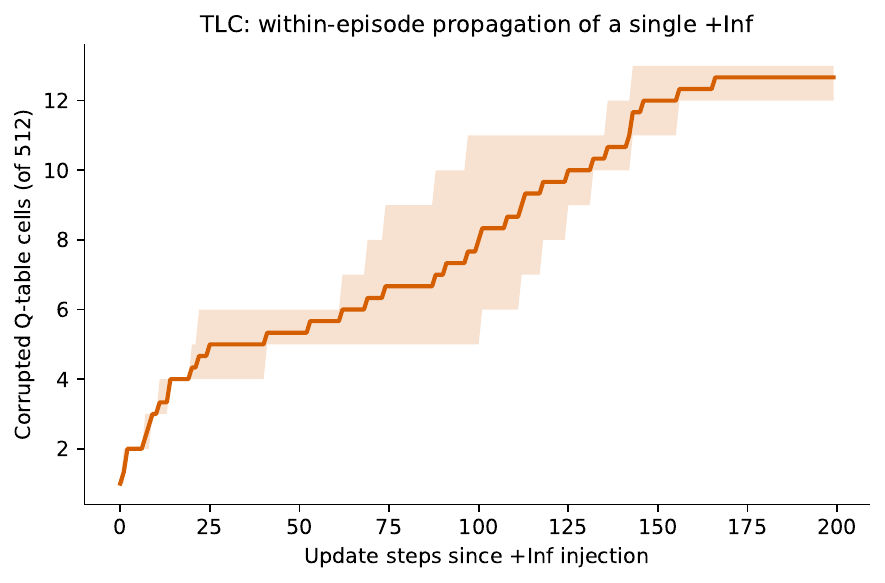}
  \caption{Within-episode propagation of a single $+\infty$. After injection, the
  number of non-finite Q-table cells is recorded per update step; the corruption
  spreads monotonically through the $\max$ operator to the states the gridlocked
  agent revisits, reaching $\sim$13 of 512 cells within the 200 update steps traced
  here and approaching the $\sim$20 cells ($\sim$4\%) observed over full episodes in
  Figure~\ref{fig:tlc-impact}. Mean of 3 traces; shaded band shows min--max.}
  \label{fig:tlc-propagation}
\end{figure}

We extend the learning-step attack of Section~\ref{sec:fault_staandard_qlearning} from the Frozen Lake benchmark to a simulated traffic-signal controller: a tabular Q-learning agent on a single four-way intersection in the SUMO microsimulator~\cite{sumorl}, with the queue state of each lane discretized into a 256-state space and two phase actions. Faults are injected into the on-chip Q-table update using the same voltage-glitching setup, with Optuna searching the width, offset, external-offset, and repeat parameters. Figures~\ref{fig:tlc-fault-value} and~\ref{fig:tlc-fault-count} characterize the successful faults. As in the tabular Frozen Lake case, the update produces both real-valued over- and under-valuations and non-finite values: real faults cluster near the correct value with a heavy tail reaching $-21.7$, while $+\infty$ faults concentrate in two narrow external-offset bands ($\approx$380--410 and $\approx$500--560 clock cycles), and a single $\approx$6-cycle window (377--383) yields $-\infty$. Notably, unlike the Frozen Lake update, this implementation produced no $\mathrm{NaN}$ values on the update path in these experiments. To measure the impact on traffic, we glitch updates during the first learning episode until a single $+\infty$ is confirmed in the table, then stop glitching and let learning proceed, averaging over three repeats. The effect is immediate and consistent across the three repeats (Figure~\ref{fig:tlc-impact}): a single fault raises the accumulated waiting time by roughly $148\times$ (from $\sim$133\,s to $\sim$19{,}500\,s) and the mean queue length by $\sim$4.9$\times$ (from $\sim$7 to $\sim$35 halted vehicles), with no recovery during the 100-episode run, while the clean agent stays at about 133\,s. The fault corrupts only $\sim$4\% of the Q-table ($\sim$20 of 512 cells): the $+\infty$ propagates through the $\max$ operator into the queue states the gridlocked controller revisits (Figure~\ref{fig:tlc-propagation}), so corrupting a small number of operationally reachable entries is enough to collapse the controller's performance.

\subsection{Faulting the Inference}
We next target the greedy policy evaluated at inference, where the trained Q-table is
fixed and the controller selects a phase by $\arg\max_a Q(s,a)$. In contrast to the
learning step, and in contrast to the Frozen Lake results of
Section~\ref{sec:fault_staandard_qlearning}, we were unable to force an incorrect phase
with voltage glitching. The glitch demonstrably reaches the targeted computation: the
trigger encloses a $221$-cycle window and the injected glitch is observed within it.
Sweeping the entire external-offset range across that window, at the width and offset settings that fault floating-point operations during the update, yields no action flips on either a state whose greedy action is $0$ or one whose action is $1$ ($0$ flips over $4{,}520$ glitches per state), while crashing the device on $99$--$100\%$ of attempts. A wider search over Width and Offset from 0 to 4400 (Table~\ref{tab:fault_search_space}) likewise produced no reproducible flip: candidate configurations that flipped an action in a single trial never reproduced when re-tested, and a reproducibility-gated re-measurement over all $256$ states found no state that flipped on a majority of repeated attempts.

One possible explanation is that with two actions, the greedy policy reduces to a single floating-point comparison, so there is no comparison chain to perturb, and a glitch strong enough to disturb that comparison may instead corrupt control flow and halt the device. In these experiments, inference-time voltage faulting acts as a denial-of-service rather than a policy-manipulation primitive: the device is halted, which is detectable, rather than induced to output an attacker-chosen phase. We did not evaluate clock glitching on this controller.

\subsection{Resilient Q-Learning TLC}
Finally we evaluate the resilient countermeasure of
Section~\ref{sec:fault_resilient_qlearning} on the traffic controller. We port the
triple-modular-redundancy design to the TLC agent: the Q-table is replicated three
times, action selection is a majority vote over the three copies, and after every
update the just-written cell is synchronized across copies by majority vote. Because
the update now writes three copies in sequence, its execution window grows from
$\sim$777 to $\sim$1{,}800--1{,}970 clock cycles, and the fault parameters that
compromise the unprotected update no longer land on a vulnerable instruction; we
therefore re-ran the parameter search against the redundant update to obtain a
configuration that corrupts a single copy.

\begin{figure*}[t]
  \centering
  \includegraphics[width=\linewidth]{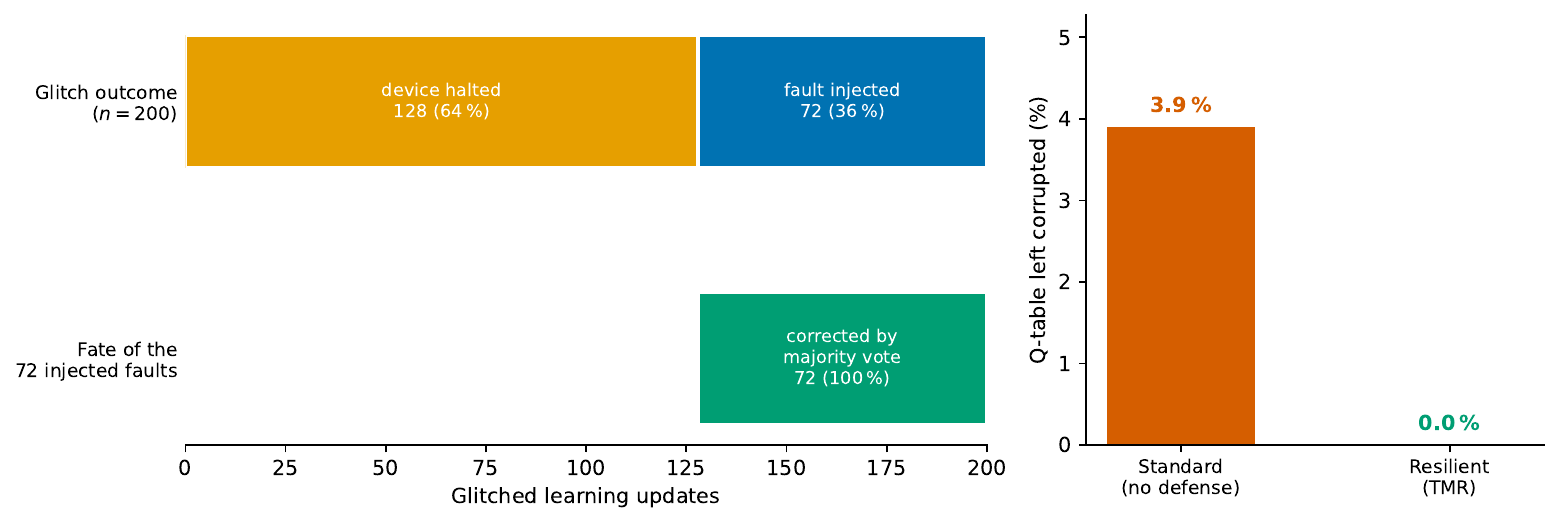}
  \caption{Resilient (TMR) TLC under the learning-step attack. Left: outcome of 200 glitched updates, injecting a fault in 72 cases ($36\%$) and halting the device in the remaining 128 ($64\%$); all 72 injected faults were restored by the post-update majority vote. Right: the standard agent has $3.9\%$ of the table non-finite ($\sim$20 of 512 cells) after the 100-episode run following one $+\infty$; the resilient value is the number of injected faults left uncorrected (0 of 72) divided by 512, checked after each update rather than after a 100-episode run, so the two bars are not the same measurement.}
  \label{fig:tlc-resilient}
\end{figure*}

With that configuration, we attempted $200$ glitched learning updates, restoring the reference table after each crash. Every glitch had an observable effect: $72$ updates ($36\%$) produced a fault flagged by the firmware's comparison of the three copies, and the remaining $128$ ($64\%$) halted the device. In all $72$ cases, the post-update majority vote left the table within our residual threshold ($10^{-3}$) of the reference table (Figure~\ref{fig:tlc-resilient}), which is consistent with corruption of a single copy. The firmware records only that a fault occurred, not its value, so this experiment does not show that these faults were $+\infty$. The unprotected agent has no such correction: there, a single uncorrected $+\infty$ persists, propagates through the $\max$ operator, and degrades the controller by $148\times$. We note four caveats. First, the injection rate varies between sessions ($30$--$48\%$ across repeated measurements of the same configuration), which is characteristic of voltage fault injection. Second, redundancy does not eliminate the attack surface but converts it: the majority of glitch attempts now halt the device, so the attacker retains a denial-of-service capability. Third, the glitch targeted only the update, not the voting step. Fourth, we did not measure traffic performance under attack, because restoring the table after crashes would confound that measurement.

\subsection{Key Findings}
The learning-step attack carries over to this controller: one $+\infty$ left in the table after glitching stops degrades control for the rest of the 100-episode run without halting the device. Our voltage-glitch searches did not reproducibly force actions in the two-action greedy policy. Triple modular redundancy corrected all $72$ faults injected in the tested configuration, while $128$ of $200$ attempts halted the device; the voting step itself was not targeted.

\section{Discussion}

In this section we highlight some of the takeaways from the results, split by the types of glitching, followed by a comparison across implementations and the limitations of this work.

\noindent Voltage Glitching:
\begin{enumerate}[label={\bfseries V\arabic*)},nosep]
    \item \textbf{Search Space} In voltage glitching we observe parameter ranges where successful glitches concentrate. In the Deep Q-Learning searches, which scanned Width and Offset up to 4000, most successful trials have Width of $1800$--$2600$, Offset of $2400$--$3000$, and Repeat of $3$--$5$. The tabular and linear searches only went up to a Width of 2280, so they cannot confirm the upper part of these ranges, and a configuration used on the traffic controller (Width $2800$, Offset $2200$, Repeat $7$) falls outside them. The External Offset is function-dependent and determines which operation is faulted. Prioritizing such ranges could reduce search time, but they should be treated as target-dependent.
    \item \textbf{Fault Targets} In voltage glitching we observe specific types of operations which successful fault models targeted. These were \textit{memory writes} and \textit{floating point operations}. By targeting memory writes a fault can impact the value stored in memory, thereby impacting not only current agent behavior but also future behavior and learning. Faults during floating point operations can result in special values such as NaN and Inf. These have the most profound impact when written back to memory, where they can severely degrade agent behavior.
\end{enumerate}

\noindent Clock Glitching:
\begin{enumerate}[label={\bfseries C\arabic*)},nosep]
    \item \textbf{Search Space} In the External Offset ranges we report, the External Offset is the dominant parameter determining fault effectiveness. For a given External Offset, a wide range of Width and Offset values produce the same faulted output. This makes the search space effectively one-dimensional, simplifying the attacker's task compared to voltage glitching. Our clock glitching figures cover only the first 200 cycles of the tabular and linear functions and about the first 4000 cycles of Deep Q-Learning inference.
    \item \textbf{Fault Targets} Our results are consistent with clock glitches disturbing individual instructions, such as comparison chains in greedy policy evaluation and arithmetic in Q-value updates, although we did not trace them and timing violations can have other effects. We observed no effect on the simple reward lookup. In our experiments clock glitching produced more repeatable faults than voltage glitching, allowing the faulted output value to be selected through the External Offset.
\end{enumerate}

\noindent Countermeasures:
\begin{enumerate}[label={\bfseries P\arabic*)},nosep]
    \item \textbf{Data Duplication} With three copies, a single glitch per update does not change the majority value of a cell when it corrupts one copy. We did not evaluate an attacker who corrupts several copies in the same update.
    \item \textbf{Fault Correction} The data duplication allows the correction of single-copy faults in the just-updated cell, which stops those faulty values from propagating during learning.
    \item \textbf{Majority Voting} Data duplication allows majority voting during inference, which mitigates the impact of corrupted data in memory. However, voltage glitches still change the selected action, although less consistently than in the unprotected implementations. Developing countermeasures which themselves are more fault-resilient should be a key future research direction.
    \item \textbf{Shared Inputs} Values read once and used for all copies, such as the reward, are not protected by replication. From the implementation, a real-valued reward fault reaches all three copies; we did not evaluate this on hardware.
    \item \textbf{Value Checks} A check that rejects NaN and Inf before they are written to the table would address non-finite faults, but not real-valued corruption, and the check itself could be faulted. We did not evaluate this countermeasure.
\end{enumerate}

\begin{table}[t]
\centering
\caption{Summary of results across implementations.}
\label{tab:summary}
\begin{adjustbox}{max width=\linewidth}
\small
\begin{tabular}{p{2.3cm}|p{3.6cm}|p{3.6cm}|p{3.6cm}}
\toprule
\textbf{Implementation} & \textbf{Faults during learning} & \textbf{Faults during inference} & \textbf{Recovery or protection} \\
\midrule
Standard & Voltage and clock faults corrupt Q-values, including NaN and Inf & Voltage and clock faults change the selected action & Real-valued errors decay with later updates; NaN and Inf can persist \\
\midrule
Robust & Same fault types as Standard & Same code as Standard, with similar faults & No protection against hardware faults was shown \\
\midrule
Resilient & Single-copy corruption is corrected; 58 of 21{,}150 full training runs still ended with a changed table & Voltage faults still change the action; clock glitching was generally unsuccessful & Protects stored Q-values, not shared inputs or the voting code \\
\midrule
Linear & Voltage faults corrupt weights, including very large and Inf values & Voltage and clock faults change the action; the most consistent voltage faults occur in the first 100 cycles & Real-valued insertions recovered in 23 of 23, 26 of 56, and 35 of 48 cases; non-real values rarely recovered \\
\midrule
Deep Q & In the best trial at each target episode, all 340 weights became NaN & Voltage and clock faults transiently change the action & Not evaluated here \\
\midrule
Traffic controller & One $+\infty$ persists for the 100-episode run and raises waiting time $148\times$ & No reproducible action flips with voltage glitching; clock glitching not tested & TMR corrected 72 of 72 injected faults; 128 of 200 attempts halted the device \\
\bottomrule
\end{tabular}
\end{adjustbox}
\end{table}

\noindent\textbf{Security Implications.} A fault written into the table can outlast the attack. On the traffic controller, glitching stops once a $+\infty$ lands, and the degradation continues for the rest of the 100-episode run without halting the device, whereas a device halt is detectable. Under the synchronization assumed in Section~\ref{sec:threat}, an attacker therefore does not need continued access once a fault has been written into the table.

\noindent\textbf{Limitations.} All experiments use one microcontroller target and our own C implementations, so fault effects and parameters may not transfer to other devices or libraries. The evaluation uses Frozen Lake and one simulated intersection rather than a physical system. Experiments rely on a firmware trigger. Search coverage differs between experiments, and the clock glitching sweeps and the Deep Q-Learning searches cover only part of the targeted windows. The recovery experiments use inserted values and differ in learning rate and map, the traffic-controller impact results use three repeats, and we do not report confidence intervals.

\section{Conclusion}
\label{sec:concl}

This study examines the vulnerabilities of Q-Learning implementations to hardware fault injection attacks, an often overlooked aspect of reinforcement learning (RL) security. We showed that faults can corrupt learning and change the selected action in standard Q-Learning, and that Robust Q-Learning, which targets environmental uncertainty, does not prevent these attacks. We also evaluated a Resilient Q-Learning variant based on triple modular redundancy. It corrects corruption of a single Q-table copy, but inputs shared by the copies and the voting code remain outside its protection. Carrying the analysis to an online traffic light controller in simulation showed that the learning-step vulnerability is not limited to small benchmarks: a single injected infinite value degraded the controller by $148\times$ in vehicle waiting time while corrupting under $4\%$ of its Q-table. The same redundancy scheme corrected the faults injected in our tests, but glitching could still halt the device, which leaves an availability attack intact.
Future work will focus on extending this resilience to other RL paradigms, evaluating attacks with realistic triggers and on other microcontrollers, and investigating the feasibility of efficient algorithmic countermeasures.

\section*{Ethical Considerations}
This work evaluates fault-injection effects on Q-learning and Deep Q-learning in controlled laboratory experiments on our own implementations and equipment, not on production systems or commercial products. The glitching techniques used are publicly documented, and we evaluate a countermeasure alongside the attacks to guide safer RL deployments.

\begin{acks}
This work was supported in part by National Science Foundation grants \nsf{2245344} and \nsf{2332406}.
This work is also partially funded by the European Union, by grant No. 101135183. Views and opinions expressed are, however, those of the author(s) only and do not necessarily reflect those of the European Union. Neither the European Union nor the granting authority can be held responsible for them.
The author(s) acknowledge the support of the USI Doctoral Mobility Grant.
\end{acks}

\bibliographystyle{ACM-Reference-Format}
\bibliography{sections/bibliography/abbrev3,sections/bibliography/crypto,sections/bibliography/biblio}

\appendix
\section*{Appendix}
\appendix

\section{Program Counter Tracing}

To understand fault mechanisms, ARM debugging features on the target device capture program counters periodically during execution. Program counters identify the assembly instruction and corresponding C code line executing when the fault occurs. Trigger synchronization between glitch injection and program counter capture enables matching faults to specific clock cycles. While program counter capture is periodic, the delay between trigger and first capture is stochastic. This randomness enables partial reconstruction of instruction timing by repeatedly executing the same code with different capture delays, building a dictionary mapping program counters to clock~cycles.

\section{Comparison to Prior Work}

Table~\ref{tab_prior_work} lists a number of exemplary works on fault injection and machine learning. Existing work has demonstrated a range of attacks, from voltage glitching and clock glitching, to Rowhammer attacks. At the same time, the target of these works has been mostly on deep neural networks, and mainly on image classification networks. In contrast, this work focuses on RL. It evaluates a number of RL implementations, from Q-Learning based on Q-Tables to Linear and Deep Q-Learning, and considers both unprotected and protected implementations.

\begin{table}[t]
\centering
\caption{Related work on fault injection attacks and machine learning.}
\begin{adjustbox}{width=\textwidth}
\label{tab_prior_work}
\footnotesize
\begin{tabular}{|p{1.3cm}|p{1.5cm}|p{1.75cm}|p{1.65cm}|p{1.65cm}|p{1.5cm}|p{1.5cm}|p{1.5cm}|}
\hline
\textbf{Paper} & \textbf{Attack Type} & \textbf{Attack Goal} & \textbf{Model/Size} & \textbf{Application Domain} & \textbf{Tests Individual Functions} & \textbf{Attacker's Knowledge} & \textbf{Target Component} \\

\hline
\cite{yao2024deepvenom}  & Rowhammer & Backdoor insertion & Medium to Large (VGG16, ResNet18/50, ViT) & Image classification & No, Full model & White-box & Model weights in memory \\

\hline
\cite{breier2018practical} & Laser fault injection & Misclassification & 3-layer NNs; sizes not provided & Image classification & Yes, (softmax, ReLU, sigmoid, tanh) & White-box & Activation function computations \\

\hline
\cite{sun2023lightning} & Fault injection & Misclassification & Medium to Large (widely-used DNNs) & DL inference on GPUs & No, Full model & White-box & GPU \\
\hline
\cite{liu2020imperceptible} & Clock glitch injection & Targeted Misclassification & Large (AlexNet, VGG16, ResNet50, etc.) & Image classification & No, Full model & White-box/Black-box & Computation units in DNN accelerator \\
\hline
\cite{khoshavi2020fiji} & Fault injection & Misclassification & Small to Medium (CNNs for MNIST, CIFAR10, etc.) & Image and speech recognition & Yes, (Layers and neurons) & White-box & Model weights and activations \\
\hline
\cite{yao2020deephammer} & Rowhammer & Indiscriminate damage & Medium to Large (ResNet, VGG, AlexNet) & Image classification & No, Full model & White-box/Black-box & Model weights in DRAM \\
\hline
\cite{hong2019terminal} & Single bit-flip fault injection & Indiscriminate damage & Medium to Large (various CNNs) & Image classification & Yes, (Parameters/layers) & White-box/Black-box & Individual bits in model weights \\
\hline
\cite{dong2023one} & Bit-flip attacks during training & Model degradation & Medium to Large (ResNet, VGG) & Image classification & No, Full model & White-box & Model weights (Last FC layer) \\
\hline
\textbf{This work} & Voltage glitch on Q-Table Q-Learning & Corrupting Q-Table or Manipulating Action Choices & Small (Q-learning) & Reinforcement learning & Yes & White-box & Computation during training and inference \\
\hline
\textbf{This work} & Clock glitch on Q-Table Q-Learning & Corrupting Q-Table or Manipulating Action Choices & Small (Q-learning) & Reinforcement learning & Yes & White-box & Computation during training and inference \\
\hline
\textbf{This work} & Voltage glitch on linear Q-Learning & Corrupting Model Parameters or Manipulating Action Choices & Small (Linear Q-learning) & Reinforcement learning & Yes & White-box & Weight updates during training and computation during inference \\
\hline
\textbf{This work} & Clock glitch on linear Q-Learning & Manipulating Action Choices & Small (Linear Q-learning) & Reinforcement learning & Yes & White-box & Computation during inference \\
\hline
\textbf{This work} & Voltage glitch on Deep Q-Learning & Corrupting Model Weights or Manipulating Action Choices & Small (340 parameters) & Reinforcement learning & No, Full model & White-box & Weight updates during training and computation during inference \\
\hline
\textbf{This work} & Clock glitch on Deep Q-Learning & Manipulating Action Choices & Small (340 parameters) & Reinforcement learning & No, Full model & White-box & Computation during inference \\
\hline
\textbf{This work} & Voltage glitch on Q-Table traffic controller & Corrupting Q-Table & Small (512 Q-table cells) & Traffic-signal control (simulated) & Yes & White-box & Q-Table update during training \\
\hline
\end{tabular}
\end{adjustbox}
\end{table}

\section{Voltage Glitching Parameter Spaces}
\label{sec:appendix_vglitch_params}

This appendix presents the voltage glitching parameter space visualizations for the Q-Learning attacks. Each figure shows the outcome of the search across the three-dimensional space of Fault Width, Fault Offset, and External Offset. Green points indicate successful faults and red points indicate device resets; trials with normal operation are not drawn.

\begin{figure}[h]
    \centering
    \includegraphics[width=0.8\linewidth]{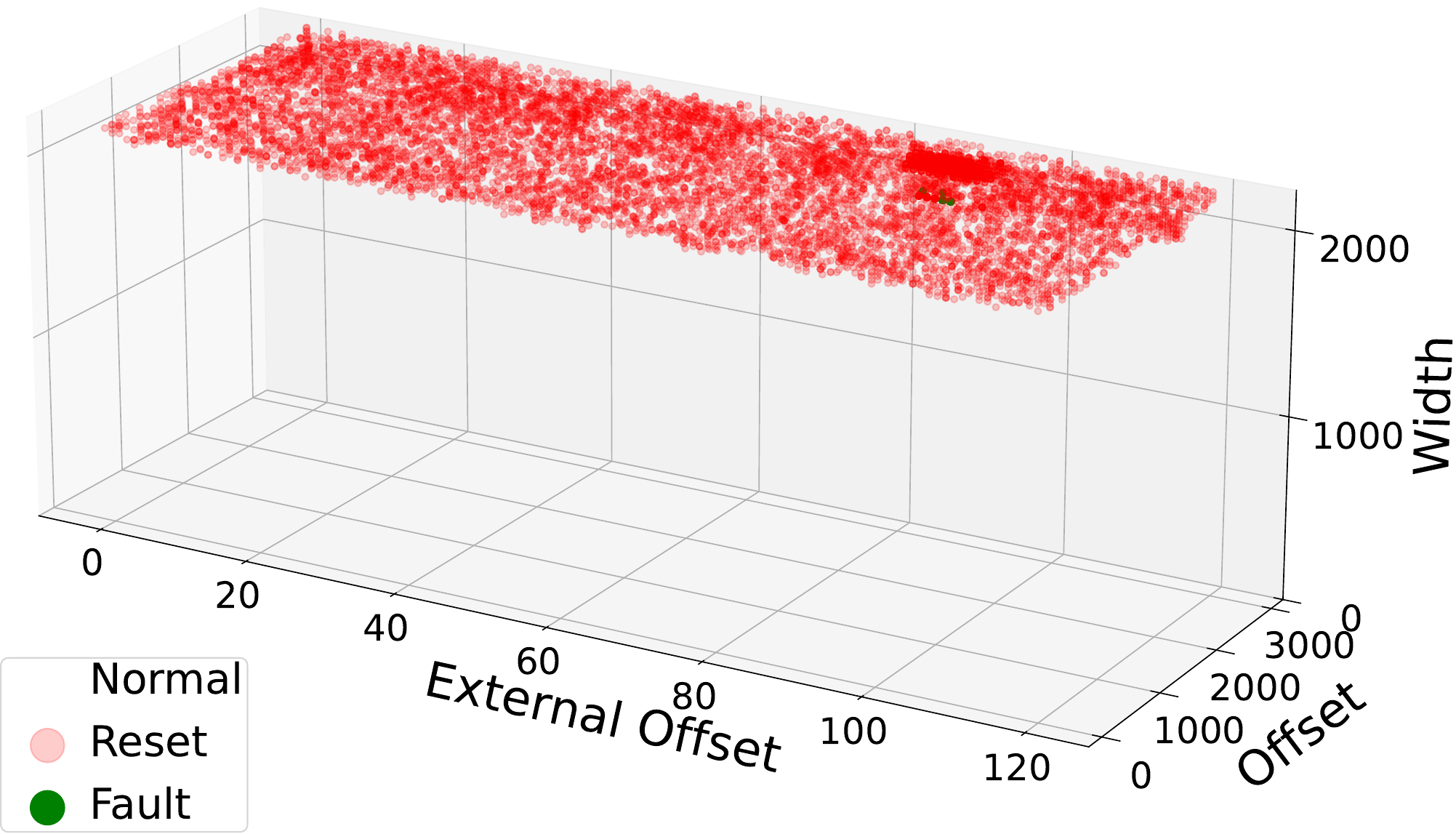}
    \caption{Voltage glitching parameter space for reward computation attacks on standard Q-Learning. Successful faults corrupt reward values during lookup, producing altered values ($\approx2.123$), NaN, or large negative values ($<-1\text{e}37$).}
    \label{fig:appendix_reward_v_space}
\end{figure}

\begin{figure}[h]
    \centering
    \includegraphics[width=0.8\linewidth]{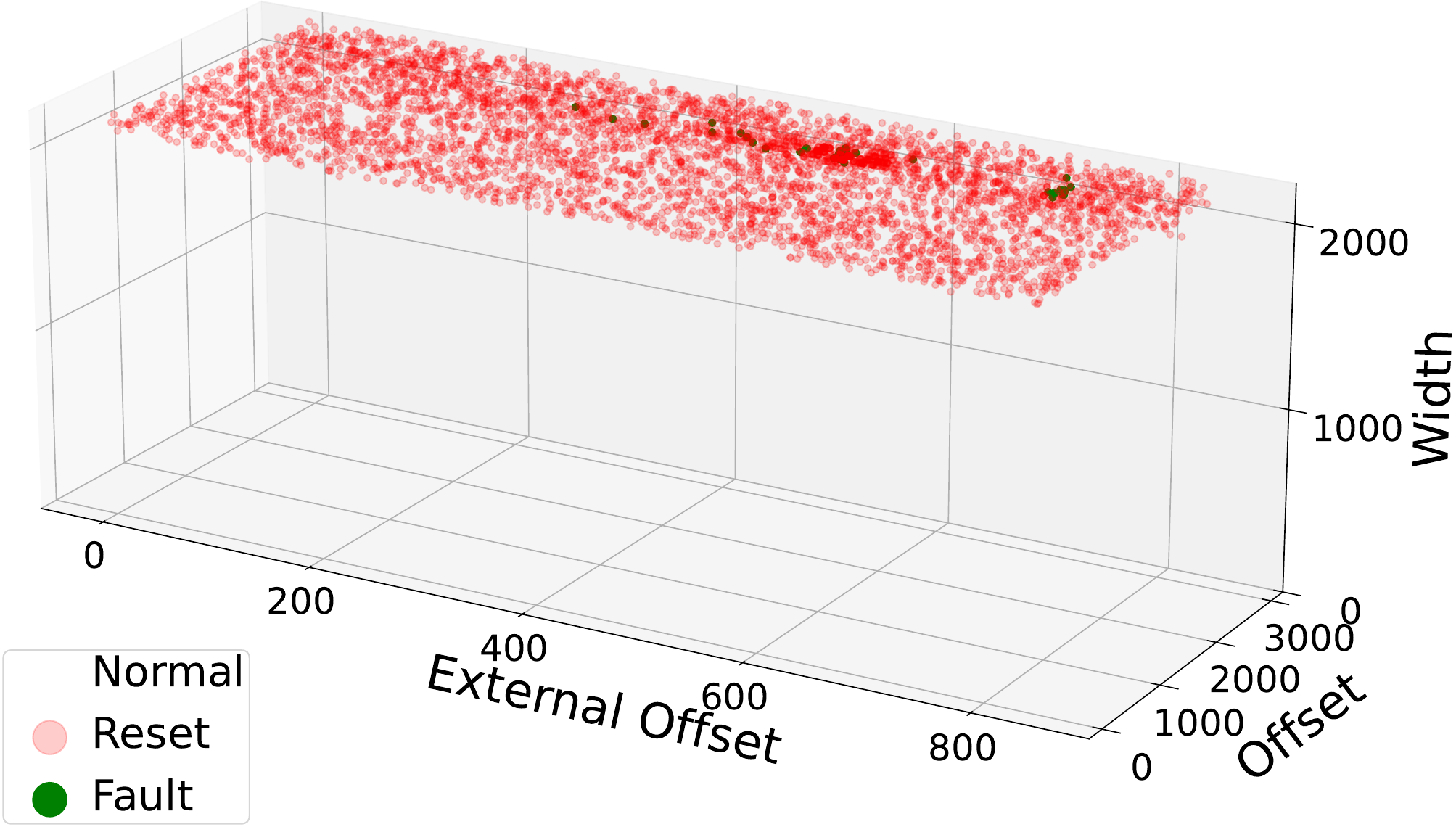}
    \caption{Voltage glitching parameter space for learning step attacks on standard Q-Learning. Successful faults corrupt Q-value updates, causing over/undervaluation and NaN/Inf injection that can propagate through the table.}
    \label{fig:appendix_update_v_space}
\end{figure}

\begin{figure}[h]
    \centering
    \includegraphics[width=0.8\linewidth]{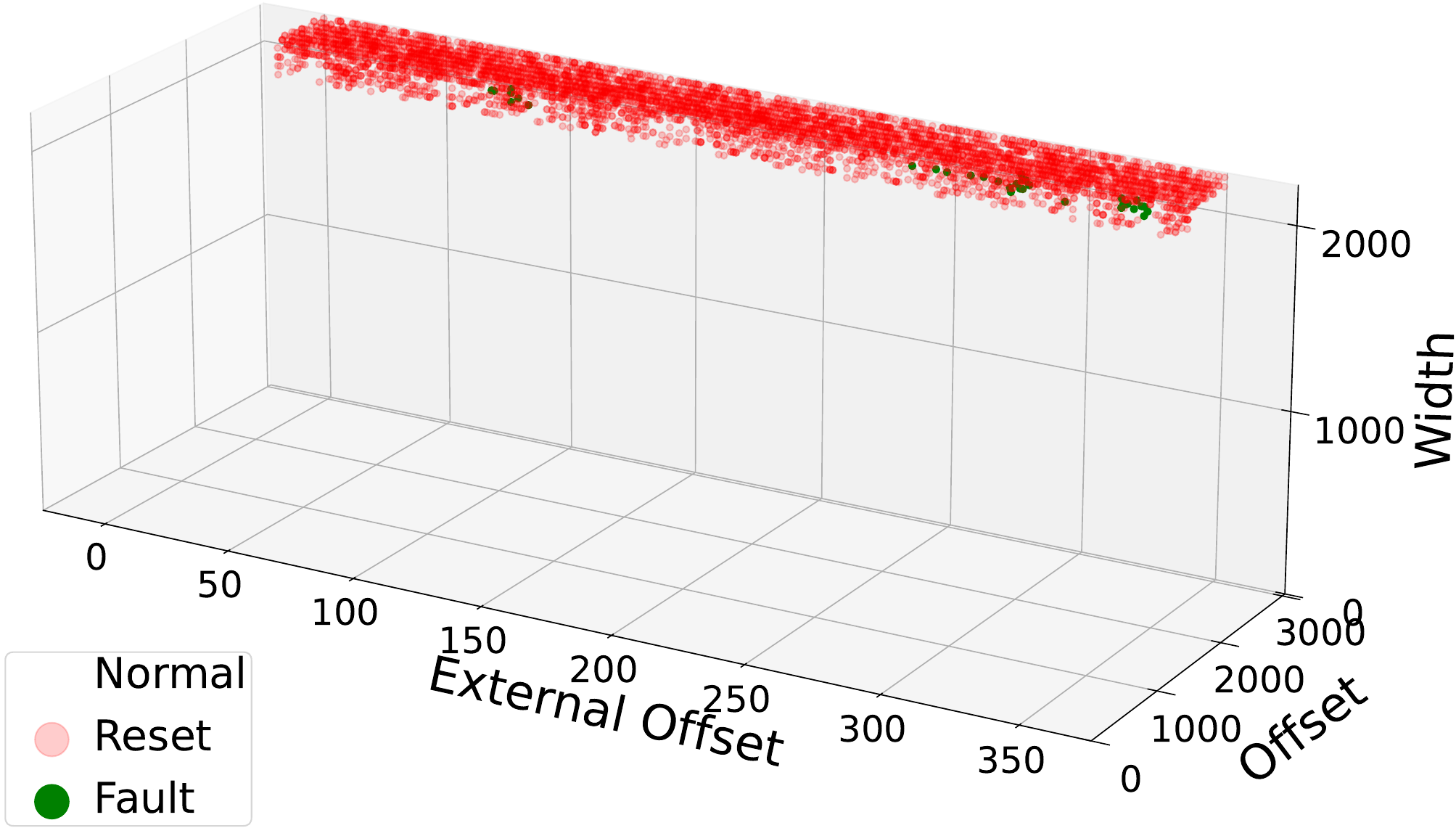}
    \caption{Voltage glitching parameter space for greedy policy inference attacks on standard Q-Learning, searched on state 8. In follow-up tests on states 0, 4, and 8, the most consistent configurations force the $\uparrow$ action in 92-100\% of non-crashing outputs.}
    \label{fig:appendix_inference_v_space}
\end{figure}

\begin{figure}[h]
    \centering
    \includegraphics[width=0.8\linewidth]{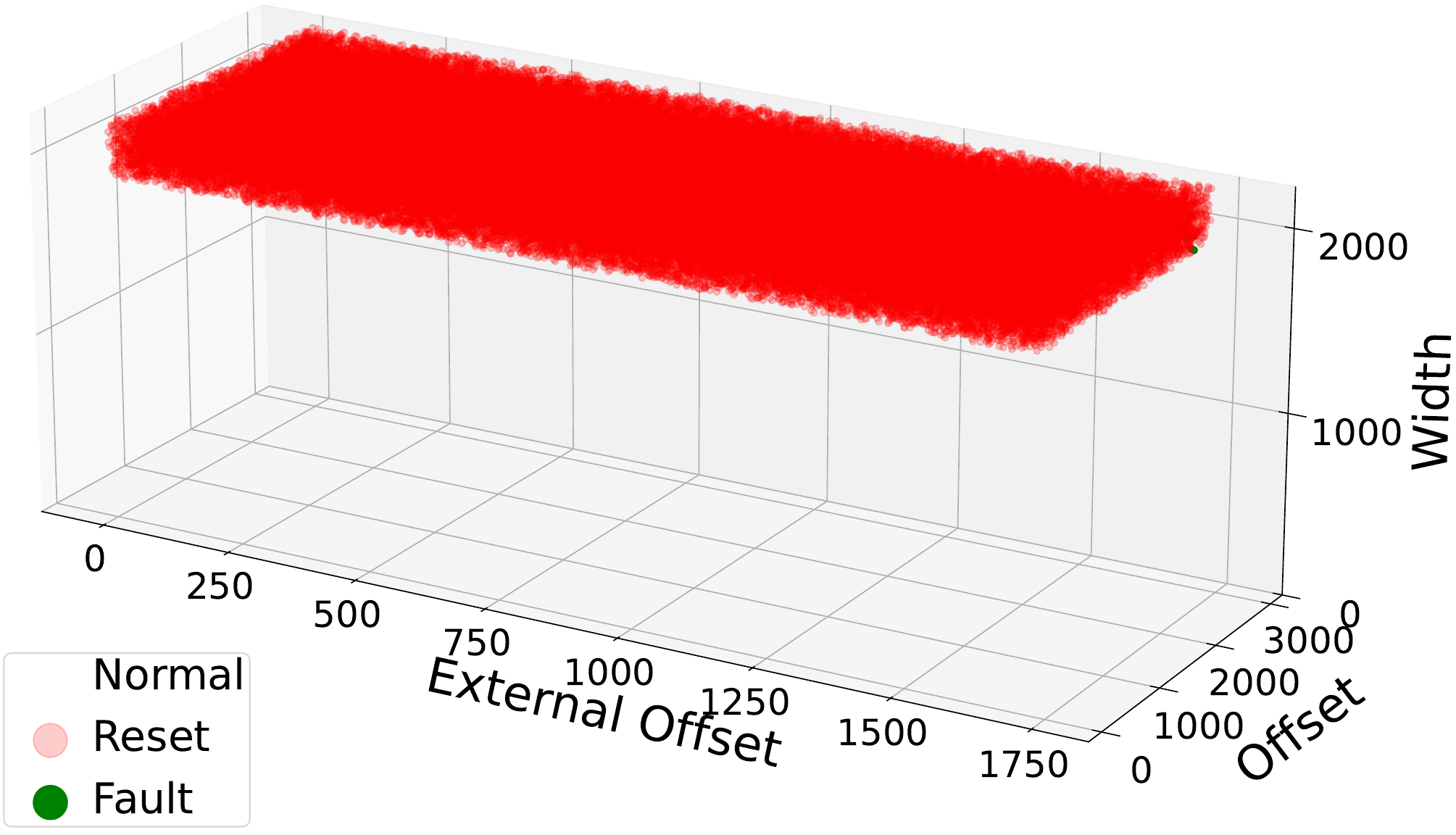}
    \caption{Voltage glitching parameter space for Robust Q-Learning update attacks.}
    \label{fig:appendix_robust_update_v_space}
\end{figure}

\begin{figure}[h]
    \centering
    \includegraphics[width=0.8\linewidth]{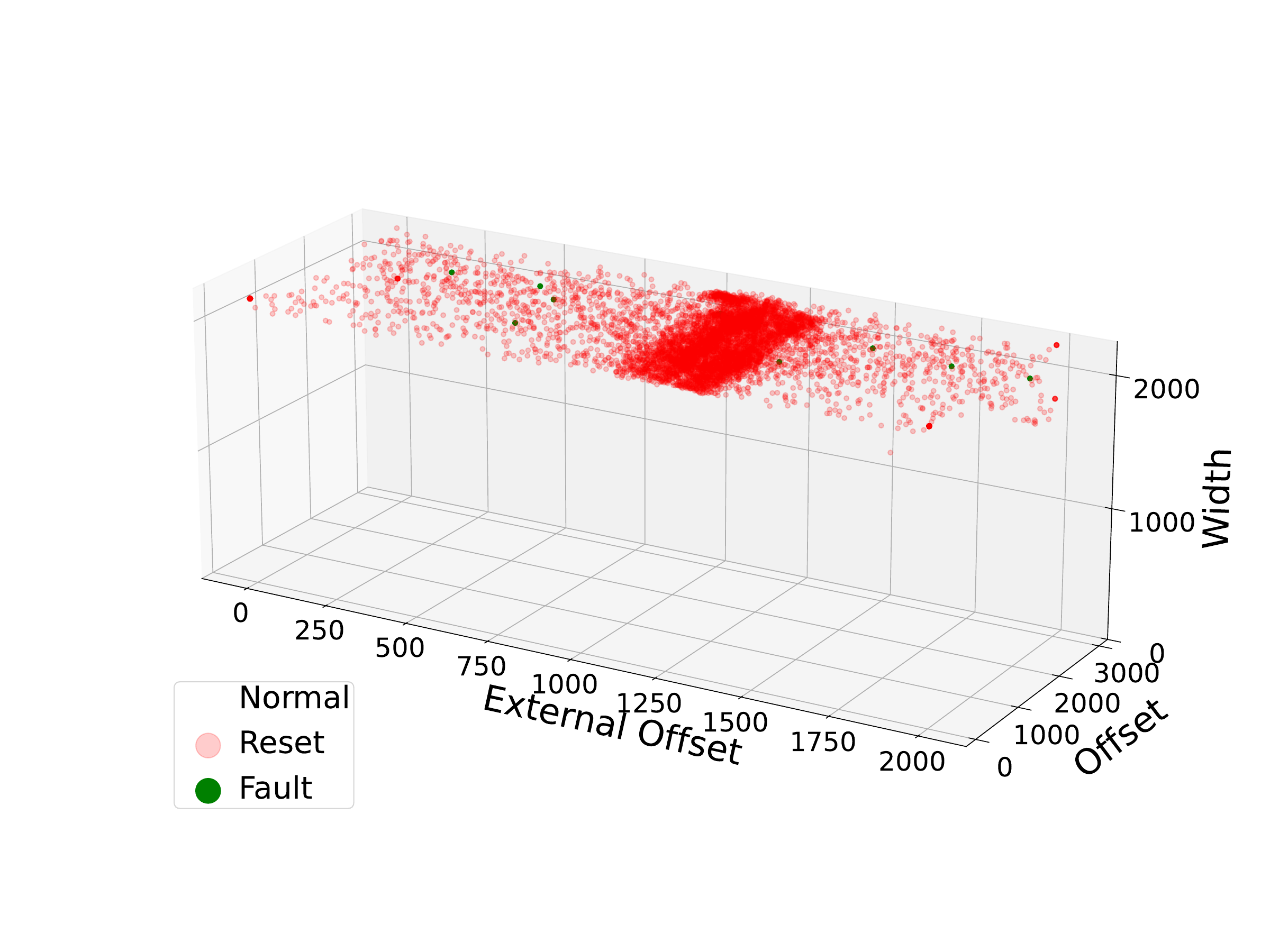}
    \caption{Voltage glitching parameter space for Resilient Q-Learning update attacks. Each trial is a full training run with the glitch armed on every update. Fault points ended with a Q-table different from the reference (58 of 21{,}150 trials, 6 with infinite values).}
    \label{fig:appendix_resilient_update_v_space}
\end{figure}

\begin{figure}[h]
    \centering
    \includegraphics[width=0.8\linewidth]{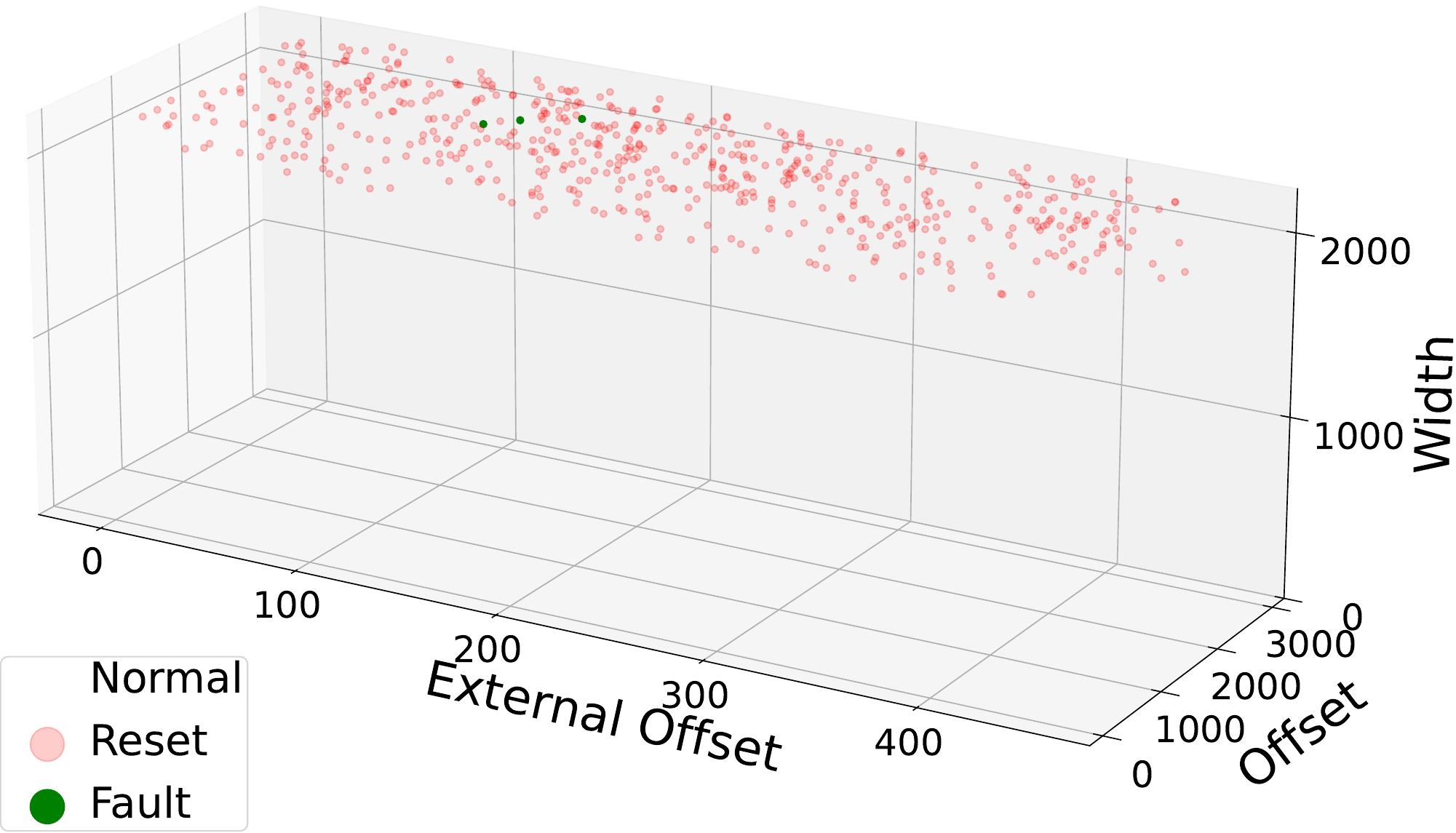}
    \caption{Voltage glitching parameter space for Resilient Q-Learning fault correction mechanism attacks. Successful faults can trigger unnecessary corrections, which reset a cell to 0 in all Q-Tables.}
    \label{fig:appendix_resilient_faultcorr_v_space}
\end{figure}

\begin{figure}[t]
    \centering
    \includegraphics[width=0.8\linewidth]{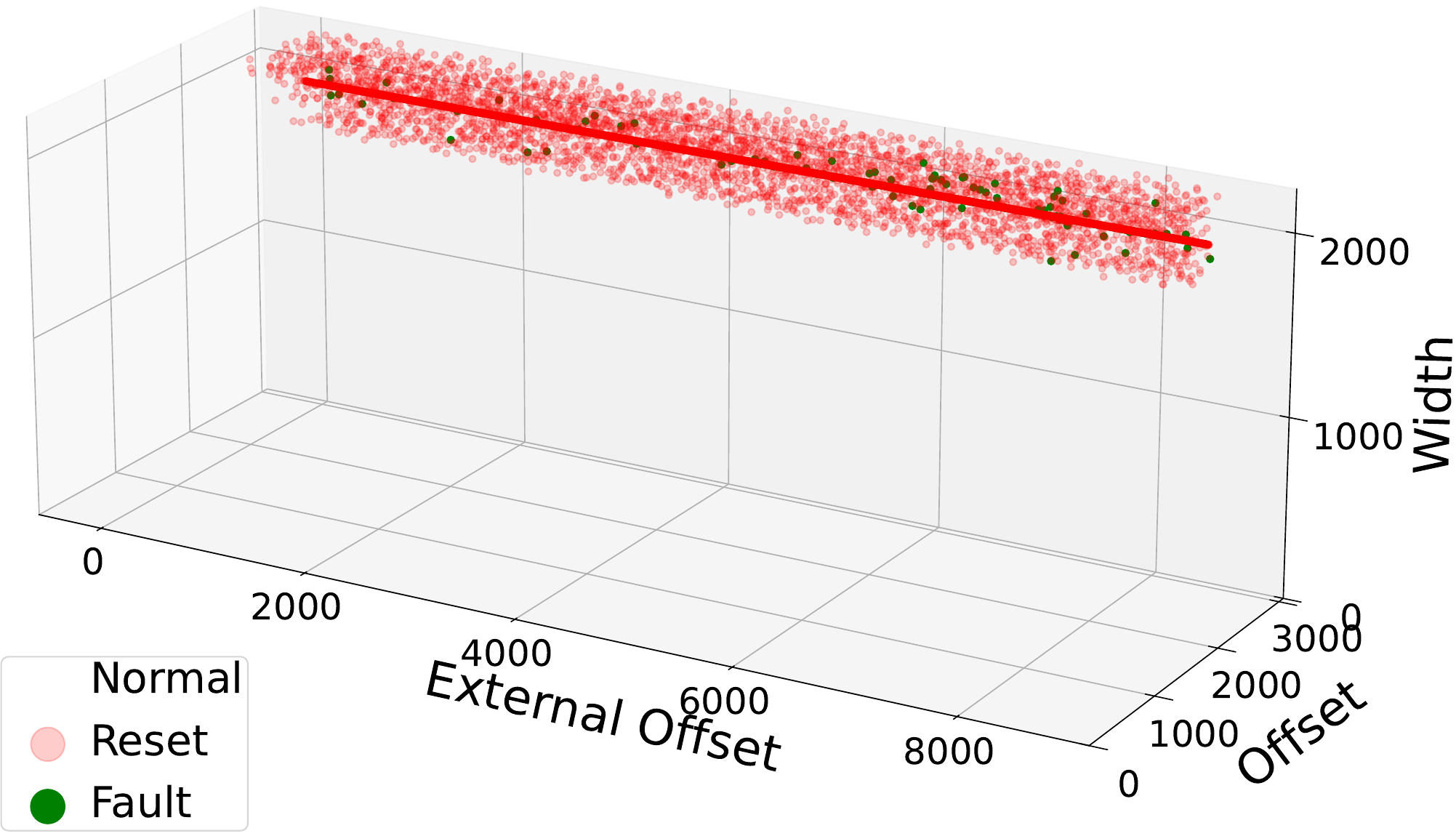}
    \caption{Voltage glitching parameter space for Linear Q-Learning weight update attacks. Successful faults can alter weights in memory.}
    \label{fig:fl_linear_update_v_space}
\end{figure}

\begin{figure}[t]
    \centering
    \includegraphics[width=0.8\linewidth]{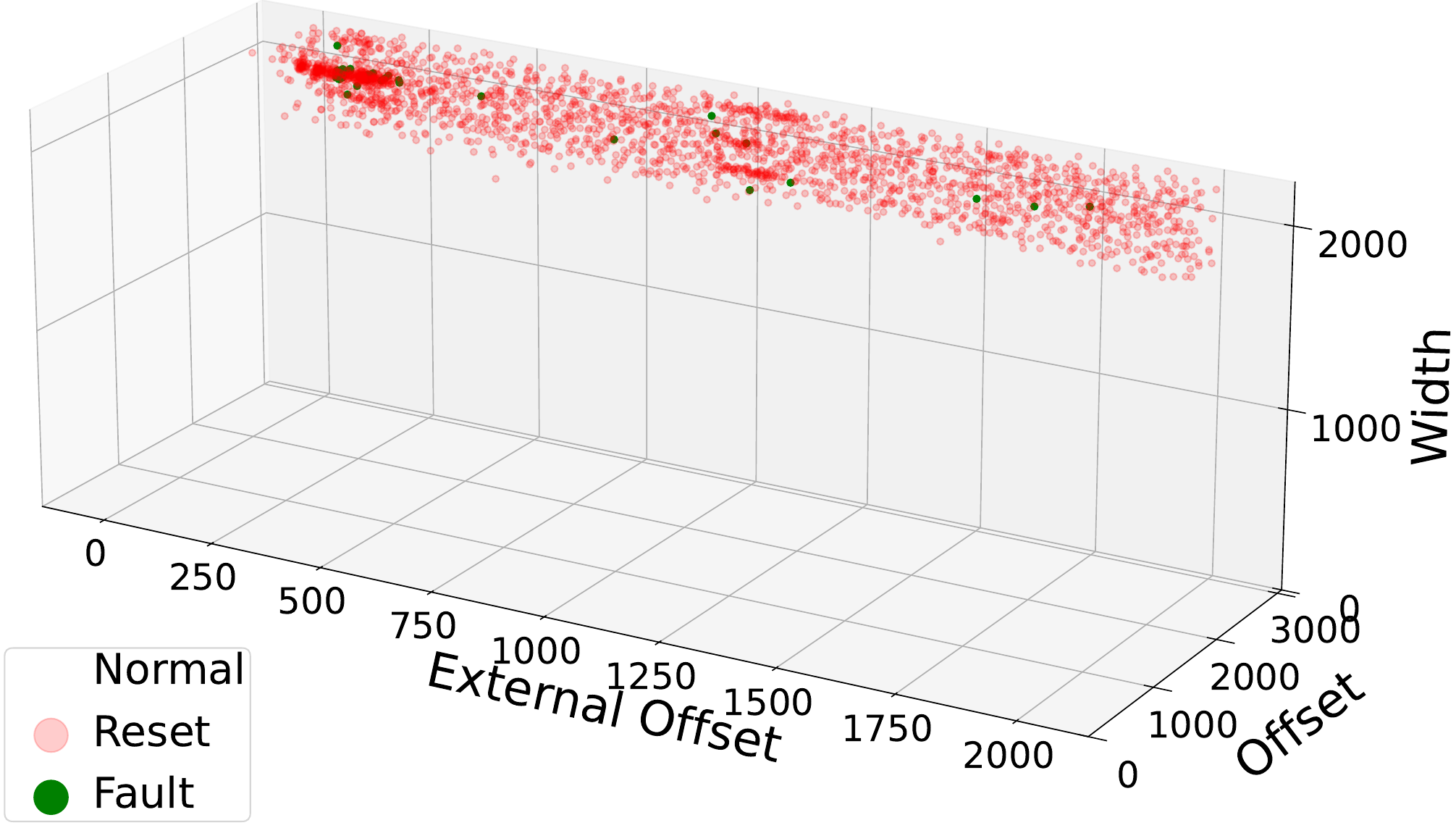}
    \caption{Voltage glitching parameter space for Linear Q-Learning inference attacks, searched on state 8.}
    \label{fig:fl_linear_inference_v_space}
\end{figure}

\section{Clock Glitching Parameter Spaces}
\label{sec:appendix_cglitch_params}

This appendix lists the visualization of clock glitching parameter search spaces for various attacks on different implementations of Q-Learning. The format and legends are similar to the figures in the previous section. Note that in some edge cases, a "normal\_reset" or "fault\_reset" behavior can be observed, representing that while the output can still be read, the device has entered a glitched state that can be detected and will lead to a reset.

\begin{figure}[t]
    \centering
    \includegraphics[width=0.8\linewidth]{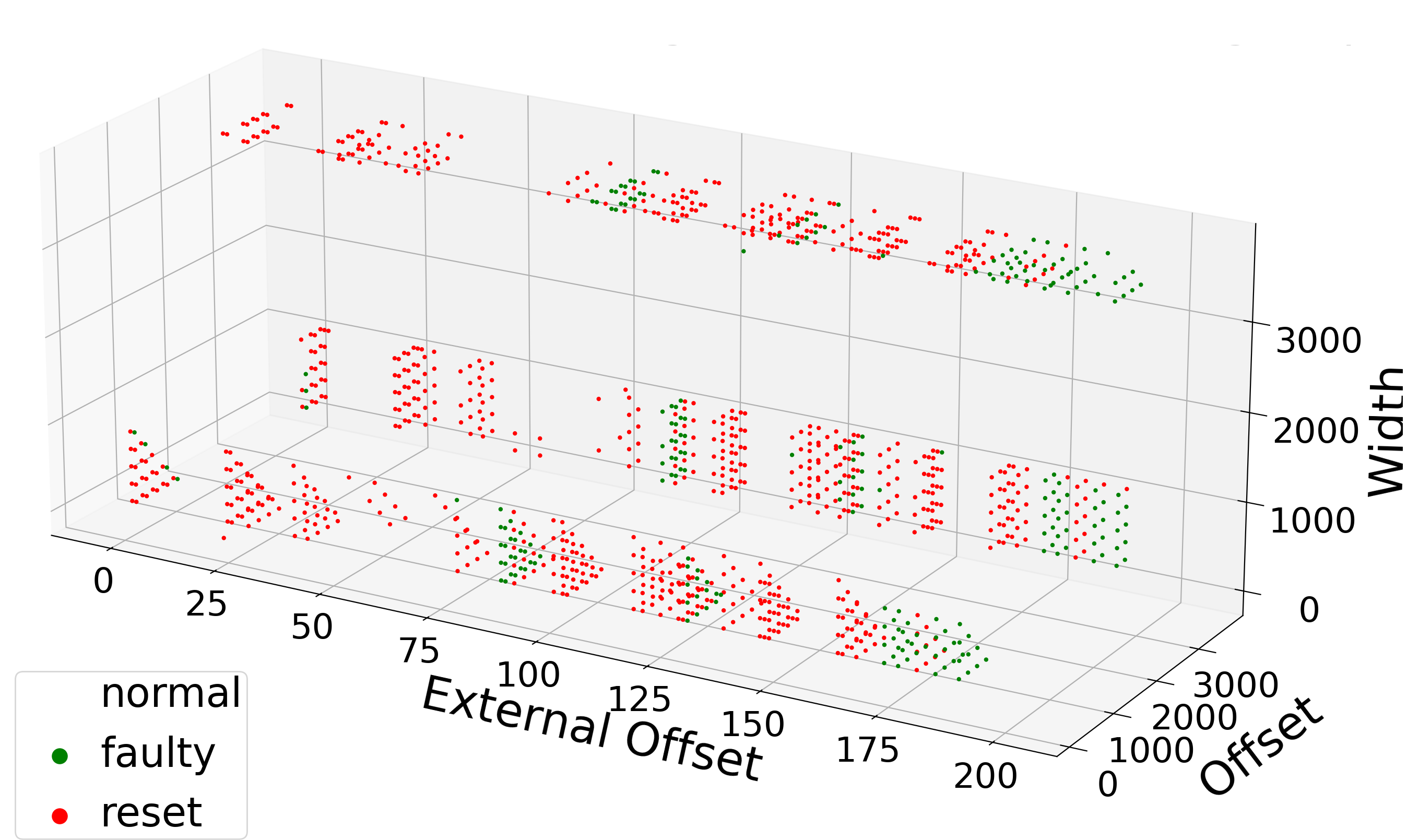}
    \caption{Clock glitching parameter space for standard Q-Learning on learning update, for External Offsets up to 200 of the 749-cycle update. Faults are widely distributed across three dimensions and were reproducible in our tests.}
    \label{fig:c1_learning_step_distribution}
\end{figure}

\begin{figure}[t]
    \centering
    \includegraphics[width=0.8\linewidth]{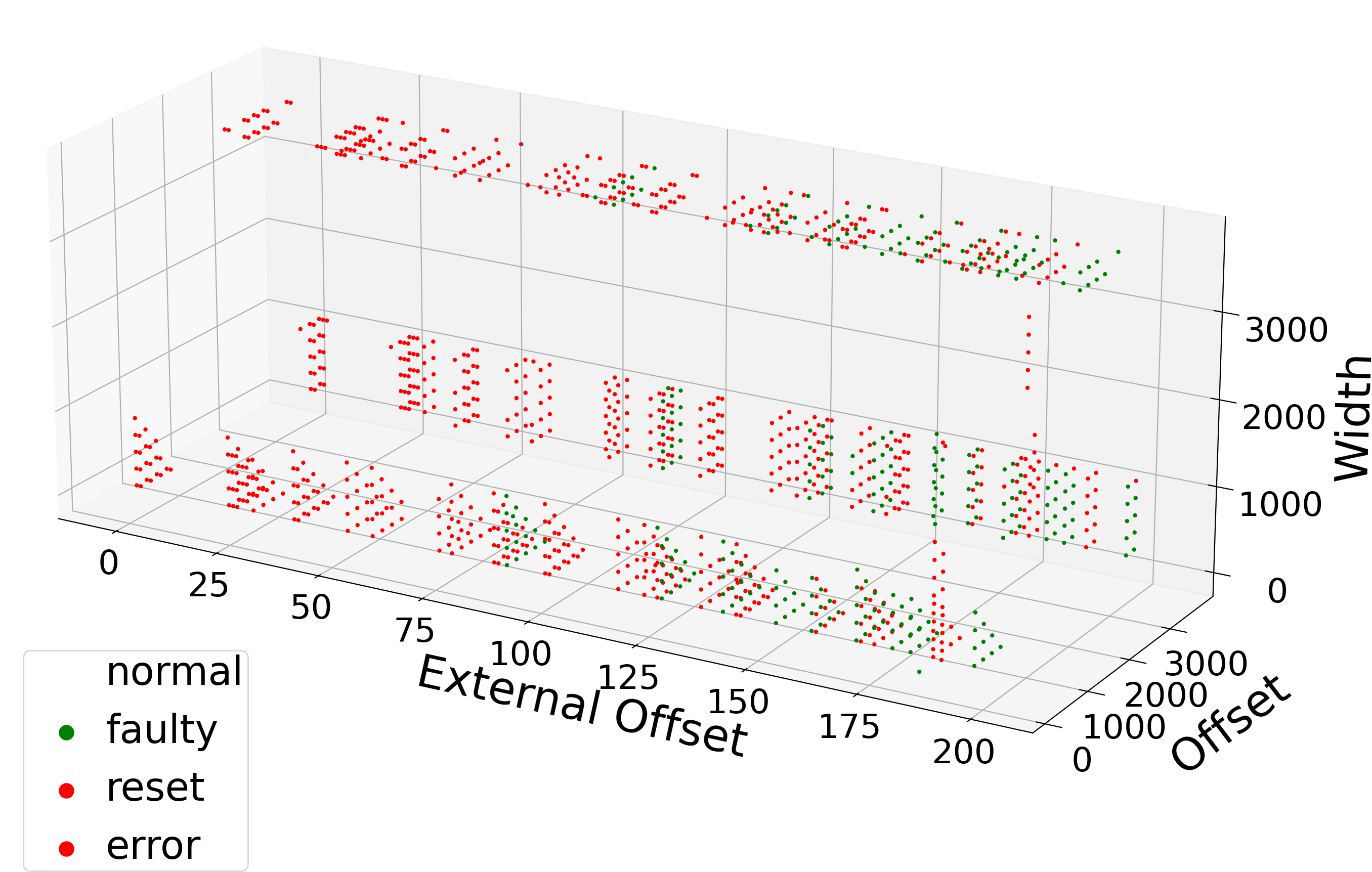}
    \caption{Clock glitching parameter space for Robust Q-Learning on learning update, for External Offsets up to 200 of the 1{,}727-cycle update. In this range the amount of successful faults is close to that of the standard implementation.}
    \label{fig:c1_robust_learning_step_distribution}
\end{figure}

\begin{figure}[t]
    \centering
    \includegraphics[width=0.8\linewidth]{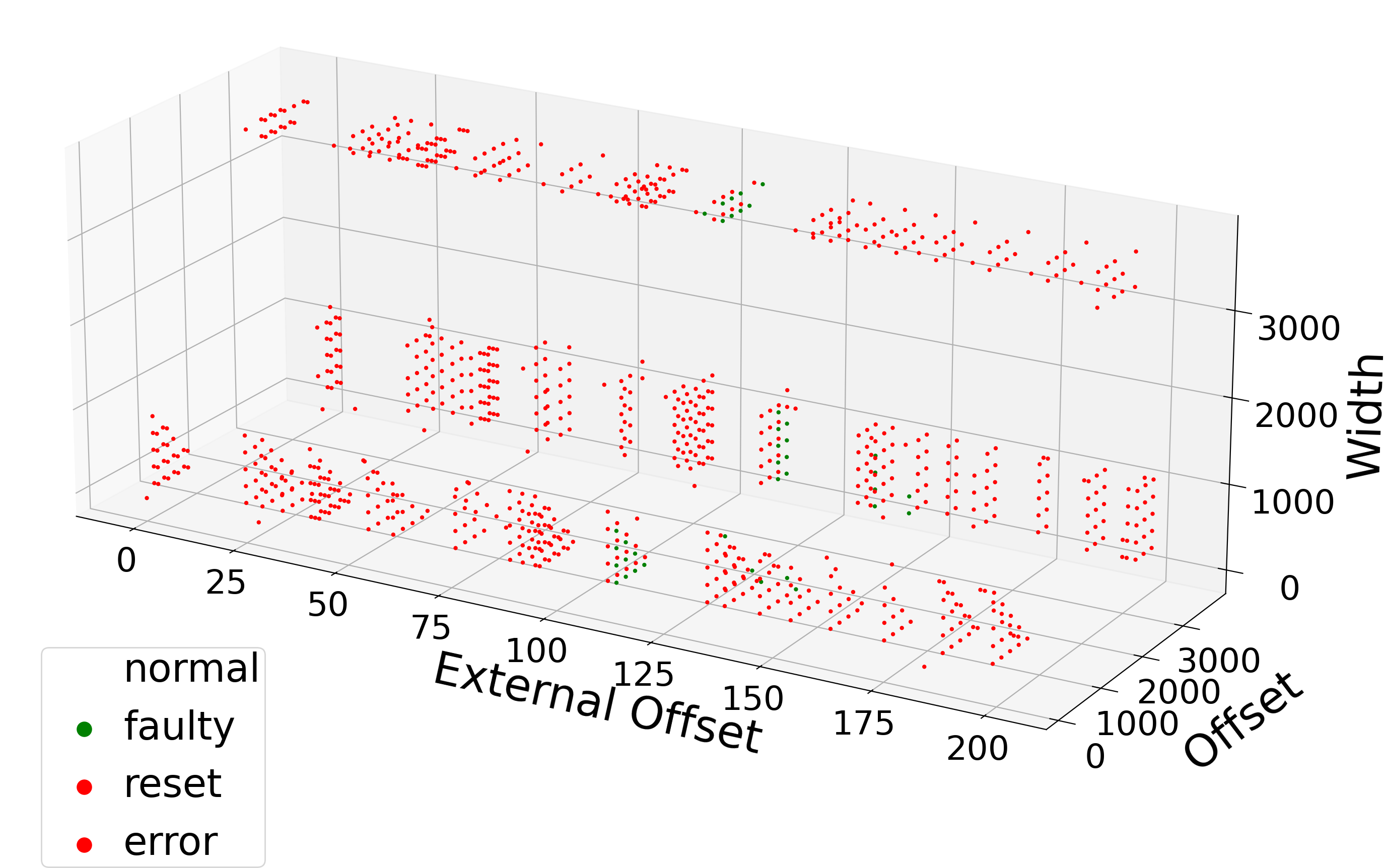}
    \caption{Clock glitching parameter space for Resilient Q-Learning on learning update, for External Offsets up to 200 of the 1{,}938-cycle update. Only two values of external offset that give a successful fault can be found in this range.}
    \label{fig:c1_resilient_learning_step_distribution}
\end{figure}

\begin{figure}[t]
    \centering
    \includegraphics[width=0.8\linewidth]{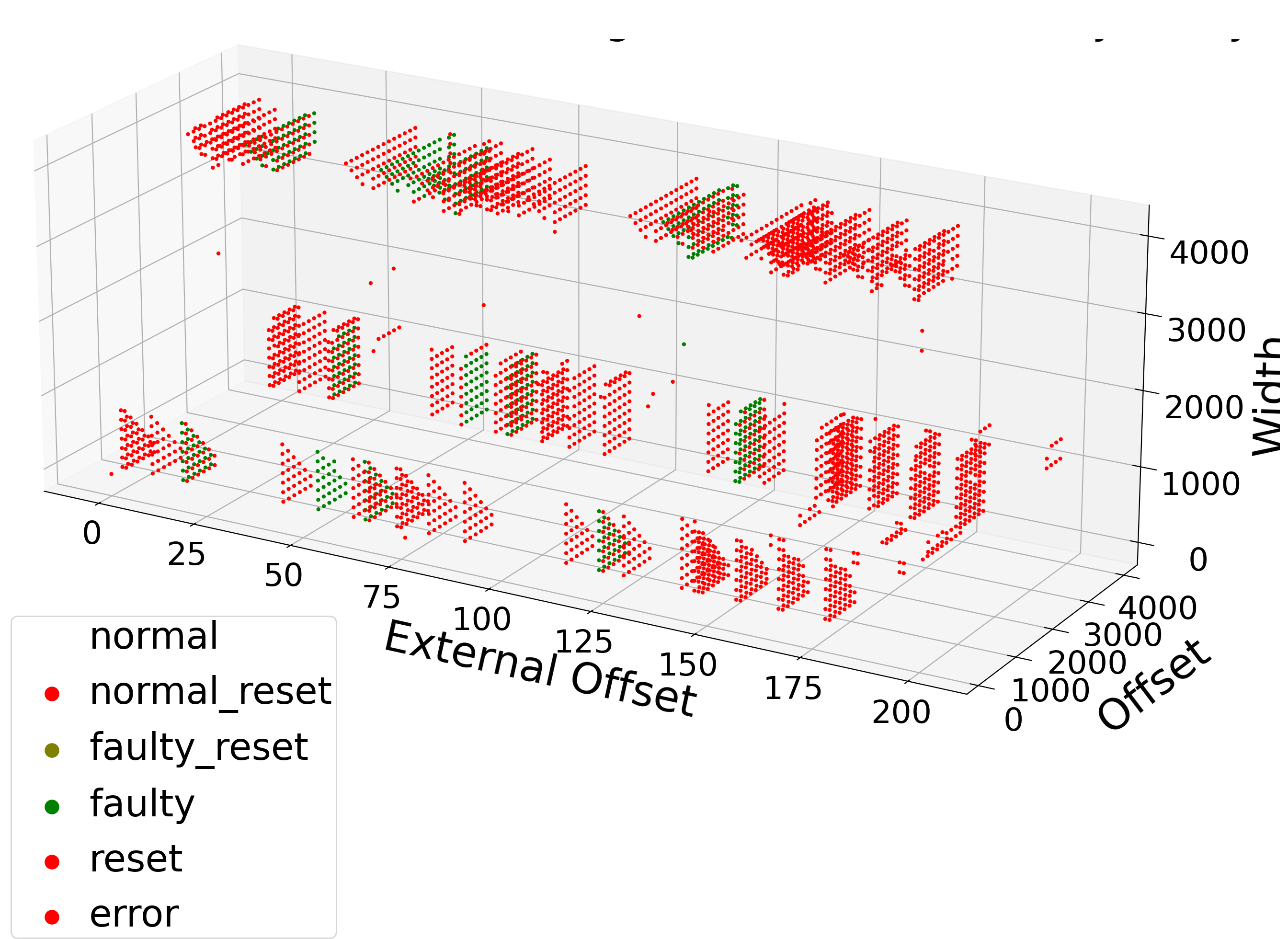}
    \caption{Clock glitching parameter space for Linear Q-Learning on greedy policy inference, for External Offsets up to 200 of the 4{,}562-cycle inference. A different pattern of faults and resets can be observed compared to the tabular implementation.}
    \label{fig:c1_linear_policy_distribution}
\end{figure}

\section{Implementation Overhead}
\label{sec:appendix_overhead}

Table~\ref{tab:resilient-overhead} shows the firmware size and cycle overhead of the Resilient Q-Learning countermeasures. With three copies, firmware size increases by 6.6\% and inference time by 2.7$\times$. The update rule takes 2.6$\times$ as long because it is computed for three copies. The post-update correction runs outside this measurement and adds up to 447 cycles (Table~\ref{tab:pipeline_resilient_learn}), which brings the learning step to about 3$\times$ the standard update rule. Q-table storage also triples, from 256 to 768 bytes.

\begin{table}[H]
\centering
\caption{Overhead of Resilient Q-Learning Implementation. Learning cycles cover the update rule only; the post-update correction is measured separately.}
\label{tab:resilient-overhead}
\begin{adjustbox}{max width=\linewidth}
\begin{tabular}{lccc}
\toprule
\textbf{Q-Table Copies} & \textbf{Firmware (Bytes)} & \textbf{Learning (Cycles)} & \textbf{Inference (Cycles)} \\
\midrule
1 (Standard)  & 19,676 & 749   & 376 \\
3 (Resilient) & 20,968 (\textbf{+6.6\%}) & 1,938 (\textbf{2.6$\times$}) & 1,029 (\textbf{2.7$\times$})\\
5 (Resilient) & 21,476 (\textbf{+9.1\%}) & 3,114 (\textbf{4.2$\times$}) & 1,611 (\textbf{4.3$\times$})\\
7 (Resilient) & 22,052 (\textbf{+12.1\%}) & 4,290 (\textbf{5.7$\times$}) & 2,182 (\textbf{5.8$\times$}) \\
\bottomrule
\end{tabular}
\end{adjustbox}
\end{table}

\section{Faulty value map and distributions}

\begin{figure}[t]
    \centering
    \includegraphics[width=0.8\linewidth]{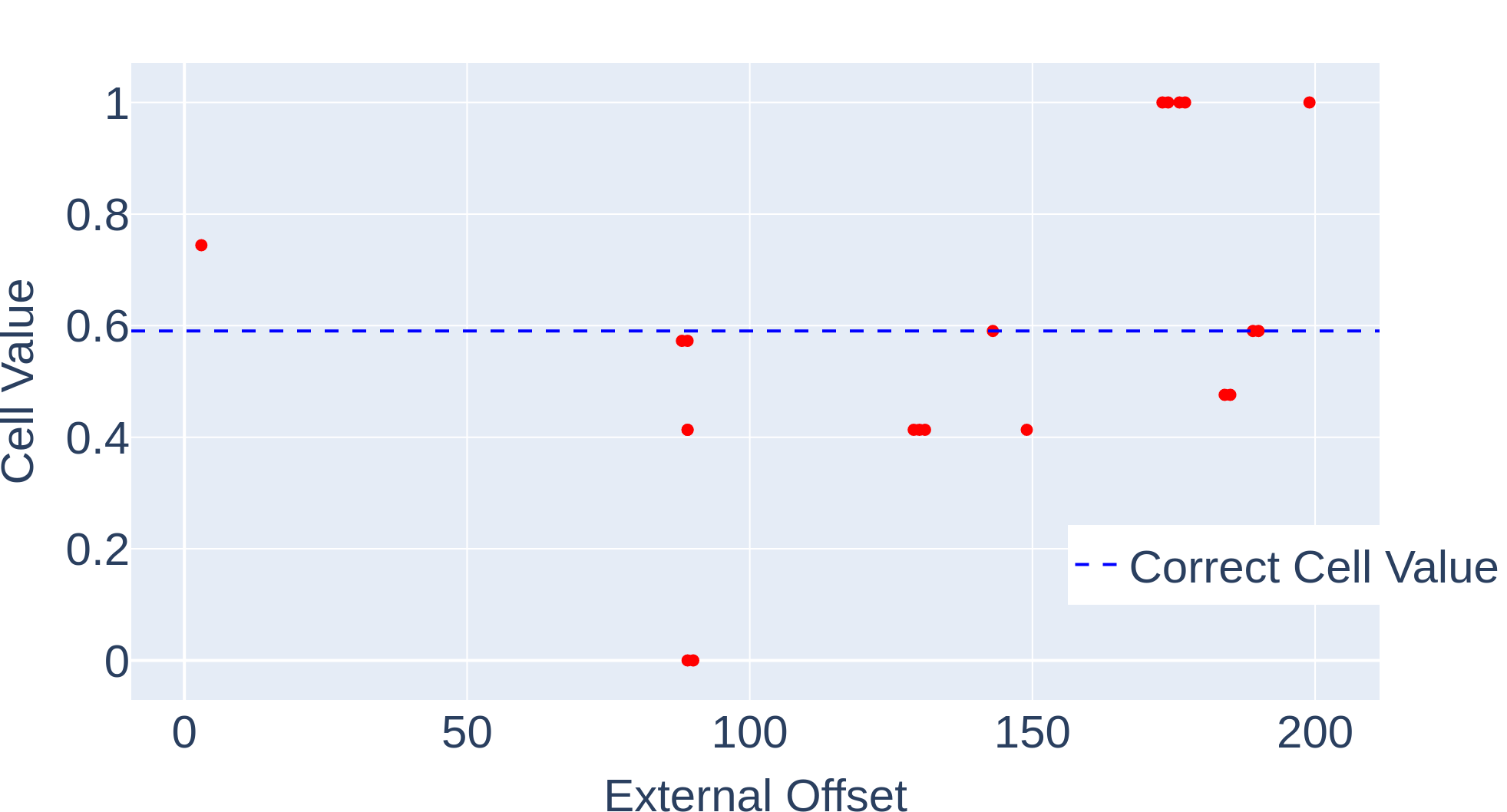}
    \caption{Clock glitching external offset and the corresponding fault output map for standard tabular Q-Learning on learning update. Since the result is reproducible, we can control the score of a given action by adjusting the external offset where we insert the clock glitch.}
    \label{fig:c1_learning_step_cell_values}
\end{figure}

\begin{figure}
    \centering
    \includegraphics[width=0.8\linewidth]{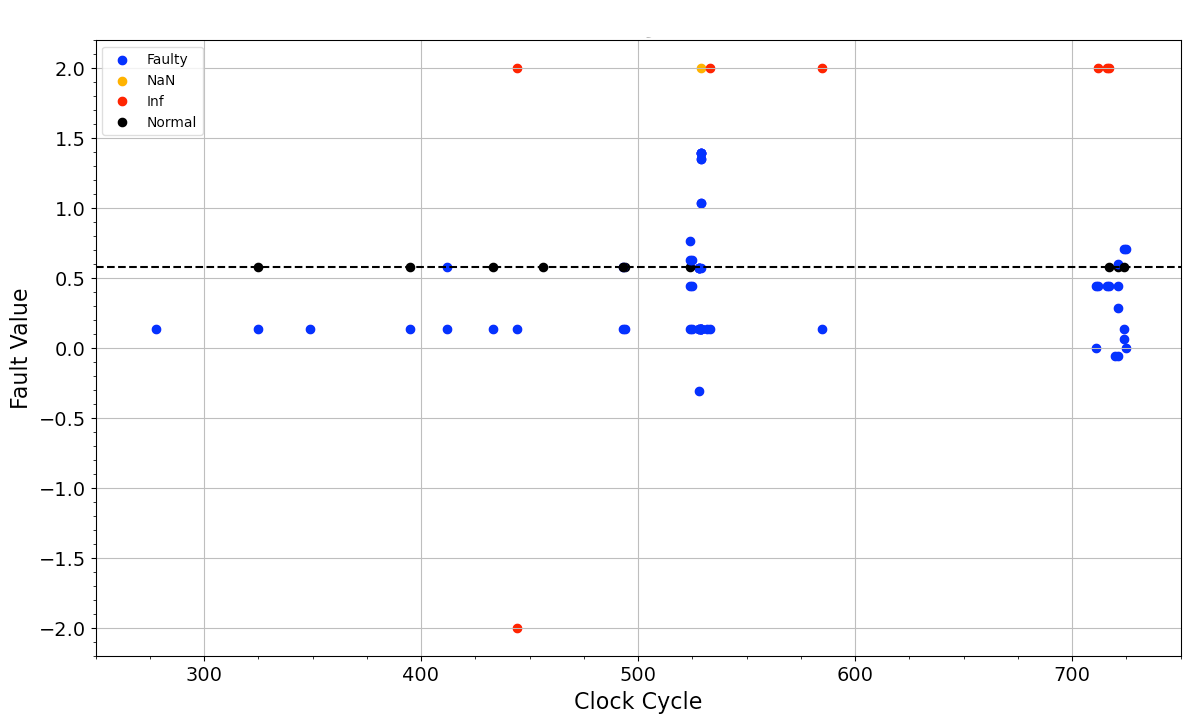}
    \caption{The faulty value introduced to the Q-Table and the corresponding clock cycle at which the voltage glitch was introduced, applied to the standard tabular Q-Learning on the update.}
    \label{fig:fl_qlearn_update_cc_distr}
\end{figure}

This appendix presents the faulty values introduced to the Q-Table for standard tabular Q-Learning on learning update at different clock cycles. Figure~\ref{fig:c1_learning_step_cell_values} shows the fault output map for the clock glitching, while Figure~\ref{fig:fl_qlearn_update_cc_distr} shows the faulty value distribution as a function of clock cycle for the voltage glitching. 

\clearpage
\section{Computational Pipeline Descriptions}
\label{sec:appendix_pipelines}

This appendix details the computational pipeline for each Q-Learning variant evaluated in this work. For each implementation, we describe the data flow through each computational step during learning and inference, including input/output sizes, data types, and where model parameters (Q-Table entries or weights) are read from or written to memory. Stored values are single-precision IEEE-754 floating point (FP32) unless otherwise noted; in the tabular implementations, the update arithmetic uses double-precision constants. The Frozen Lake environment uses 16 states and 4 actions.

These tables serve two purposes: (1)~they contextualize the External Offset values reported in the fault injection experiments by mapping the overall cycle budget to specific computational steps, and (2)~they identify which steps involve model parameter reads~(R) or writes~(W), highlighting where faults can cause persistent vs.\ transient corruption. Steps marked ``--'' in the R/W column do not access model parameters (Q-Table entries or weights), though they may access other data such as the reward array or random number generator.

\paragraph{Measurement methodology.} Cycle counts were measured on the target board described in Section~\ref{sec:exp}, clocked at 7.37~MHz, using the ChipWhisperer Husky's trigger-duration counter with the capture clock synchronised to the target clock, so that the reported count is exactly the number of CPU cycles for which the on-chip trigger was asserted. Per-step values were obtained by holding a single trigger from the start of the pipeline and releasing it after step~$k$, sweeping $k$ over the pipeline; the \textbf{Cycles} column is therefore \emph{cumulative} through each step, and successive differences give the cost of each individual step. Values are medians over 25 repetitions.

Two conventions apply throughout. First, each firmware asserts its trigger around a specific region, which does not always span the entire logical pipeline: steps that fall outside that region are marked with an asterisk~(*) and were measured on their own separate trigger, so they are \emph{standalone} per-step costs rather than part of the cumulative running total. Second, the \textbf{Total} row reports the native, uninstrumented measurement, which is the figure to cite. The step decomposition differs from it slightly in either direction: on the learning pipelines the step boundaries add $+0.7\%$ to $+3.5\%$, so the cumulative column ends marginally above the total, whereas on the inference pipelines the decomposition inlines calls that the native code makes through a function call or framework wrapper and so ends $3.7\%$ to $8.0\%$ below it. In both cases the relative cost of each step is unaffected. The first measured step in each table includes the trigger assertion and function prologue ($\approx$130~cycles), analogous to the ``Entering function'' rows in Tables~\ref{tab:pipeline_resilient_infer}, \ref{tab:pipeline_linear_learn} and~\ref{tab:pipeline_linear_infer}.

\subsection{Standard Q-Learning}

The standard Q-Table stores $16 \times 4 = 64$ FP32 values. The reward array stores $16$ FP32 values separately on the agent device. Table~\ref{tab:pipeline_qtable_learn} shows the learning pipeline and Table~\ref{tab:pipeline_qtable_infer} shows the inference pipeline.

\begin{table}[H]
\centering
\caption{Standard Q-Learning: Learning pipeline (749 cycles total, measured). Steps~3--7 are cumulative within the measured window; steps~1--2 (*) are measured on separate triggers.}
\begin{adjustbox}{max width=\linewidth}
\label{tab:pipeline_qtable_learn}
\small
\begin{tabular}{r|l|l|l|r|c}
\toprule
\textbf{\#} & \textbf{Step} & \textbf{Input} & \textbf{Output} & \textbf{Cycles} & \textbf{R/W} \\
\midrule
1 & Reward lookup & uint8\_t s & float & 128\rlap{*} & -- \\
2 & $\epsilon$-greedy & uint8\_t s, float eps & uint8\_t & 603\rlap{*} & R \\
3 & Read $Q(s,a)$ & uint8\_t s, uint8\_t a & float & 139 & R \\
4 & Compute $\max_{a'} Q(s',a')$ & uint8\_t s\_next & float, uint8\_t & 396 & R \\
5 & TD target: $r + \gamma \cdot \max$ & float r, float gamma, float max & float & 560 & -- \\
6 & Update: $(1{-}\alpha) Q + \alpha \cdot \text{target}$ & float alpha, float q, float target & float & 767 & -- \\
7 & Write $Q'(s,a)$ to table & float & float & 772 & W \\
\midrule
  & \textbf{Total (measured)} & & & \textbf{749} & \\
\bottomrule
\end{tabular}
\end{adjustbox}
\end{table}

\begin{table}[H]
\centering
\caption{Standard Q-Learning: Inference pipeline (376 cycles total, measured).}
\begin{adjustbox}{max width=\linewidth}
\label{tab:pipeline_qtable_infer}
\small
\begin{tabular}{r|l|l|l|r|c}
\toprule
\textbf{\#} & \textbf{Step} & \textbf{Input} & \textbf{Output} & \textbf{Cycles} & \textbf{R/W} \\
\midrule
1 & Read $Q(s, \cdot)$ & uint8\_t s & float & 131 & R \\
2 & Argmax comparison loop & float, uint8\_t & uint8\_t & 346 & R \\
3 & Return action & - & uint8\_t & 346 & -- \\
\midrule
  & \textbf{Total (measured)} & & & \textbf{376} & \\
\bottomrule
\end{tabular}
\end{adjustbox}
\end{table}

\noindent \textit{Notes:} Step~1 reads from the reward array (not the Q-Table), so faults here corrupt reward values without affecting stored Q-values. Step~2 reads Q-Table values when the greedy branch is taken. Faults in Steps~3--7 can corrupt the Q-Table persistently (Step~7 writes the updated value). During inference, faults in Step~2 (the FP32 comparison loop
) are the primary target; these produce transient action changes without corrupting stored values.

The measurements show that the $\epsilon$-greedy action selection (Step~2, 603~cycles) is comparable in cost to the entire update rule it precedes (749~cycles), because \texttt{random\_float()} performs a software floating-point division. The update-rule sweeps do not cover this window; action selection was targeted on its own trigger in Section~\ref{sec:fault_staandard_qlearning}, with External Offsets up to 396 of these 603 cycles.

\subsection{Robust Q-Learning}

Robust Q-Learning uses the same $16 \times 4$ Q-Table plus an additional $V$-array of 16~FP32 values. In the reward-based variant used in this work, $V(s) = \max_a Q(s,a)$ is computed for each state. The critical difference from standard Q-Learning is in the update rule: the Robust update considers $V$-array values across \emph{all} states (via the uncertainty set), rather than only considering the next state $s'$. This additional computation dominates the cost of a robust learning step: rebuilding the whole $V$-array (Step~4) takes 3{,}885~cycles, more than twice the 1{,}727~cycles of the update rule it feeds.

\begin{table}[H]
\centering
\caption{Robust Q-Learning: Learning pipeline (1{,}727 cycles total, measured). Steps~3 and~5--7 are cumulative within the measured window; step~4 (*) is measured on a separate trigger and dominates the overall cost.}
\begin{adjustbox}{max width=\linewidth}
\label{tab:pipeline_robust_learn}
\small
\begin{tabular}{r|l|l|l|r|c}
\toprule
\textbf{\#} & \textbf{Step} & \textbf{Input} & \textbf{Output} & \textbf{Cycles} & \textbf{R/W} \\
\midrule
1 & Reward lookup & uint8\_t s & float & 128\rlap{*} & -- \\
2 & $\epsilon$-greedy & uint8\_t s, float eps & uint8\_t & -- & R \\
3 & Read $Q(s,a)$ & uint8\_t s, uint8\_t a & float & 138 & R \\
4 & Compute $V(s)$ for all $s$ & float qtable{[}16{]}{[}4{]} & float vtable{[}16{]} & 3{,}885\rlap{*} & R, W \\
5 & Robust TD target (uses $\min V$) & float r, float R, float* vtable & float & 1{,}250 & R \\
6 & Update: $(1{-}\alpha) Q + \alpha \cdot \text{target}$ & float alpha, float q, float target & float & 1{,}535 & -- \\
7 & Write $Q'(s,a)$ to table & float & float & 1{,}739 & W \\
\midrule
  & \textbf{Total (measured)} & & & \textbf{1{,}727} & \\
\bottomrule
\end{tabular}
\end{adjustbox}
\end{table}

\begin{table}[H]
\centering
\caption{Robust Q-Learning: Inference pipeline (376 cycles total, measured; identical to standard).}
\begin{adjustbox}{max width=\linewidth}
\label{tab:pipeline_robust_infer}
\small
\begin{tabular}{r|l|l|l|r|c}
\toprule
\textbf{\#} & \textbf{Step} & \textbf{Input} & \textbf{Output} & \textbf{Cycles} & \textbf{R/W} \\
\midrule
1 & Read $Q(s, \cdot)$ & uint8\_t s & float & 131 & R \\
2 & Argmax comparison loop & float, uint8\_t & uint8\_t & 346 & R \\
3 & Return action & - & uint8\_t & 346 & -- \\
\midrule
  & \textbf{Total (measured)} & & & \textbf{376} & \\
\bottomrule
\end{tabular}
\end{adjustbox}
\end{table}

\noindent \textit{Notes:} The key difference from standard Q-Learning is Step~4: computing $V(s) = \max_a Q(s,a)$ for all 16 states requires reading all 64 Q-Table entries and rewriting all 16 $V$ entries. In this implementation the $V$-array is rebuilt wholesale on every learning step rather than being incrementally updated for the visited state, which is why Step~4 alone costs more than twice the update rule. The Robust update rule (Step~5) then uses $V$ values across states through the minimization in Equation~\ref{eq:robust_update}. Because $V(s)$ takes the maximum over actions, a single $-Inf$ cell does not change $V(s)$ unless every action of that state is $-Inf$. If a state value does become $-Inf$, every subsequent update uses it regardless of the state being updated, whereas in standard Q-Learning corruption only spreads when the agent visits the affected state. Step~4 runs outside the trigger window of the update experiments, so faults in it were not evaluated. The inference pipeline is identical to standard Q-Learning, producing identical vulnerability.

\subsection{Resilient Q-Learning}

Resilient Q-Learning maintains three independent copies of the Q-Table ($3 \times 16 \times 4 = 192$ FP32 values). Majority voting is applied during both the learning update (to correct faulted writes) and inference (to tolerate corrupted reads).

\begin{table}[H]
\centering
\caption{Resilient Q-Learning: Learning pipeline (1{,}938 cycles total, measured). Steps~3--6 are cumulative within the measured window; step~7 (*) is the correction pass, measured on a separate trigger.}
\begin{adjustbox}{max width=\linewidth}
\label{tab:pipeline_resilient_learn}
\small
\begin{tabular}{r|l|l|l|r|c}
\toprule
\textbf{\#} & \textbf{Step} & \textbf{Input} & \textbf{Output} & \textbf{Cycles} & \textbf{R/W (Table \#)} \\
\midrule
1 & Reward lookup & uint8\_t s & float & -- & -- \\
2 & $\epsilon$-greedy (3 tables) & uint8\_t s, float eps & uint8\_t & -- & R (0,1,2) \\
3 & Read $Q_k(s,a)$, $Q_k(s',\cdot)$ & uint8\_t s, uint8\_t a & float{[}3{]} & 140 & R (0,1,2) \\
4 & Compute $\max_{a'}$ per copy & uint8\_t s\_next, uint8\_t j & float{[}3{]} & 932 & R (0,1,2) \\
5 & TD target + update per copy & float r, float{[}3{]}, float{[}3{]} & float{[}3{]} & 1{,}987 & -- \\
6 & Write $Q'_k(s,a)$, $k{=}0{,}1{,}2$ & float{[}3{]} & float{[}3{]} & 2{,}005 & W (0,1,2) \\
7 & Majority vote + correction & uint8\_t s, uint8\_t a & uint8\_t & 447\rlap{*} & R, W (0,1,2) \\
\midrule
  & \textbf{Total (measured)} & & & \textbf{1{,}938} & \\
\bottomrule
\end{tabular}
\end{adjustbox}
\end{table}

\begin{table}[H]
\centering
\caption{Resilient Q-Learning: Inference pipeline (1{,}029$\pm$3 cycles total, measured).}
\begin{adjustbox}{max width=\linewidth}
\label{tab:pipeline_resilient_infer}
\small
\begin{tabular}{r|l|l|l|r|c}
\toprule
\textbf{\#} & \textbf{Step} & \textbf{Input} & \textbf{Output} & \textbf{Cycles} & \textbf{R/W (Table \#)} \\
\midrule
1 & Entering function & uint8\_t s & - & 145 & - \\
2 & Initialize best actions array & - & uint8\_t* & 153 & - \\
3 & Calculate best\_action{[}0{]} & uint8\_t, uint8\_t & uint8\_t & 405-407 & R (0) \\
4 & Calculate best\_action{[}1{]} & uint8\_t, uint8\_t & uint8\_t & 667-671 & R (1) \\
5 & Calculate best\_action{[}2{]} & uint8\_t, uint8\_t & uint8\_t & 923-929 & R (2) \\
6 & Calculate majority vote & uint8\_t* & uint8\_t & 1003-1009 & - \\
7 & Exiting function & - & uint8\_t & 1026-1032 & - \\
\midrule
  & \textbf{Total (measured)} & & & \textbf{1{,}029$\pm$3} & \\
\bottomrule
\end{tabular}
\end{adjustbox}
\end{table}

\noindent \textit{Notes:} The steps in Table~\ref{tab:pipeline_resilient_learn} group each operation across the copies so that it can be measured; the attack firmware instead computes and writes each copy in turn, so the first copy is written well before cycle 2{,}005. Step~7 of the learning pipeline (majority voting on the updated cell) is the fault correction mechanism, costing up to 447~cycles including trigger overhead. It reads all three copies of the just-updated cell and overwrites any minority value with the majority. This stops NaN/Inf propagation \emph{if} the fault affects only one copy. However, faulting the majority voting code itself can trigger unnecessary resets (setting a cell to 0 across all copies). Note that the correction pass runs outside the trigger window used for the update-rule experiments, and is targeted by a separate trigger. During inference, voltage glitching of the combined greedy policy and voting code still finds some successful faults; we did not determine which step each fault affects.

\subsection{Linear Q-Learning (16 Features)}

Linear Q-Learning replaces the Q-Table with a weight matrix of $4 \times 16 = 64$ FP32 weights (one weight vector per action). The state is represented as a feature vector $\mathbf{f}(s) \in \mathbb{R}^{16}$. For the 16-feature implementation, $\mathbf{f}(s)$ is the one-hot encoding of the state. Q-values are computed as $Q(s,a) = \sum_{i=0}^{15} w^a_i f_i(s)$. Unlike the tabular implementations where rewards are stored on the agent device, function approximation implementations receive state features and rewards from the host computer via serial.

\begin{table}[H]
\centering
\caption{Linear Q-Learning (16 features): Learning pipeline (8487 cycles total, measured).}
\begin{adjustbox}{max width=\linewidth}
\label{tab:pipeline_linear_learn}
\small
\begin{tabular}{r|l|l|l|r|c}
\toprule
\textbf{\#} & \textbf{Step} & \textbf{Input} & \textbf{Output} & \textbf{Cycles} & \textbf{R/W} \\
\midrule
1 & Entering function & float* f, uint8\_t a, float r, float* f\_next & - & 161 & - \\
2 & Calculate Q(s', 0) & float* f, uint8\_t a & float & 1206 & R \\
3 & Calculate Q(s', 1) & float* f, uint8\_t a & float & 2323 & R \\
4 & Calculate Q(s', 2) & float* f, uint8\_t a & float & 3445 & R \\
5 & Calculate Q(s', 3) & float* f, uint8\_t a & float & 4543 & R \\
6 & calculate td\_target & float r, float GAMMA, float best\_q & float & 4725 & - \\
7 & calculate tdiff & float td\_target, float Q\_hat(f, a) & float & 5797 & R \\
8 & loop through weights and update & float ALPHA,   float tdiff, float f{[}j{]} & float & 8484 & R+W \\
9 & Exiting function & - & - & 8487 & - \\
\midrule
  & \textbf{Total (measured)} & & & \textbf{8487} & \\
\bottomrule
\end{tabular}
\end{adjustbox}
\end{table}

\begin{table}[H]
\centering
\caption{Linear Q-Learning (16 features): Inference pipeline ($4562$ cycles total, measured).}
\begin{adjustbox}{max width=\linewidth}
\label{tab:pipeline_linear_infer}
\small
\begin{tabular}{r|l|l|l|r|c}
\toprule
\textbf{\#} & \textbf{Step} & \textbf{Input} & \textbf{Output} & \textbf{Cycles} & \textbf{R/W} \\
\midrule
1 & Entering function & float* & - & 152 & - \\
2 & Initialize best action & - & uint8\_t & 152 & - \\
3 & Calculate Q(f, 0) & float*, uint8\_t & float & 1195 & R \\
4 & Calculate Q(f, 1) & float*, uint8\_t & float, uint8\_t & 2310 & R \\
5 & Calculate Q(f, 2) & float*, uint8\_t & float, uint8\_t & 3432 & R \\
6 & Calculate Q(f, 3) & float*, uint8\_t & float, uint8\_t & 4527 & R \\
7 & Exit loop & float*, uint8\_t & float, uint8\_t & 4547 & R \\
8 & Exiting function & - & uint8\_t & 4562 & - \\
\midrule
  & \textbf{Total (measured)} & & &\textbf{4562} & \\
\bottomrule
\end{tabular}
\end{adjustbox}
\end{table}

\noindent \textit{Notes:} Unlike the tabular implementation where Q-value lookup is a single memory read, Linear Q-Learning computes each Q-value as a 16-element dot product (Steps~3--6 of Table~\ref{tab:pipeline_linear_infer}). Each dot product involves 16 floating-point multiply-accumulate operations, making the execution time per action longer than a simple table lookup. As noted in Section~\ref{sec:linear_16}, the timing of this implementation is not constant because it depends on the state, since different feature values lead to different floating-point operands with potentially different execution times on the ARM software FP library. As shown in Figure~\ref{fig:fault_consistency_ext_offset}, consistent faults cluster in the first $\sim$100 cycles, before the first dot product in Step~3, while later faults targeting Steps~3--6 are less consistent, possibly because their cumulative timing offset varies with the state.

Faults during Step~8 of the learning pipeline write corrupted weights to memory; due to weight sharing across states, a single corrupted weight affects $Q(s,a)$ for \emph{all} states where $f_i(s) \neq 0$. For the 16-feature one-hot encoding, each weight maps to exactly one state, so a fault affects only one (state, action) pair; for the 8-feature and 4-feature variants with shared features, the impact is broader (see Table~\ref{tab:recovery_comparison}).

\subsection{Deep Q-Learning}

The Deep Q-Network uses a fully connected neural network with architecture: input (16) $\rightarrow$ Dense($H$) + ReLU $\rightarrow$ Dense(4) + Linear, where $H = 16$ is the hidden layer size. The network is implemented using the AIfES embedded ML framework~\cite{wulfert2024aifes}. The total parameter count is $16H + H$ (hidden layer weights and biases) $+ 4H + 4$ (output layer weights and biases) $= 21H + 4 = 340$. The table below covers inference, where the network is pre-trained and frozen; training-time faults are described in Section~\ref{sec:dqn_fault_results}.

\begin{table}[H]
\centering
\caption{Deep Q-Learning: Inference pipeline (32{,}182 cycles total, measured).}
\begin{adjustbox}{max width=\linewidth}
\label{tab:pipeline_dqn_infer}
\small
\begin{tabular}{r|l|l|l|r|c}
\toprule
\textbf{\#} & \textbf{Step} & \textbf{Input} & \textbf{Output} & \textbf{Cycles} & \textbf{R/W} \\
\midrule
1 & One-hot encode $s$ & uint8\_t s & float* & 222 & -- \\
2 & Dense 1: $\mathbf{z} = W_1 \mathbf{x} + \mathbf{b}_1$ & float* x, aimodel\_t* policy & float* & 21{,}191 & R \\
3 & ReLU: $\mathbf{h} = \max(0, \mathbf{z})$ & float* z & float* & 23{,}520 & -- \\
4 & Dense 2: $\mathbf{q} = W_2 \mathbf{h} + \mathbf{b}_2$ & float* h, aimodel\_t* policy & float* & 30{,}732 & R \\
5 & Argmax over Q-values & float* & uint8\_t & 30{,}998 & -- \\
6 & Return action & - & uint8\_t & 30{,}998 & -- \\
\midrule
  & \textbf{Total (measured)} & & & \textbf{32{,}182} & \\
\bottomrule
\end{tabular}
\end{adjustbox}
\end{table}

Weight reads occur in Steps~2 and~4. Since the network is frozen during the inference experiments, these faults produce transient errors without updating the stored weights. The attack command asserts its trigger after the one-hot encoding, whereas Step~1 here includes it, so External Offsets do not map exactly onto this column. Fault counts are given in Section~\ref{sec:dqn_fault_results}.

The per-step measurements show that the first dense layer dominates inference, accounting for 20{,}969 of the 32{,}182 cycles (65\%), compared with 7{,}212 cycles for the second dense layer, which is consistent with the first layer having more output neurons (16 versus 4).

\end{document}